\documentclass[12pt]{article}

\usepackage{amsmath, cite, graphicx}
\usepackage{multirow}
\usepackage{amsfonts}
\usepackage{amssymb}
\usepackage{amscd}
\usepackage{cite}
\usepackage{amsmath}
\usepackage{bbm}
\def\hybrid{\topmargin -20pt    \oddsidemargin 0pt
        \headheight 0pt \headsep 0pt
        \textwidth 6.25in       % A4 paper
        \textheight 9.5in       % A4 paper
        \marginparwidth .875in
        \parskip 5pt plus 1pt   \jot = 1.5ex}
\hybrid

\numberwithin{equation}{section}
\numberwithin{table}{section}

\newcommand{\be}{\begin{equation}}
\newcommand{\ee}{\end{equation}}

\newcommand{\bea}{\begin{eqnarray}}
\newcommand{\eea}{\end{eqnarray}}

\renewcommand{\Re}{\operatorname{Re}}
\renewcommand{\Im}{\operatorname{Im}}

\newcommand\e{\mathrm{e}}
\newcommand\iu{\operatorname{i}}
\newcommand\diff{\mathrm{d}}
\newcommand\vol{\operatorname{vol}}

\newcommand{\ba}{\begin{eqnarray}}
\newcommand{\ea}{\end{eqnarray}}
\newcommand{\ban}{\begin{eqnarray}}
\newcommand{\ean}{\end{eqnarray}}
\newcommand{\IZ}{\mathbb{Z}}

\begin{document}

%%%%%%%%%%%%%%%%%%%%%%%%%%%%%%%%%%%%%%%%%%
\begin{titlepage}
%%%%%%%%%%%%%%%%%%%%%%%%%%%%%%%%%%%%%%%%%%

\begin{center}

\rightline{\small IPhT-t13/005}
\rightline{\small LPTENS 13/01}

\vskip 2cm

{\Large \bf Enhanced supersymmetry from vanishing Euler number}

\vskip 1.7cm

{\bf Amir-Kian Kashani-Poor$^{a}$, Ruben Minasian$^{b}$ and Hagen Triendl$^{b}$}\\

\vskip 1.0cm

{}$^{a}${\em Laboratoire de Physique Th\'eorique de l'\'Ecole Normale Sup\'erieure, \\
24 rue Lhomond, 75231 Paris, France}

\vskip 0.7cm

{}$^{b}${\em Institut de Physique Th\'eorique, CEA Saclay,\\
Orme de Merisiers, F-91191 Gif-sur-Yvette, France}

\vskip 1.6cm

{\tt kashani@lpt.ens.fr, ruben.minasian@cea.fr, hagen.triendl@cea.fr} \\

\end{center}

\vskip 2cm

\begin{center} {\bf ABSTRACT } \end{center}

We argue that compactifications on  Calabi-Yau threefolds  with vanishing Euler number yield effective four dimensional theories exhibiting (spontaneously broken) $N=4$ supersymmetry. To this end, we derive the low-energy effective action for general $SU(2)$ structure manifolds in type IIA string theory and show its consistency with gauged $N=4$ supergravity.  Focusing on the special case of Calabi-Yau manifolds with vanishing Euler number, we explain the absence of perturbative corrections at the two-derivative level. In addition, we conjecture that all non-perturbative corrections are governed and constrained by the couplings of $N=4$ massive gravitino multiplets.

\vfill

January 2013

%%%%%%%%%%%%%%%%%%%%%%%%%%%%%%%%%%%%%%%%%
\end{titlepage}
%%%%%%%%%%%%%%%%%%%%%%%%%%%%%%%%%%%%%%%%%

\tableofcontents

%%%%%%%%%%%%%%%%%%%%%%%%%%
\section{Introduction}
%%%%%%%%%%%%%%%%%%%%%%%%%%

Dimensional reductions of string theory on compact manifolds produce the most promising candidates for gravitational theories with UV completions in dimensions lower than ten. While many such backgrounds are well understood at the classical level, a complete understanding of quantum string corrections remains a challenging task. Often, supersymmetry provides a powerful tool for controlling these corrections, at least at the two-derivative level.

In this paper, we focus on a class of theories which, while exhibiting an $N=2$ vacuum, admit further (spontaneously broken) supersymmetries. Since these provide non-renormalization theorems for some of the quantities in the off-shell supersymmetric action, they can help to constrain quantum corrections. Upon supersymmetry breaking, masses generated by the super-Higgs mechanism give rise to additional quantum corrections. However, they also are constrained by the spontaneously broken supersymmetry.

$G$-structure manifolds which admit nowhere vanishing spinors (see e.g.\ \cite{Hitchin, Hitchin:2001rw, CS, waldram, Kaste:2003dh, Gauntlett:2003cy}) are natural candidates for internal spaces on which to reduce the supergravity action to four dimensions in a supersymmetric fashion.\footnote{A priori, a $G$-structure does not necessarily refer to the reduction of the structure group of the frame bundle, but can be defined with regard to a larger bundle, such as  the generalized tangent bundle, cf. \cite{Hitchin:2004ut, Gualtieri:2003dx}.} It has been argued in \cite{ Gurrieri:2002wz, Grana:2005ny}
that such reductions can yield supersymmetric effective theories which do not necessarily possess supersymmetric vacua.\footnote{For conditions for the existence of four-dimensional supersymmetric vacua, see \cite{Grana:2004bg, Witt, Jeschek:2004wy, Grana:2005sn}.}
We want to illustrate in this work how such off-shell supersymmetries can help in understanding quantum corrections.

In the following, our main focus will be on compactifications on Calabi-Yau threefolds, an extensively studied class of backgrounds in string theory. These manifolds have holonomy group $SU(3)$ and hence permit a covariantly constant spinor $\eta$, as opposed to a spinor which is merely nowhere vanishing. Compactifications of type II string theory on Calabi-Yau threefolds give rise to $N=2$ supergravity in four dimensions. The general form of perturbative and worldsheet instanton corrections is completely known at the two-derivative level. While the worldsheet instanton corrections are computed by Gromov-Witten invariants \cite{Candelas:1990rm}, the only perturbative $\alpha'$ correction arises at cubic order and corrects the cubic holomorphic prepotential of the complexified K\"ahler structure $B+ \iu J$ by a constant proportional to the Euler number of the Calabi-Yau threefold. Similarly, the hypermultiplet sector gets corrected at one loop string order by a contribution which is also proportional to the Euler number \cite{ Antoniadis:1997eg, Antoniadis:2003sw}.
This means in particular that if the Euler number vanishes, there are no perturbative corrections at the two-derivative level. The vanishing of these corrections suggests the presence of an additional symmetry. Indeed, we will argue that this additional symmetry is a consequence of the Hopf theorem. It states that any manifold with vanishing Euler number admits a nowhere vanishing vector field $\hat v$ (and vice versa).\footnote{In fact, the nowhere vanishing vector is complex, as can be seen by using the non-degenerate complex structure of the Calabi-Yau, see the discussion at the beginning of Section~\ref{sec:CY}. More generally, the vanishing Euler number on any compact six-dimensional manifold is a necessary and sufficient condition for the existence of a pair of vectors that are linearly independent at any point of the manifold \cite{ethomas}.} Applied to Calabi-Yau threefolds, this implies that we can define a second nowhere vanishing spinor $(\hat v^m \Gamma_m) \eta$ that is everywhere linearly independent of $\eta$. Therefore, Calabi-Yau threefolds of vanishing Euler number exhibit $SU(2)$ structure, in addition to $SU(3)$ holonomy. One might therefore expect that compactification on such distinguished Calabi-Yau manifolds yields a four-dimensional action admitting twice the conventional number of supercharges, albeit off-shell.

$SU(2)$ structures and reductions on $SU(2)$ structure manifolds have been discussed in \cite{Bovy:2005qq, ReidEdwards:2008rd, Lust:2009zb, Triendl:2009ap, Louis:2009dq, Danckaert:2011ju, Schulz:2012uj}. In particular, it was shown that the dimensional reduction of the type II string action on an $SU(2)$ structure manifold gives rise to $N=4$ gauged supergravity in four dimensions. These reductions take subclasses of $SU(2)$ structure manifolds as a starting point, for which certain torsion classes are absent \cite{ReidEdwards:2008rd,Louis:2009dq,Danckaert:2011ju}. As we will explain below, Calabi-Yau threefolds of vanishing Euler number always lie outside of these subclasses, hence require a more general treatment of $SU(2)$ structure backgrounds. In this work, we will perform a reduction of the type IIA action on manifolds of general $SU(2)$ structure. The resulting gauged $N=4$ supergravity will admit $N=2$ Minkowski vacua, where half of the supersymmetries are spontaneously broken. The effective action with eight unbroken supercharges around such vacua yields the $N=2$ supergravity of a standard Calabi-Yau reduction in type IIA.

 As $N=4$ gauged supergravity does not permit perturbative corrections at the two-derivative level, this would explain their absence for Calabi-Yau threefolds with vanishing Euler number. Furthermore, even non-perturbative corrections at the two-derivative level are restricted by $N=4$ supersymmetry. One can thus expect simplifications even at the non-perturbative level when compactifying on such Calabi-Yau manifolds, perhaps in the form of relations among their Gromov-Witten invariants. In principle, all ten-dimensional modes could be rewritten in terms of $N=4$ massive multiplets in four dimensions. The quantum corrections coming from these modes would then all have to fit into the framework of $N=4$ gauged supergravity, yielding further constraints on quantum corrections for Calabi-Yau threefold backgrounds with vanishing Euler number.

\begin{figure}
\centering
\includegraphics[width=7cm]{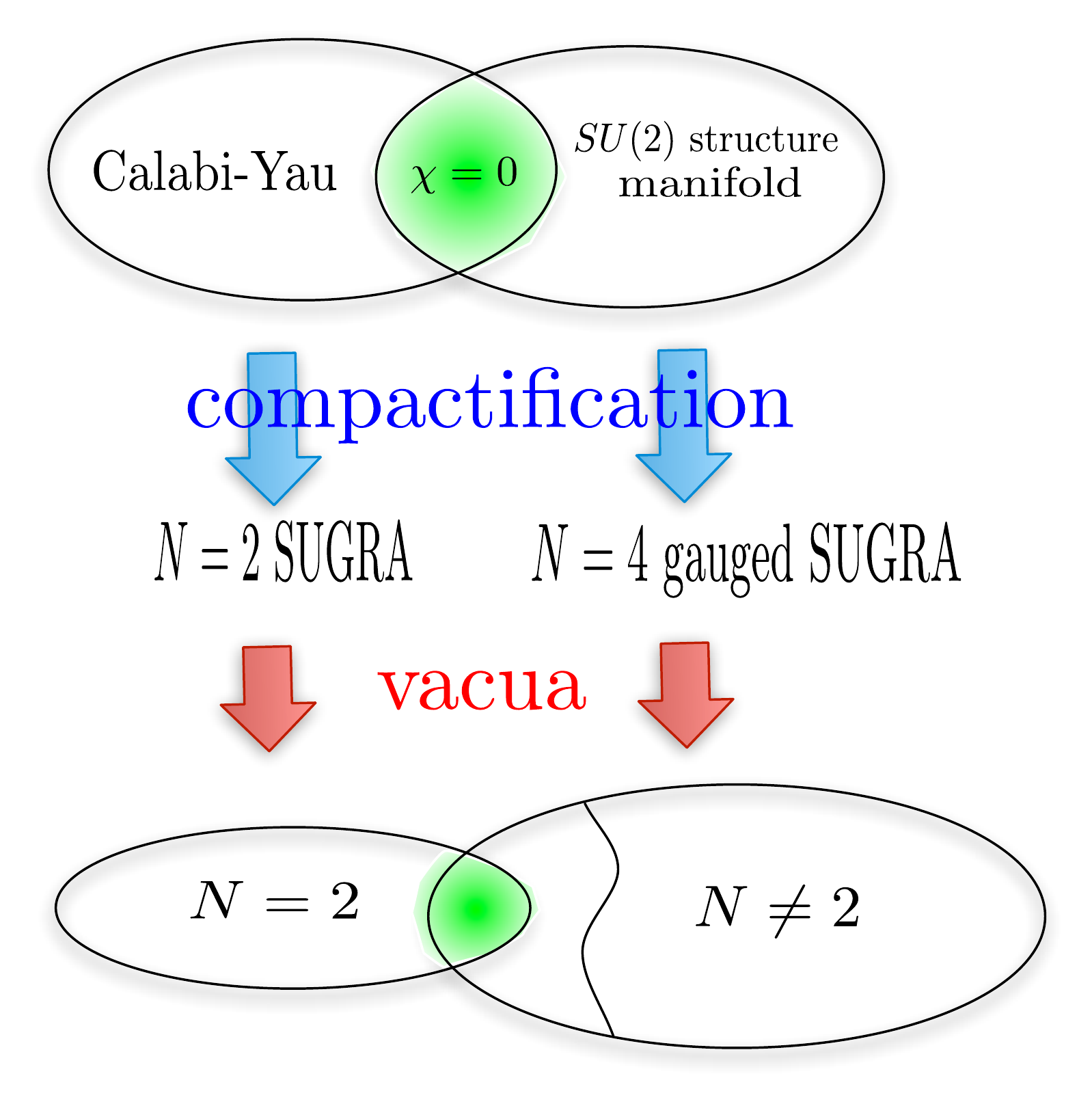}
\caption{\small{Compactifications on Calabi-Yau manifolds with vanishing Euler number yield $N=4$ gauged supergravities with $N=2$ vacua.}}
\end{figure}

This work is organized as follows. In Section~\ref{sec:basics}, we will review basic facts concerning $SU(2)$ structures. Section~\ref{sec:reduction} contains the main work on the $SU(2)$-structure reduction of the type IIA action to four dimensions, yielding an $N=4$ gauged supergravity.  Section~\ref{sec:cons_trunc} might be of particular interest, where we discuss the truncation ansatz.
In Section~\ref{sec:supergravity}, we identify the $N=4$ gauged supergravity in terms of its gauging parameters, discuss the four-dimensional gauge group and identify supersymmetric vacua.
Section~\ref{sec:CYEulerzero} contains the discussion of Calabi-Yau manifolds of Euler number zero. The corresponding spontaneous breaking of supersymmetry to $N=2$ is discussed in Section~\ref{sec:4to2}. In Section~\ref{sec:conclusions}, we collect and further discuss our results. Details of the reduction have been assembled in the rather technical Appendix~\ref{app:Ricci}, and the derivation of the gravitino mass matrix is performed in Appendix~\ref{sec:SUSYvar}.

%%%%%%%%%%%%%%%%%%%%%%%%%%
\section{Basics on $SU(2)$ structure manifolds}
%%%%%%%%%%%%%%%%%%%%%%%%%%
\label{sec:basics}
Let us start by introducing the concept of an $SU(2)$ structure on a six-dimensional manifold $Y$. This section is meant as a brief review of the basic properties of $SU(2)$ structures. For more information, see for instance \cite{Bovy:2005qq, ReidEdwards:2008rd, Lust:2009zb, Triendl:2009ap}.

A six-dimensional $SU(2)$ structure manifold $Y$ admits a pair of nowhere vanishing spinors $\eta_i$, $i=1,2$, whose norm we fix by imposing
\begin{equation}\label{eq:spinor_norm}
\bar \eta^i \eta_j = \delta^i_j\ .
\end{equation}
Based on these, one can introduce a $SU(2)$ triple of real two-forms $J^a$, $a=1,2,3$, and a holomorphic one-form $K$, via
\begin{equation} \label{eq:def_JaK}
 J^a = \sqrt{\tfrac{3}{2}} \iu(\sigma^a)^{ i}_j \bar \eta^j_+ \gamma_{mn} \eta_{+\, i} \diff x^m \wedge \diff x^n \ , \qquad K = - \epsilon^{ij} \bar \eta_{-\, i} \gamma_m \eta_{+\, j} \diff x^m \ .
\end{equation}
We have here introduced the notation
\begin{equation}
\eta_{+\,i} = \eta_i \,, \quad \eta_{-\,i} = \eta_i^c \ ,
\end{equation}
where the superscript $c$ indicates charge conjugation.
The Fierz identities for these spinors can now  be parametrized as follows:
\begin{equation} \label{spinor_bilinears} \begin{aligned}
(\Psi_{+})^j_{i} \equiv  \eta_{+\, i} \otimes \bar \eta_+^j  = &  \tfrac18 (1 + \tfrac12 (K \wedge \bar K)_{mn} \gamma^{mn})\left(\delta_i^j  (1 - \gamma_{(4)}) +  \sqrt{\tfrac{2}{3}}\iu(\sigma_a)_i^j J^a_{pq} \gamma^{pq} \right) \ , \\
(\Psi_{-})^{i}_j \equiv  \eta_{-}^i \otimes \bar \eta_{-\, j}  = &  \tfrac18 (1 - \tfrac12 (K \wedge \bar K)_{mn} \gamma^{mn})\left(\delta_j^i  (1 - \gamma_{(4)}) -  \sqrt{\tfrac{2}{3}}\iu (\sigma_a)_j^i J^a_{pq} \gamma^{pq} \right) \ , \\
(\Psi_{0})_{ij} \equiv   \eta_{+\, i} \otimes \bar \eta_{-\, j} = & \tfrac18 K_m \gamma^m \left(\epsilon_{ij} (1 - \gamma_{(4)}) + \sqrt{\tfrac{2}{3}}\iu(\sigma_a)_{ij} J^a_{pq} \gamma^{pq}\right) \ , \\
(\bar \Psi_{0})^{ij} \equiv   \eta_{-}^{i} \otimes \bar \eta_{+}^j = & \tfrac18 \bar K_m \gamma^m \left(\epsilon^{ij} (1 - \gamma_{(4)}) -  \sqrt{\tfrac{2}{3}}\iu(\sigma_a)^{ij} J^a_{pq} \gamma^{pq}\right) \ ,
\end{aligned}\end{equation}
where $(\sigma_a)_{ij} = (\sigma_a)_i^k\epsilon_{kj}$ and $\gamma_{(4)}= \vol_{4\, pqrs} \gamma^{pqrs}$. Taking the products of these bilinears and using \eqref{eq:spinor_norm} yields the relations
\begin{equation}\label{eq:JaJb}
J^a \wedge J^b = \delta^{ab} \vol_4
\end{equation}
and
\begin{equation}\label{eq:K_compatible}
K \cdot K =0 \ , \qquad \bar{K} \cdot K = 2 \ , \qquad \iota_K J^a =0 \ .
\end{equation}
Alternatively, one can define an $SU(2)$-structure with no reference to spinors, by specifying an $SU(2)$ triple of real two-forms $J^a$, $a=1,2,3$, and a holomorphic one-form $K$ which satisfy the conditions \eqref{eq:JaJb} and \eqref{eq:K_compatible}.

The existence of the one-form $K$ permits the introduction of an almost product structure $P: TY \rightarrow TY$ on the manifold, defined locally via
\begin{equation}
\label{eq:almost_product_structure}
{P_m}^n = K_m \bar{K}^n + \bar{K}_m K^n - \delta_m^{\phantom{m}n} \ .
\end{equation}
The eigenspaces $T_2 Y$ and $T_4 Y$ of $P$ to the eigenvalues $+1$ and $-1$ respectively yield a global decomposition of the tangent space,
\begin{equation}\label{eq:tangent_space_splitting}
T Y = T_2 Y \oplus T_4 Y \ .
\end{equation}
The subbundle $T_2 Y$ is trivial, spanned by $K^1=\Re K$ and $K^2=\Im K$.

Let us now discuss the frame bundle over spacetime $\times$ $Y$. We choose a section (vielbein)\footnote{The first two equations in \eqref{eq:K_compatible} imply that the $K^i$ can be chosen as components of the vielbein.}
\begin{equation}\label{vielbein}
 e^A = (e^\mu, K^i, e^\alpha) \ ,
\end{equation}
where the $e^\mu$ live in spacetime and depend only on spacetime coordinates. In contrast, the $K^i = k^i_j (v^j + G^j)$ consist of a one-form $v^i$ in $T^*_2$ and two spacetime gauge fields $G^i$ (the Kaluza-Klein vectors) that parameterize the fibration of $T^*_2$ over spacetime, as well as the coefficient $k^i_j$, which is a spacetime scalar.\footnote{Due to the mixed spacetime/internal components of the ten-dimensional metric, the components $K^i$ of the vielbein are not purely internal. We will nevertheless retain the same nomenclature as in (\ref{eq:def_JaK}) for simplicity.} Furthermore, the $e^\alpha$ are one-forms on $T^*_4$ such that
\begin{equation}
 J^a = \tfrac12 I^a_{\alpha \beta} e^\alpha \wedge e^\beta \ ,
\end{equation}
with constant coefficients $I^a_{\alpha \beta}$ that span the $SU(2)$ algebra of complex structures on the frame bundle, i.e.\
\begin{equation}
 (I^a)^{\alpha}_{\gamma} (I^b)^{\gamma}_{\beta} = \epsilon^{abc} (I^c)^{\alpha}_{\beta} - \delta^{ab} \delta^{\alpha}_{\beta} \ .
\end{equation}
The dual vielbein to \eqref{vielbein} is
\begin{equation}
 \hat e_A = (\hat e_\mu, \hat K_i, \hat e_\alpha) = (\partial_\mu- G_\mu^i \hat v_i, (k^{-1})^j_i \hat v_j, \hat e_\alpha)\ ,
\end{equation}
where $\hat v_i$ is the vector field dual to the vielbein component $v^i$.

Next, we consider the Levi-Civita connection one-form $\Omega$, which is the unique torsion-free connection satisfying the Maurer-Cartan equation
\begin{equation}\label{eq:MaurerCartan}
D e = \diff e + \Omega \wedge e = 0 \ .
\end{equation}
The corresponding curvature two-form is defined by
\begin{equation} \label{eq:curv}
R = \diff \Omega + \Omega \wedge\Omega \ .
\end{equation}
The Ricci tensor (in flat indices) is defined by contraction with the dual vielbein,
\begin{equation} \label{eq:Ricci}
{\rm Ric}_{AB} = R^C_A (\hat e_C,\hat e_B) \ ,
\end{equation}
and the Ricci scalar as its trace
\begin{equation} \label{eq:rscalar}
r_{10} = {\rm Ric}_{AB}\delta^{AB} \ .
\end{equation}
Let us decompose the ten-dimensional connection under $SO(1,9) \to SO(1,3) \times SO(2)\times SO(4)$ as
\begin{equation}\begin{aligned}
 {\bf 45} &&= && ({\bf 6},{\bf 1},{\bf 1}) && \oplus &\ ({\bf 4},{\bf 1},{\bf 4}) && \oplus &\ ({\bf 4},{\bf 2},{\bf 1}) && \oplus &\ (({\bf 1},{\bf 1},{\bf 1}) \oplus ({\bf 1},{\bf 2},{\bf 4}) \oplus ({\bf 1},{\bf 1},{\bf 6}))\ ,\\
 \Omega &&= && \omega &&+&\quad [\lambda] &&+&\quad [\gamma] &&+&\quad \Theta \ ,
\end{aligned}\end{equation}
where we have called the full $SO(6)$ connection $\Theta$.
From the decomposition of the adjoint representation of $SO(6)$ under the breaking $SO(6) \to SO(2) \times SO(4) \equiv SO(2) \times (SU(2) \times SU(2))/\mathbb{Z}_2$, we find
\begin{equation}\label{so6decomp_connection}\begin{aligned}
 so(6) &&= && so(2)&& \oplus &\ su(2)& \oplus & \ su(2)'& \oplus &\ ({\bf 2},{\bf 2},{\bf 2}) \ ,\\
 \Theta &&= && [\phi^0] &&+&\quad  \theta &+& \quad [\phi^a] &+&\quad  [\tau] \ ,
\end{aligned}\end{equation}
where $su(2)$ is the adjoint of the $SU(2)$ structure group and $su(2)'$ is spanned by the $I^a$. The $so(2)$ is generated by the almost complex structure $I^0$ on $T^*_2$, given by $(I^0)^i_{\,j} = \epsilon_{ij}$. The component $\theta$ is the torsionful $SU(2)$ connection. Its internal torsion on $T_2 Y$ is given by $T^i = \diff K^i$ and the component on $T_4$ is
\begin{equation} \label{torsion4d}
 T^\alpha = \diff e^\alpha + \theta^\alpha_\beta \wedge e^\beta =  - (I^a)^\alpha_\beta \phi^a \wedge e^\beta - \tau^\alpha_i \wedge K^i \ .
\end{equation}
We can also decompose the Ricci tensor group-theoretically. In particular, we are interested in the `symmetric' representation $S^2 T^* Y$, which decomposes as
\begin{equation}\label{symmetric_decomp}\begin{aligned}
 S^2 T^* Y= & S^2_0 T^*_2 Y \oplus \mathbb{R} g_2^{(0)}\oplus S^2_0 T^*_4 Y \oplus \mathbb{R} g_4^{(0)} \oplus (T^*_2 Y \otimes T^*_4 Y) \\
 = & ({\bf 2},{\bf 1},{\bf 1}) \oplus ({\bf 1},{\bf 1},{\bf 1}) \oplus ({\bf 1},{\bf 3},{\bf 3})\oplus ({\bf 1},{\bf 1},{\bf 1})\oplus ({\bf 2},{\bf 2},{\bf 2}) \ .
\end{aligned}\end{equation}
Here, the $({\bf 1},{\bf 3},{\bf 3})$ representation $S^2_0 T^*_4$ is spanned by the products of generators of the two $su(2)$. In other words, since the elements of $su(2)$ and $su(2)'$ commute, the representation can be written as
\begin{equation}\label{S2T4representation}
 S^2_0 T^*_4 = \{ I^\alpha_\gamma \tilde I^\gamma_\beta | \tilde I \in su(2), \ I \in su(2)' \} \ .
\end{equation}
The Maurer-Cartan equations \eqref{eq:MaurerCartan} in components now read
\begin{equation}\label{10d_connection}\begin{aligned}
 \diff e^\mu + \omega^\mu_\nu \wedge e^\nu + \lambda^\mu_\alpha \wedge e^\alpha + \gamma^\mu_i \wedge K^i & =  0 \ , \\
 \diff K^i + \epsilon_{ij} \phi^0 \wedge K^j + \tau^i_\alpha \wedge e^\alpha + \gamma_\mu^i \wedge e^\mu & =  0 \ , \\
 \diff e^\alpha + \theta^\alpha_\beta \wedge e^\beta + (I^a)^\alpha_\beta \phi^a \wedge e^\beta + \tau^\alpha_i \wedge K^i + \lambda_\mu^\alpha \wedge e^\mu & =  0 \ .
\end{aligned}\end{equation}

%%%%%%%%%%%%%%%%%%%%%%%%%%
\section{Dimensional reduction}
%%%%%%%%%%%%%%%%%%%%%%%%%%
\label{sec:reduction}

In this section, we will reduce the ten-dimensional type IIA action
\begin{equation} \label{eq:actionIIA} \begin{aligned}
S_{\rm IIA} = & \ \tfrac12  \int_{10} \e^{-2\Phi} ((\ast_{10} 1) r_{10} + 4 \diff \Phi \wedge \ast_{10} \diff \Phi) -\frac12 \e^{-2 \Phi} H_3 \wedge \ast_{10} H_3 \\ & -\tfrac14 \int_{10} \tilde F_4 \wedge \ast_{10} \tilde F_4 + F_2 \wedge \ast_{10} F_2 + F_4 \wedge F_4 \wedge B_2
\end{aligned} \end{equation}
to four dimensions. Here, $H_3 = \diff B_2$ is the form field strength of $B_2$, $F_{p+1} = \diff C_p$ and $\tilde F_4 = F_4 - C_1 \wedge H_3$. The scalar $\Phi$ is the ten-dimensional dilaton. We will discuss the truncation ansatz for manifolds of $SU(2)$-structure that we will use in Section \ref{sec:cons_trunc}. We next study the reduction of the metric sector and of the form fields in Sections \ref{sec:metric} and \ref{sec:forms}. To obtain the four-dimensional action in standard form, various field dualizations must be performed. We discuss these in Section \ref{sec:dualization}.

%%%%%%%%%%%%%%%%%%%%%%%%%%
\subsection{The reduction ansatz}
%\subsection{Consistent truncation}
%%%%%%%%%%%%%%%%%%%%%%%%%%
\label{sec:cons_trunc}
We now discuss the reduction ansatz which is to yield $N=4$ gauged supergravity in four dimensions. We will perform the reduction at the level of the action. Any reduction ansatz parametrizes a subset in the space of ten-dimensional fields. A ten-dimensional solution lying in this subset will necessarily be the lift of a solution of the reduced four-dimensional action. By contrast, a field configuration which is the lift of a four-dimensional solution yields a minimum of the action on this subset, but the evaluation of the action might decrease further as we move off of the subset. A reduction ansatz which excludes this possibility is called a consistent truncation. Under such happy circumstances, all four-dimensional solutions lift. If we imagine deriving four-dimensional equations of motion for {\it all} modes of the theory, dividing these into $\Phi_{\rm keep}$ and $\Phi_{\rm discard}$ (and the corresponding internal modes into $\Omega_{\rm keep}$ and $\Omega_{\rm discard}$), the subscript indicating their future fate, the requirement of a consistent truncation translates into the vanishing of all source terms for the fields belonging to $\Phi_{\rm discard}$, once the fields themselves are set to zero. We must hence exclude that linear terms in $\Phi_{\rm discard}$ occur in the action. As a first requirement, let us demand that $\Omega_{\rm keep}$ be closed under wedge product, and choose $\Omega_{\rm discard}$ in an orthogonal complement to $\Omega_{\rm keep}$. Mixed terms between $\Phi_{\rm keep}$ and $\Phi_{\rm discard}$ in the action that are linear in $\Phi_{\rm discard}$ will then be traceable to the action of $\diff$ and $\ast$.\footnote{For non-linear terms, we would also have to study the behavior of $\Omega_{discard}$ under wedge product.} In such terms, integration by parts or $\ast$ being proportional to its adjoint operator will allow us to restrict attention to the action of these operators on $\Omega_{\rm keep}$. Requiring that $\Omega_{\rm keep}$ be closed under $\diff$ and $\ast$ therefore insures the vanishing of these terms, and the consistency of the truncation. In the following, in contrast to the reduction ansatz in conventional Calabi-Yau compactifications, we will hence choose to reduce in a set of forms that close under the three operations of exterior derivative $\diff$, wedge product $\wedge$, and Hodge star $\ast$.

It has been extensively discussed in \cite{Triendl:2009ap} how the ten-dimensional fields decompose into representations of the $SU(2)$ structure group and then automatically assemble into four-dimensional $N=4$ multiplets over each point of the internal manifold. The components of the type IIA fields in ten-dimensions that are singlets under $SU(2)$ assemble into the gravity multiplet and three vector multiplets. Similarly, the $SU(2)$ triplet representation gives exactly one (triplet of) vector multiplet(s). This triplet representation forms a non-trivial bundle over $Y$ from which we will choose a finite number ($n-3$) of sections in our reduction ansatz. The remaining degrees of freedom all sit in $SU(2)$ doublet representations. These modes assemble into two (doublets of) $N=4$ gravitino multiplets. As we do not expect to be able to preserve more than sixteen supercharges in the reduction of the action, all of these gravitino multiplets should correspond to towers of only massive modes in four dimensions. This means that these multiplets cannot contribute on the $N=4$ massless level. We will hence exclude $SU(2)$ doublet representations from our ansatz. This restriction has important consequences when considering reductions on Calabi-Yau manifolds with vanishing Euler number. We will discuss these consequences in Sections \ref{sec:CYEulerzero} and \ref{sec:4to2}.

The almost product structure \eqref{eq:almost_product_structure} on $Y$ will play a central role in the choice of our reduction ansatz. $T_2$ has trivial structure group and is therefore parallelizable. We hence introduce a basis of two global one-forms $v^i$, $i=1,2$, on this subbundle, yielding two one-forms and a two-form (their wedge product) as expansion forms. On $T_4$, as discussed above, our ansatz is to only contain $SU(2)$ singlets and triplets. It is easily checked that $SU(2)$ doublets exactly correspond to odd forms on $T_4$. Therefore, the ansatz will consist of two-forms $\omega^I, I=1,\dots, n,$ that all square to the same volume form $\vol^{(0)}_4$ on $T_4$, i.e.\
\begin{equation} \label{eq:omegavol4}
 \omega^I \wedge \omega^J = \eta^{IJ} \vol^{(0)}_4 \ ,
\end{equation}
where $\eta$ is a metric with signature $(3,n-3)$, reflecting the number of singlet and triplet representations as discussed above. Furthermore, we include all wedge products of $\omega^I$ and $v^i$ in the reduction ansatz. For instance, we expand the forms $J^a$ and $K$ of \eqref{eq:def_JaK} that specify the $SU(2)$ structure in the set of modes $\omega^I, I=1,\dots, n,$ and $v^i$, $i=1,2$, i.e.\
\begin{equation} \label{param_forms}
 J^a = \e^{\rho_4/2} \zeta^a_I \omega^I \ , \qquad K = (k^1_i + \iu k^2_i) (v^i + G^i) \ ,
\end{equation}
where $-\iu k_i \epsilon^{ij} \bar k_j = 2 \operatorname{det}(k)>0$ with $k_i = k_i^1 + \iu k_I^2$ and $k=(k_i^j)$. Equivalently, by using the parameterization
\begin{equation} \label{matrix_k}
k= \e^{\rho_2/2} (\Im \tau)^{-1/2} \left(  1\, , \ \tau \right) \ ,
\end{equation}
with $\Im \tau > 0$  and such that $\operatorname{det}(k)= \e^{\rho_2}$, we obtain
\begin{equation}\label{param_forms_K}
K = \e^{\rho_2/2} (\Im \tau)^{-1/2} ((v^1 + G^1) + \tau (v^2 + G^2)) \ .
\end{equation}
Note that the presence of internal one-forms in our ansatz gives rise to Kaluza-Klein vectors $G^i$, i.e. mixed spacetime and internal components of the ten-dimensional metric. The expansion coefficients $\zeta^a_I$, $\rho_4$, $\rho_2$ and $\tau$ depend on the spacetime coordinates and give rise to scalar fields in four dimensions.
Furthermore, \eqref{eq:JaJb} yields the relations
\begin{equation}\label{zeta_ab}
\zeta^a_I \eta^{IJ} \zeta^b_J = \delta^{ab} \ ,
\end{equation}
and
\begin{equation}
\vol_4 = \e^{\rho_4} \vol^{(0)}_4 \ .
\end{equation}
The four-dimensional fields $\rho_{2/4}$ describe the volume moduli of $T_{2/4}$ while the $\zeta^a_I$ describe the $SU(2)$-structure geometry.

The form fields $B_2$, $C_1$ and $C_3$ of type IIA supergravity must be expanded in the same set of forms, giving
\begin{equation}\label{form_fields_expansion_IIA} \begin{aligned}
 B_2 = & B + B_i (v^i + G^i) + b_{12} (v^1 + G^1)\wedge (v^2 + G^2) + b_I \omega^I \ ,\\
 C_1 = & A + a_{i} (v^i + G^i)\ , \\
 C_3 = &  (C - A \wedge B) + (C_i - A \wedge B_i) \wedge (v^i + G^i)+ (C_I- b_I A)\wedge \omega^I\\ & + (C_{12}-b_12 A) \wedge (v^1 + G^1)\wedge (v^2 + G^2)  + c_{iI} (v^i + G^i)\wedge \omega^I \ ,
\end{aligned}\end{equation}
where we have shifted the components of $C_3= \tilde C_3 - A \wedge B_2$ by some combination of the components of $B_2$ and $C_1$ for later convenience.

The Hodge star splits into a purely space-time component and two components $\ast_2$ and $\ast_4$, defined with regard to the respective component of the tangent space \eqref{eq:tangent_space_splitting}.  Both are completely determined by the $SU(2)$ structure via\footnote{We choose our conventions such that $\epsilon_{12}=1$.}
\begin{equation}
\ast_2 K^i = \epsilon_{ij} K^j \ , \qquad \ast_2 1 = K^1 \wedge K^2 = \e^{\rho_2} (v^1 + G^1)\wedge (v^2 + G^2) \ ,
\end{equation}
and by the requirement that the $J^a$ are self-dual under $\ast_4$, which implies for the ansatz \eqref{param_forms} that
\begin{equation}
 \ast_4 \omega^I = (2 \zeta^{a\, I} \zeta^a_J - \delta^I_J) \omega^J = H^{I}_J \omega^J \ , \qquad \ast_4 1 = \vol_4 = \e^{\rho_4} \vol_4^{(0)} \ .
\end{equation}
In the following reduction, we will assume that the internal volume is normalized,
\begin{equation}
\int_6 v^1 \wedge v^2 \wedge \vol_4^{(0)} = 1 \ .
\end{equation}

To perform the reduction, we must next specify the differentials of the expansion forms $\{ v^i, \omega^I \}$. As remarked above, we will require that the differential algebra of modes they span closes, i.e.
\begin{equation} \label{eq:truncation_ansatz}\begin{aligned}
 \diff v^i = & t^i v^1 \wedge v^2 + t^i_I \omega^I \ , \\
 \diff \omega^I = & T^I_{iJ} v^i \wedge \omega^J \ .
\end{aligned}\end{equation}
Note in particular that we exclude any terms on the right hand side of the above equations involving $SU(2)$ doublets.

The $t^i$, $t^i_I$ and $T^I_{iJ}$ specify the torsion classes of $Y$. We choose them and hence the torsion classes of $Y$ constant. These constants are constrained by the fact that the exterior derivative squares to zero and the integral of $\diff (v^i\wedge \omega^I \wedge \omega^J)$ over Y should vanish. The constraints are encapsulated by algebraic relations, given by
\begin{equation}\label{eq:quadratic_constraints} \begin{aligned}
 t^i t^k_I \epsilon_{kj} + t^i_J T^J_{jI} & = 0 \ , \\
 T^I_{iJ} \eta^{JK} t^i_K & = 0 \ , \\
 T^I_{iJ} t^i - T^I_{iK} \epsilon_{ij} T^K_{jJ} & = 0 \ , \\
 t^i \eta^{IJ} - \epsilon_{ij} T^I_{jK}\eta^{KJ} - \epsilon_{ij} T^J_{jK}\eta^{KI} & = 0 \ .
\end{aligned}\end{equation}
The last equation determines the symmetric part of $T^I_{jK}\eta^{KJ}$, $j=1,2$, so that
\begin{equation}
 T^I_{iK}\eta^{KJ} = \tilde T^I_{iK}\eta^{KJ}- \tfrac12 \epsilon_{ij} t^j \eta^{IJ} \ ,
\end{equation}
where $\tilde T^I_{jK}$ is a pair of $so(3,n-3)$ matrices, i.e.\
\begin{equation}
\tilde T^I_{jK} \eta^{KJ} + \tilde T^J_{jK}\eta^{KI} = 0 \ .
\end{equation}
The third condition just states that the $\tilde T^I_{iJ}$ form a solvable $so(3,n-3)$ subalgebra $S$ defined by
\begin{equation}
 \tilde T^I_{iK} \epsilon_{ij} \tilde T^K_{jJ} =  \tilde T^I_{iJ} t^i \ .
\end{equation}
In particular, $S$ is Abelian if $t^i=0$.
The remaining condition is
\begin{equation}
   \tilde T^J_{jI} t^i_J = \epsilon_{jk} (t^i t^k_I + \tfrac12 t^i_I t^k )\ .
\end{equation}
If $t^i$ is zero, the $t^i_J$ are invariant under $\tilde T^J_{jI}$. If $t^i$ is non-zero, the $t^i_J$ form a non-trivial representation under $S$.

Before we close this Section, we want to stress that the main conditions we impose on the reduction ansatz are \eqref{eq:omegavol4} and \eqref{eq:truncation_ansatz}. These conditions must be checked case by case for each $SU(2)$-structure compactification individually.
In Section~\ref{sec:Enriques}, we will construct a set of modes on the Enriques Calabi-Yau that satisfies all of these conditions.

%%%%%%%%%%%%%%%%%%%%%%%%%%
\subsection{Reducing gravity to four dimensions}  \label{sec:metric}
%%%%%%%%%%%%%%%%%%%%%%%%%%
We start with the task of dimensionally reducing the gravitational term in the ten-dimensional supergravity action \eqref{eq:actionIIA},
\begin{equation}
S_{\rm grav} = \tfrac12  \int_{10} \e^{-2\Phi} (\ast_{10} 1) r_{10} \ ,
\end{equation}
where $r_{10}$ is the ten-dimensional Ricci scalar and $\Phi$ is the ten-dimensional dilaton with the kinetic term
\begin{equation}
S_{\rm \Phi} = 2  \int_{10} \diff \Phi \wedge \ast_{10} \diff \Phi \ .
\end{equation}
The main task is the computation of the ten-dimensional Ricci scalar in terms of the ansatz of Section~\ref{sec:cons_trunc}. This is performed in detail in Appendix~\ref{app:Ricci}. We first compute the ten-dimensional Levi-Cevita connection in Appendix~\ref{app:connection}  from \eqref{10d_connection} for the ansatz \eqref{eq:truncation_ansatz}, up to the $SU(2)$ connection $\theta$, which cannot be computed explicitly, but also does not appear in the four-dimensional expressions. The connection components read
\begin{equation} \label{eq:connection_explicit}\begin{aligned}
\omega^\mu_\nu =\, \, & \tilde \omega^\mu_\nu + D_{[\mu} G_{\nu]} \cdot k^T \cdot K \ , \\
\gamma^i_\mu = &  k^i_j D_{[\mu} G^j_{\nu]} e^\nu + \tfrac12 D_\mu \rho_2 K^i  - \tfrac12 (\Im \tau)^{-1} D_\mu \Im \tau (\sigma^3)^i_j  K^j  \\ & + \tfrac12 (\Im \tau)^{-1} D_\mu \Re \tau (\sigma^1)^i_j K^j  \ , \\
\lambda^\alpha_\mu  = &  \tfrac14 e^\alpha D_\mu \rho_4  + \tfrac12 \e^{\rho_4/2} (I^a)^\beta_\alpha \omega^I_{\beta \gamma}  e^\gamma (\delta^J_I - \zeta^b_I \zeta^{b J})D_\mu \zeta^a_J \ , \\
\phi^0 = & - \e^{-\rho_2} t \cdot k^T \cdot K  - \frac{\diff \Re \tau}{2\Im \tau} + \tfrac12 t \cdot g^\tau \cdot G \ , \\
\phi^a = & (\tfrac14 \epsilon^{abc} \zeta^b_I \zeta^{c\,J} \tilde T^I_{jJ}(k^{-1})^j_i - \tfrac12 \e^{-\rho_4 /2} \zeta^{aI} t^j_I k^i_j) K^i -\tfrac14 \epsilon^{abc} \zeta^b_I \zeta^{c\,J} \tilde T^I_{iJ}G^i     \ , \\
\tau^i_{\alpha}  = &  \tfrac12 k^i_j t^j_I \omega^I_{\alpha \beta} e^\beta + \tfrac14 \e^{-\rho_2} \epsilon_{ij} k^j_k t^k e^\alpha \\ &  +  \tfrac12 \e^{\rho_4/2} (k^{-1})_i^j \zeta^a_K \tilde T^K_{jI} (\delta^I_J - \zeta^{b\, I} \zeta^b_{\ J})\omega^J_{\alpha \gamma}(I^a)^\gamma_\beta e^\beta  \ ,
\end{aligned}\end{equation}
where we have defined the covariant derivatives as
\begin{equation} \label{eq:cov_metric} \begin{aligned}
 D\rho_2 = & \diff \rho_2 - G \cdot \epsilon \cdot t \ , \\
 D\rho_4 = & \diff \rho_4 +G \cdot \epsilon \cdot t \ , \\
 D \tau = & \diff \tau - ((1 ,\, \tau) \cdot G)  ((1 ,\, \tau) \cdot t) \ , \\
 D \zeta^a_I = & \diff \zeta^a_I - G^i \tilde T^J_{iI} \zeta^a_J \ , \\
 D G^i = & \diff G^i + \tfrac12 t^i G \cdot \epsilon \cdot G \ .
\end{aligned}\end{equation}
These covariant derivatives will appear in the four-dimensional action and are related to the gaugings of the theory.
From the explicit form of the connection in \eqref{eq:connection_explicit}, we can compute the ten-dimensional Ricci scalar $r_{10}$ in a straight-forward way. The computation is performed in the Appendix. Its result reads
\begin{equation}\begin{aligned}
r_{10} = & \hat r_4 - \e^{\rho_2} D_{[\mu} G_{\nu]} \cdot g^\tau \cdot D^{[\mu} G^{\nu]} - 2\nabla^\mu D_\mu \rho_2 - \tfrac32(D^\mu \rho_2) (D_\mu \rho_2)  \\ & - \tfrac12 (\Im \tau)^{-2} (D_\mu \tau)(D^\mu \bar \tau) - 2 \nabla^\mu D_\mu \rho_4 -\tfrac{5}{4} D_\mu \rho_4 D^\mu \rho_4 -\tfrac54  \e^{-\rho_2} t \cdot g^\tau \cdot t \\ & -4 \e^{\rho_2-\rho_4} t_I \cdot g^\tau \cdot t_J \eta^{IJ}
+\e^{- \rho_2} (\eta^{IJ} - \zeta^{b\, I}\zeta^{b\, J})\zeta^{a}_K \zeta^{a}_L \tilde T^K_I \cdot (g^\tau)^{-1} \cdot \tilde T^L_J  \\ & - 2 D_\mu \rho_2 D^\mu \rho_4     +  D_\mu \zeta^a_I (\eta^{IJ} - \zeta^{b\, I}\zeta^{b\, J}) D^\mu \zeta^a_J \ ,
\end{aligned}\end{equation}
where we have defined
\begin{equation}
 g^{\tau} = (\Im \tau)^{-1} \left( \begin{aligned} 1 && \Re \tau \\ \Re \tau && |\tau|^2 \end{aligned} \right) \ .
\end{equation}
With this, the reduction of the ten-dimensional action
\begin{equation}
S_{{\rm grav}, \Phi} = S_{\rm grav}+S_{\Phi}= \tfrac12  \int_{10} \e^{-2\Phi} ((\ast_{10} 1) r_{10} + 4 \diff \Phi \wedge \ast_{10} \diff \Phi)
\end{equation}
yields
\begin{equation}\begin{aligned}
S_{{\rm grav}, \Phi} = & \tfrac12  \int_{4} \e^{-2\phi}(\ast_{4} 1) \big( \hat r_{4}  - \e^{\rho_2} D_{[\mu} G_{\nu]} \cdot g^\tau \cdot D^{[\mu} G^{\nu]} - \tfrac12(D^\mu \rho_2) (D_\mu \rho_2)  \\ & - \tfrac12 (\Im \tau)^{-2} (D_\mu \tau)(D^\mu \bar \tau) -\tfrac{1}{4} D_\mu \rho_4 D^\mu \rho_4 -\tfrac54  \e^{-\rho_2} t \cdot g^\tau \cdot t \\ & -4 \e^{\rho_2-\rho_4} t_I \cdot g^\tau \cdot t_J \eta^{IJ}
+\e^{- \rho_2} (\eta^{IJ} - \zeta^{b\, I}\zeta^{b\, J})\zeta^{a}_K \zeta^{a}_L \tilde T^K_I \cdot (g^\tau)^{-1} \cdot \tilde T^L_J  \\ &   +  D_\mu \zeta^a_I (\eta^{IJ} - \zeta^{b\, I}\zeta^{b\, J}) D^\mu \zeta^a_J + 4 \partial_\mu \phi \partial^\mu \phi \big)  \ ,
\end{aligned}\end{equation}
where we have defined the four-dimensional dilaton $\phi = \Phi - \tfrac12 (\rho_4 + \rho_2)$. In order to obtain the action in the four-dimensional Einstein frame, we perform the Weyl rescaling
\begin{equation}\label{eq:Weyl}
e^\mu \to \e^{\phi} e^\mu \ ,
\end{equation}
which leads to the final result
\begin{equation}\begin{aligned}
S_{{\rm grav}, \Phi} = & \tfrac12  \int_{4} (\ast_{4} 1) ( r_{4}  - \e^{-2 \phi+ \rho_2} D_{[\mu} G_{\nu]} \cdot g^\tau \cdot D^{[\mu} G^{\nu]} - \tfrac12(D^\mu \rho_2) (D_\mu \rho_2)  \\ & - \tfrac12 (\Im \tau)^{-2} (D_\mu \tau)(D^\mu \bar \tau) -\tfrac{1}{4} D_\mu \rho_4 D^\mu \rho_4 -\tfrac54  \e^{2 \phi-\rho_2} t \cdot g^\tau \cdot t \\ & -4 \e^{2 \phi+\rho_2-\rho_4} t_I \cdot g^\tau \cdot t_J \eta^{IJ}
+\e^{2 \phi- \rho_2} (\eta^{IJ} - \zeta^{b\, I}\zeta^{b\, J})\zeta^{a}_K \zeta^{a}_L \tilde T^K_I \cdot (g^\tau)^{-1} \cdot \tilde T^L_J  \\ &   +  D_\mu \zeta^a_I (\eta^{IJ} - \zeta^{b\, I}\zeta^{b\, J}) D^\mu \zeta^a_J -2 \partial_\mu \phi \partial^\mu \phi) \ .
\end{aligned}\end{equation}
We see that several fields appear through their covariant derivative, defined in \eqref{eq:cov_metric}. These scalars are gauged under the Kaluza-Klein vectors $G^i$. This is already a first indication that the reduced action will include gaugings as a standard feature.

%%%%%%%%%%%%%%%%%%%%%%%%%%
\subsection{The form fields}  \label{sec:forms}
%%%%%%%%%%%%%%%%%%%%%%%%%%
Now let us complete the dimensional reduction of the action \eqref{eq:actionIIA} by reducing the terms involving the form fields $B_2$, $C_1$ and $C_3$. The ansatz for these fields has been given in \eqref{form_fields_expansion_IIA}.
Let us start with the NS-NS two-form $B_2$. From \eqref{form_fields_expansion_IIA}, we find
\begin{equation}\label{eq:H} \begin{aligned}
 H_3 =& (\diff B - B_i \wedge D G^i) +(DB_i- \epsilon_{ij} b_{12} D G^j) \wedge (v^i +G^i)   + Db_I \wedge \omega^I \\ & + D b_{12} \wedge (v^1 +G^1) \wedge (v^2 +G^2) + (b_J T^J_{iI} + b_{12} t^j_I \epsilon_{ji}) (v^i +G^i) \wedge \omega^I \ ,
\end{aligned}\end{equation}
where the covariant derivative of $G^i$ has been given in \eqref{eq:cov_metric} and we have further defined the covariant derivatives
\begin{equation}\label{eq:cov_B}\begin{aligned}
 DB_i = & \diff B_i  - \epsilon_{ij}t^k G^j \wedge B_k   \ , \\
 D b_{12} = & \diff b_{12} + b_{12} t \cdot \epsilon \cdot G - B_i t^i \ , \\
 D b_{I} = & \diff b_{I} - B_i t^i_I  - b_J T^J_{iI} G^i \ .
\end{aligned}\end{equation}
In the derivation of \eqref{eq:H}, we have used \eqref{eq:truncation_ansatz} and in particular the thereby implied identity
\begin{equation}
 \diff (v^i + G^i) = DG^i + t^i (v^1+G^1) \wedge (v^2 + G^2) - t^i \epsilon_{jk} (v^j+G^j) \wedge G^k +t^i_I \omega^I \ .
\end{equation}
The ten-dimensional kinetic term $S_{H}= -\frac14 \int_{10} \e^{-2 \Phi} H_3 \wedge \ast_{10} H_3$ becomes
\begin{equation}\begin{aligned}
S_{H} =  -\tfrac14 \int_4 \e^{-2 \phi} \big( & (\diff B - B_i D G^i) \wedge \hat \ast_4 (\diff B - B_i D G^i)   \\ & +  \e^{-\rho_2} (g^\tau)^{-1\, ij} (DB_i- \epsilon_{ik} b_{12} D G^k) \wedge \hat \ast_4 (DB_j- \epsilon_{jl} b_{12} D G^l)  \\ &
+ \e^{-\rho_4} H^{IJ} D b_{I} \wedge \hat \ast_4 D b_{J}+ \e^{-2 \rho_2} D b_{12} \wedge \hat \ast_4 Db_{12} \\ &
 +(\ast_4 1)\e^{-\rho_4-\rho_2} (g^\tau)^{-1\, ij} H^{IJ} (b_K T^K_{iI} + b_{12} t^k_I \epsilon_{ki})(b_L T^L_{jJ} + b_{12} t^l_J \epsilon_{lj}) \big) \ .
\end{aligned}\end{equation}
The Weyl rescaling \eqref{eq:Weyl}  of the metric by $\phi$ gives
\begin{equation}\begin{aligned}
S_{H} =  -\tfrac14 \int_4  \big( & \e^{-4 \phi}(\diff B - B_i\wedge D G^i) \wedge \ast_4 (\diff B - B_i\wedge D G^i)  \\ & +  \e^{-2 \phi-\rho_2} (g^\tau)^{-1\, ij} (DB_i- \epsilon_{ik} b_{12} D G^k) \wedge \ast_4 (DB_j- \epsilon_{jl} b_{12} D G^l)\\ &
 + \e^{-\rho_4} H^{IJ} D b_{I} \wedge \ast_4 D b_{J} +\e^{-2 \rho_2}D b_{12} \wedge \ast_4 Db_{12}  \\ &
   + (\ast_4 1)\e^{2\phi-\rho_4-\rho_2} (g^\tau)^{-1\, ij} H^{IJ} (b_K T^K_{iI} + b_{12} t^k_I \epsilon_{ki})(b_L T^L_{jJ} + b_{12} t^l_J \epsilon_{lj}) \big) \ .
\end{aligned}\end{equation}
We can combine $S_{H}$ with $S_{{\rm grav}, \Phi}$ to find the reduced action for the entire NS-NS sector,
\begin{equation}\begin{aligned}
S_{\rm NS} = \tfrac12  \int_{4} \big( &(\ast_{4} 1)  r_{4} -\tfrac12 \e^{-4 \phi}(\diff B - B_i \wedge D G^i) \wedge \ast_4 (\diff B - B_i\wedge D G^i) \\ &
    - \tfrac12 \e^{-2 \phi+ \rho_2} g^\tau_{ij} DG^i \wedge \ast_4 DG^j \\ &
     -\tfrac12 \e^{-2 \phi-\rho_2} (g^\tau)^{-1\, ij} (DB_i- \epsilon_{ik} b_{12} D G^k) \wedge \ast_4 (DB_j- \epsilon_{jl} b_{12} D G^l)\\ &
     -\diff \phi \wedge \ast_4 \diff \phi -\tfrac12 \e^{-2 \rho_2}(D b_{12} \wedge \ast_4 Db_{12} + D \e^{\rho_2} \wedge \ast_4 D \e^{\rho_2}) \\ &
      - \tfrac14 (\Im \tau)^{-2} D \tau \wedge \ast_4 D\bar \tau -\tfrac{1}{8} D \rho_4 \wedge \ast_4 D \rho_4 -\tfrac14    (H^{IJ} - \eta^{IJ} ) D \zeta^a_I \wedge \ast_4 D \zeta^a_J ) \\ &
      -\tfrac12 \e^{-\rho_4} H^{IJ} D b_{I} \wedge \ast_4 D b_{J} -\tfrac54  \e^{2 \phi-\rho_2} t \cdot g^\tau \cdot t \\ &
      +\e^{2 \phi- \rho_2} (\eta^{IJ} - \zeta^{b\, I}\zeta^{b\, J})\zeta^{a}_K \zeta^{a}_L \tilde T^K_I \cdot (g^\tau)^{-1} \cdot \tilde T^L_J
        -4 \e^{2 \phi+\rho_2-\rho_4} t_I \cdot g^\tau \cdot t_J \eta^{IJ} \\ &
        + (\ast_4 1)\e^{2\phi-\rho_4-\rho_2} (g^\tau)^{-1\, ij} H^{IJ} (b_K T^K_{iI} + b_{12} t^k_I \epsilon_{ki})(b_L T^L_{jJ} + b_{12} t^l_J \epsilon_{lj}) \big)  \ .
\end{aligned}\end{equation}

Let us now turn to the Ramond-Ramond fields $C_1$ and $C_3$. From \eqref{form_fields_expansion_IIA}, we find
\begin{equation}\begin{aligned}
 F_4 =& (\diff C- \diff A \wedge B + A\wedge dB + (C_i-A \wedge B_i) \wedge DG^i) \\ &
 + (DC_i - \diff A \wedge B_i + A \wedge DB_i + \epsilon_{ij} (C_{12} - b_{12} A )\wedge DG^j) \wedge (v^i + G^i) \\ &
 + (D C_{12} + A \wedge Db_{12}  - b_{12} \diff A) \wedge (v^1 + G^1)\wedge (v^2 + G^2) \\ &
 + (DC_I + A \wedge D b_I  - b_I \diff A + c_{iI} D G^i) \wedge \omega^I \\ &
 + (D c_{iI} + (T^J_{iI} b_J-\epsilon_{ij} t^j_I b_{12})A  )\wedge (v^i + G^i) \wedge \omega^I \\&
 +(c_{iI} t^i - c_{iJ} \epsilon_{ij} T^J_{jI} ) (v^1 + G^1)\wedge (v^2 + G^2)\wedge \omega^I  + c_{iI} t^i_J \eta^{IJ} \vol^{(0)}_4 \ , \\
 F_2 =& (\diff A + a_i DG^i) + Da_i \wedge (v^i + G^i) + a_i t^i (v^1 + G^1)\wedge (v^2 + G^2) + a_i t^i_I \omega^I \ ,
\end{aligned}\end{equation}
where we have defined the covariant derivatives
\begin{equation}\label{eq:Cfields_covder}\begin{aligned}
 D C_i = & \diff C_i + \epsilon_{ij} G^j \wedge t^k C_k    \ , \\
 D C_{12} = & \diff C_{12} + C_i t^i + C_{12} \wedge G \cdot \epsilon \cdot t   \ , \\
 D C_{I} = & \diff C_{I} + C_i t^i_I + T^J_{iI} C_J \wedge G^i  \ , \\
 D c_{iI} = & \diff c_{iI} + \epsilon_{ij} t^j_I C_{12} - C_J T^J_{iI} + \epsilon_{ij} G^j c_{kI} t^k  - c_{iJ} T^J_{jI} G^j \ , \\
 D a_{i} = & \diff a_{i} + \epsilon_{ij} G^j t^k a_k   \ .
\end{aligned}\end{equation}
Furthermore, the twisted field strength $\tilde F_4 = F_4 - C_1 \wedge H_3$ is given by
\begin{equation}\begin{aligned}
 \tilde F_4 =& (\diff C- \diff A \wedge B  +C_i \wedge DG^i) \\ &
 + (DC_i +a_i \diff B - \diff A \wedge B_i + (\epsilon_{ij} C_{12} - a_i B_j)\wedge DG^j) \wedge (v^i + G^i) \\ &
 + (D C_{12} - b_{12} \diff A - a_i (\epsilon_{ij} Db_j + b_{12} DG^i) ) \wedge (v^1 + G^1)\wedge (v^2 + G^2) \\ &
 + (DC_I  - b_I \diff A + c_{iI} D G^i) \wedge \omega^I  + (D c_{iI} +a_i Db_I)\wedge (v^i + G^i) \wedge \omega^I \\&
 +(c_{iI} t^i - a_i b_{12} t^i_I - (c_{iJ}+ a_i b_J )\epsilon_{ij} T^J_{jI} ) (v^1 + G^1)\wedge (v^2 + G^2)\wedge \omega^I \\ &  + c_{iI} t^i_J \eta^{IJ} \vol^{(0)}_4 \ .
\end{aligned}\end{equation}
We can now insert these results into the ten-dimensional action for the RR fields, consisting of the kinetic term
\begin{equation}
S_{\rm RR}=-\tfrac14 \int_{10} ( \tilde F_4 \wedge \ast_{10} \tilde F_4 + F_2 \wedge \ast_{10} F_2) \ ,
\end{equation}
and the Chern-Simons term
\begin{equation} \label{eq:CS10d}
S_{\rm CS}=-\tfrac14 \int_{10} F_4 \wedge F_4 \wedge B_2 \ .
\end{equation}
The kinetic term can be reduced in a straight-forward way. After taking into account the Weyl rescaling (\ref{eq:Weyl}), we find
\begin{equation}\begin{aligned}
S_{\rm RR}=    -\tfrac14 \int_{4} \Big( &
\e^{-4 \phi+ \rho_2 + \rho_4} |\diff C - \diff A \wedge B + C_i \wedge DG^i|^2 \\ &
+ \e^{-2 \phi+ \rho_4} (g^\tau)^{-1 \, ij} (DC_i +a_i \diff B - \diff A \wedge B_i + (\epsilon_{ik} C_{12} - a_i B_k)\wedge DG^k)\\ &
\quad \wedge \ast_4 (DC_j +a_j \diff B - \diff A \wedge B_j + (\epsilon_{jl} C_{12} - a_j B_l)\wedge DG^l) \\ &
+\e^{-\rho_2+ \rho_4} |DC_{12} - b_{12} \diff A - a_i (\epsilon_{ij} D b_j+b_{12} DG^i)|^2 \\ &
 + \e^{\rho_2} H^{IJ} (DC_I - b_I \diff A + c_{iI} DG^i) \wedge \ast_4 (DC_I - b_I \diff A + c_{iI} DG^i) \\ &
+\e^{\rho_2 + \rho_4} |\diff A + a_i DG^i|^2+\e^{2 \phi + \rho_4} (g^\tau)^{-1\, ij} Da_i \wedge \ast_4 Da_j \\ &
+ \e^{2\phi} H^{IJ} (g^\tau)^{-1\, ij} (D c_{iI} + a_i Db_I) \wedge \ast_4 (D c_{jJ} + a_j Db_J) \\ &
 + \e^{4\phi-\rho_2} H^{IJ} (c_{iI} t^i - a_i b_{12} t^i_I - (c_{iK} +a_i b_K) \epsilon_{ik} T^K_{kI})\\ &
\quad (c_{jJ} t^j - a_j b_{12} t^j_J - (c_{jL} +a_j b_L) \epsilon_{jl} T^L_{lJ})(\ast_4 1) \\ &
+\e^{4\phi + \rho_2 - \rho_4} |c_{iI} t^i_J \eta^{IJ}|^2(\ast_4 1) +\e^{4\phi - \rho_2 + \rho_4} |a_i t^i|^2(\ast_4 1)\\ &
 + \e^{4\phi+\rho_2} (a_i t^i_I)H^{IJ} (a_j t^j_J) (\ast_4 1) \Big) \ .
\end{aligned}\end{equation}

Let us now compute the Chern-Simons term \eqref{eq:CS10d}.
First, we use partial integration to write it as
\begin{equation}
S_{\rm CS}=-\tfrac14 \int_{10} (B_2  \wedge \diff \tilde C_3 \wedge \diff \tilde C_3 - B_2 \wedge B_2 \wedge \diff A \wedge \diff \tilde C_3 + \tfrac13 B_2 \wedge B_2 \wedge B_2 \wedge \diff A \wedge \diff A )\ .
\end{equation}
Then, inserting the expansions \eqref{form_fields_expansion_IIA} and using \eqref{eq:quadratic_constraints} yields
\begin{equation}\begin{aligned}
S_{\rm CS} = & -\tfrac12 \int_4 (\diff C - \diff A\wedge B + C_i \wedge DG^i ) (b_{12} c_{jI} t^j_J \eta^{IJ} + b_I c_{jK} \eta^{KJ} \epsilon_{jk} T^I_{kJ}) \\ &
- \tfrac12 \int_4 (DC_i + \epsilon_{ij} C_{12} \wedge DG^j - \diff A \wedge B_i)\epsilon_{ik}(B_k c_{lI} t^l_J \eta^{IJ} + Dc_{kI} \eta^{IJ} b_J) \\ &
+ \tfrac14 \int_4 (\diff B - B_k \wedge DG^k) \wedge c_{iI}\eta^{IJ}(\epsilon_{ij}Dc_{jJ} + \epsilon_{ij} T^K_{jJ} C_K +t^i_J C_{12}) \\ &
+ \tfrac14 \int_4 B_k c_{iI}\eta^{IJ} \wedge (DG^k \wedge (\epsilon_{ij}Dc_{jJ} + \epsilon_{ij} T^K_{jJ} C_K +t^i_J C_{12}) - \epsilon_{kj}  t^i_J  B_j \wedge \diff A ) \\ &
- \tfrac12 \int_{4} \epsilon_{ik} B_k \wedge   (DC_I + c_{jI} DG^j)\wedge D c_{iJ} \eta^{IJ} + \tfrac12 \int_4 b_{12} b_I \eta^{IJ} c_{iJ} \diff A \wedge DG^i \\ &
-\tfrac14 \int_{4} b_{12}\eta^{IJ}(DC_I + c_{iI} DG^i) \wedge (DC_J + c_{jJ} DG^j) \\ &
- \tfrac12 \int_4 b_I  \eta^{IJ} (DC_{12} - b_{12} \diff A) \wedge (DC_J -\tfrac12 b_J \diff A + c_{iJ} \wedge DG^i)  \ .
\end{aligned}\end{equation}

Note that the dimensional reduction of the action for the form fields yields, beyond four-dimensional fields in canonical form, also a three-form $C$ and three two-form fields $B$ and $C_i$, $i=1,2$. In order to compare the action we have obtained to a four-dimensional supergravity in a conventional presentation, we must eliminate these form fields or tranform them into standard four-dimensional fields. In four dimensions, a three-form $C$ has no dynamical degrees of freedom, as its four-form field strength is a top form. We can thus eliminate this field simply by replacing it by its equations of motion. The 2-form fields however are dynamical. They are dual to scalar fields, as a three-form field strength is Hodge-dual to a one-form, the field strength of a scalar.

In the next section, we will integrate out $C$ and perform field dualizations for the tensor fields $B$ and $C_i$.

%%%%%%%%%%%%%%%%%%%%%%%%%%
\subsection{Field dualizations}     \label{sec:dualization}
%%%%%%%%%%%%%%%%%%%%%%%%%%
In this section, we will follow the strategy of \cite{Danckaert:2011ju} of eliminating the three-form $C$ via its equation of motion and of dualizing all three tensor fields $B$ and $C_i$ to dual scalars $\beta$ and $\gamma_i$. The latter task involves a subtlety in the generality in which we are performing the reduction which did not arise in the treatment of \cite{Danckaert:2011ju}. We will explain this below.

Let us start with the three-form $C$. It does not carry any degrees of freedom and can be integrated out. The part of the action involving it reads\footnote{Here and in the following, all integrals and Hodge operations will be in four-dimensional spacetime. We will hence simplify our notation by dropping this specification.}
\begin{equation}\begin{aligned}
S_C = -\tfrac14 \int  \Big( &
\e^{-4 \phi+ \rho_2 + \rho_4} |\diff C - \diff A \wedge B + C_i \wedge DG^i|^2 \\ & + 2 (\diff C - \diff A\wedge B + C_i \wedge DG^i ) (b_{12} c_{jI} t^j_J \eta^{IJ} + b_I c_{jK} \eta^{KJ} \epsilon_{jk} T^I_{kJ}) \Big) \ .
\end{aligned}\end{equation}
The corresponding equation of motion is
\begin{equation}
\diff C - \diff A\wedge B + C_i \wedge DG^i = - \e^{4 \phi- \rho_2 - \rho_4} (b_{12} c_{jI} t^j_J \eta^{IJ} + b_I c_{jK} \eta^{KJ} \epsilon_{jk} T^I_{kJ}) \ast 1 \ .
\end{equation}
Substituting this back into the above action gives the potential term
\begin{equation}
\tilde S_C = \tfrac14 \int  \e^{4 \phi- \rho_2 - \rho_4} |b_{12} c_{jI} t^j_J \eta^{IJ} + b_I c_{jK} \eta^{KJ} \epsilon_{jk} T^I_{kJ}|^2   \ .
\end{equation}
Next, we want to dualize the two-forms $C_i$ and $B$ to scalars $\gamma_i$ and $\beta$. However, the $C_i$ appear in the action not only through their covariant derivatives, but also in the covariant derivative of the vectors $C_{12}$ and $C_I$, cf.\ \eqref{eq:Cfields_covder}, giving a St\"uckelberg-like mass term. In \cite{Danckaert:2011ju}, this complication was addressed by also dualizing these vectors. At first blush, the more general gaugings \eqref{eq:Cfields_covder} that arise in our reduction do not allow us to stop here, as the vectors $C_{12}$ and $C_I$ in turn appear in the covariant derivative of the $c_{iI}$ -- the dualizing of which would reintroduce two-form fields into the action. Fortunately, the quadratic constraints \eqref{eq:quadratic_constraints} imply that the combinations of $C_{12}$ and $C_I$ appearing in the covariant derivative of the $c_{iI}$ are orthogonal to those combinations that need to be dualized due to the two-forms $C_i$ appearing in their covariant derivative. To compute the contributions to the action involving $\gamma_i$, we will for simplicity introduce the magnetic duals to {\it all} vectors $C_{12}$ and $C_I$ as Lagrange multipliers for the respective Bianchi identities. The dualization procedure for $C_i$ will automatically pick out the linear combinations of the vectors in whose covariant derivative the $C_i$ appear; the action we will obtain for $\gamma_i$ will only depend on these linear combinations.

With regard to the fields $C_{12}$ and $C_I$, we will follow a different path than the one pursued in \cite{Danckaert:2011ju}. While we will introduce the magnetic duals of the appropriate linear combinations of these vectors, we will not integrate out the corresponding field strengths. The action will hence depend on both electric and magnetic potentials, in the fashion of the embedding tensor formalism \cite{deWit:2005ub}, with the magnetic duals remaining non-dynamical. In a final step, we will rewrite our results in terms of the embedding tensor formalism \cite{deWit:2005ub} and find that they fit into its $N=4$ supergravity version, discussed in \cite{Schon:2006kz}.

Let us now proceed with the dualization of $C_i$ and $B$. First, we replace these tensor fields by their field strengths $F_i$ and $H$, defined by
\begin{equation}\begin{aligned}
 F_i =& DC_i + \epsilon_{ij} C_{12} \wedge DG^j \\
     = &\diff C_i +\epsilon_{ij} G^j \wedge t\cdot C + \epsilon_{ij} C_{12} \wedge DG^j \ , \\
 H = & \diff B - B_i \wedge DG^i \ ,
 \end{aligned} \end{equation}
respectively. Their Bianchi identities
\begin{equation}\begin{aligned}
 \diff F_i = & t \cdot F \wedge \epsilon_{ij} G^j + \epsilon_{ij} DC_{12} \wedge DG^j \ , \\
 \diff H =  & - DB_i \wedge DG^i \ ,
 \end{aligned} \end{equation}
will be imposed via Lagrange multipliers $\gamma_i \epsilon_{ij} $ and $\beta$.
As discussed above, we will also introduce the field strengths
\begin{equation}\begin{aligned}
 F_I =& DC_I + c_{iI} DG^i = \diff C_I + t_I^i C_i + T^J_{iI} C_J \wedge G^i + c_{iI} DG^i \ , \\
 F_{12} =& D C_{12} = \diff C_{12} + t^i C_i + C_{12} \wedge G \cdot \epsilon \cdot t   \ .
 \end{aligned} \end{equation}
For their Bianchi identities, we find
\begin{equation}\begin{aligned}
 \diff F_I =& t_I^i F_i + T^J_{iI} F_J \wedge G^i + Dc_{iI} \wedge DG^i \ , \\
 \diff F_{12} =&  t^i F_i + F_{12} \wedge G \cdot \epsilon \cdot t   \ .
\end{aligned} \end{equation}
Now we introduce Lagrange multipliers $\tilde C_I$ and $\tilde C_{12}$ together with the before-mentioned $\gamma_i \epsilon_{ij} $ and $\beta$ in the action, by adding
\begin{equation}\begin{aligned}
 S_{BI} =& \tfrac12 \int \gamma_i (\epsilon_{ij} \diff F_j + t \cdot F \wedge G^i + F_{12}\wedge DG^i ) + \tfrac12 \int \beta (\diff H + DB_i \wedge DG^i) \\ &
 + \tfrac12 \int \tilde C_K \eta^{KI} \wedge (\diff F_I - t^i_I F_i - T^J_{iI} F_J \wedge G^i + Dc_{iI} \wedge DG^i )\\ &
 + \tfrac12 \int \tilde C_{12} \wedge (\diff F_{12} - t^i F_i - F_{12} \wedge G \cdot \epsilon \cdot t)
\end{aligned} \end{equation}
to the action. We furthermore replace the two-form fields $B$ and $C_i$ by their field strengths throughout. The action for the $F_i$ then reads
\begin{equation}\label{eq:actionFi}\begin{aligned}
 S_{F_i} = & -\tfrac14 \int \e^{-2\phi+\rho_4} (g^\tau)^{-1\, ij} (F_i + a_i H - \diff A \wedge B_i) \wedge \ast (F_j + a_j H - \diff A \wedge B_j) \\ &
 - \tfrac12 \int (F_i - \diff A \wedge B_i) \epsilon_{ik} L_k  - \tfrac12 \int F_i \wedge \epsilon_{ij} \tilde D \gamma_j + \tfrac12 \int \gamma_i  F_{12}\wedge DG^i \ ,
\end{aligned} \end{equation}
where we have used the abbreviations
\begin{equation}\begin{aligned}
 \tilde D\gamma_i =& \diff \gamma_i - \epsilon_{ij} t^j \gamma \cdot G + \epsilon_{ij} t^j_I \eta^{IJ} \tilde C_J + \epsilon_{ij} t^j \tilde C_{12} \ , \\
 L_i =& Dc_{iI} \eta^{IJ} b_J + B_i t^j_I \eta^{IJ} c_{jJ} \ .
\end{aligned} \end{equation}
The equations of motion for the $F_i$ are
\begin{equation}\label{eq:dualityFi}
 F_i = \diff A \wedge B_i - a_i H - \e^{2\phi-\rho_4} g^\tau_{ij} \epsilon_{jk} \ast (\tilde D \gamma_k +L_k) \ .
\end{equation}
We can insert this into \eqref{eq:actionFi} and find
\begin{equation}\begin{aligned}
 S_{\gamma_i} = & - \tfrac14 \int \e^{2\phi-\rho_4} (g^\tau)^{-1\, ij} (D\gamma_i + b_I \eta^{IJ} Dc_{iJ}) \wedge \ast (D\gamma_j + b_K \eta^{KL} Dc_{jL}) \\ &
 + \tfrac12 \int H \wedge a_i \epsilon_{ij} (D\gamma_j +b_I \eta^{IJ} Dc_{jJ}) - \tfrac12 \int \diff A \wedge B_i \wedge \epsilon_{ij} (D\gamma_j - B_j t^k_I \eta^{IJ} c_{kJ}) \\ &
 + \tfrac12 \int \gamma_i  F_{12}\wedge DG^i\ ,
\end{aligned} \end{equation}
where we have defined the covariant derivative
\begin{equation}\label{eq:covder_gamma}
 D\gamma_i =\diff \gamma_i - \epsilon_{ij} t^j \gamma \cdot G + \epsilon_{ij} t^j_I \eta^{IJ} \tilde C_J + \epsilon_{ij} t^j \tilde C_{12} + B_i t^j_I \eta^{IJ} c_{jJ} \ .
\end{equation}
Having obtained this action, we can also take a different point of view, generalizing the example of \cite{Cassani:2012pj}. Starting with the action for the two-form fields $C_i$ and the related vectors $C_I$ and $C_{12}$, we can also directly introduce new scalar fields $\gamma_i$ and auxiliary vector fields $\tilde C_I$, $\tilde C_{12}$ by replacing the action for the $C_i$ by
\begin{equation}\label{eq:Ci_embedding_tensor} \begin{aligned}
 S_{\gamma_i} =& - \tfrac14 \int \e^{2\phi-\rho_4} (g^\tau)^{-1\, ij} (D\gamma_i + b_I \eta^{IJ} Dc_{iJ}) \wedge \ast (D\gamma_j + b_K \eta^{KL} Dc_{jL}) \\ &
 + \tfrac12 \int H \wedge a_i \epsilon_{ij} (D\gamma_j +b_I \eta^{IJ} Dc_{jJ}) - \tfrac12 \int \diff A \wedge B_i \wedge \epsilon_{ij} (D\gamma_j - B_j t^k_I \eta^{IJ} c_{kJ}) \\ &
  - \tfrac12 \int (D C_i +\epsilon_{ij} C_{12} \wedge DG^j) \wedge \epsilon_{ij} ( t^j_I \eta^{IJ} \tilde C_J + t^j \tilde C_{12}) + \tfrac12 \int \gamma_i  F_{12}\wedge DG^i \ ,
\end{aligned} \end{equation}
where the covariant derivative of $\gamma_i$ is given by \eqref{eq:covder_gamma}. Note that by removing the kinetic term, the dynamical two-form fields $C_i$ as they descend from string theory are rendered non-dynamical, befitting the two-form fields of the embedding formalism. Given the action in the form (\ref{eq:Ci_embedding_tensor}), the equations of motion for the vectors $\tilde C_J$ and $\tilde C_{12}$ give the duality equation \eqref{eq:dualityFi}, while the equations of motion for $C_i$ imply in turn that the vectors $\tilde C_J$ and $\tilde C_{12}$ are the magnetic duals of $C_I$ and $C_{12}$. Inserting these equations of motion back into the action brings one back to the original action for the $C_i$ and the vectors $C_I$ and $C_{12}$ as obtained directly from string theory. This action is reformulated in \eqref{eq:Ci_embedding_tensor} in terms of the embedding tensor formalism of \cite{deWit:2005ub,Schon:2006kz}.

Dualizing $B$ proceeds along the same lines as the dualization procedure for the $C_i$. The starting point is the action
\begin{equation}\label{eq:actionH}\begin{aligned}
 S_H = & - \tfrac14 \int \e^{-4\phi} H \wedge \ast H + \tfrac12 \int \beta DB_i \wedge DG^i \\ &
 + \tfrac14 \int H \wedge (D \beta - a_i \epsilon_{ij} D\gamma_j - \epsilon_{ij} (\tfrac12 c_{iI} - a_i b_I) \eta^{IJ} D c_{jJ}) \ ,
\end{aligned} \end{equation}
where we have defined the covariant derivative
\begin{equation}\label{eq:covder_beta}
 D \beta = \diff \beta - \tfrac12  c_{iI} \eta^{IJ} ( \epsilon_{ij} T^K_{jJ} C_K + t^i_j C_{12}) \ .
\end{equation}
As $B$ does not appear in any covariant derivatives, no further fields need to be dualized in its wake.
Note that the combinations of $C_I$ and $C_{12}$ appearing in this covariant derivative are orthogonal to those that required dualization above. Integrating out $H$ in \eqref{eq:actionH} finally yields the action
\begin{equation}\begin{aligned}
 S_{\beta}  = & - \tfrac14 \int \e^{4\phi} (D \beta - a_i \epsilon_{ij} D\gamma_j - \epsilon_{ij} (\tfrac12 c_{iI} - a_i b_I) \eta^{IJ} D c_{jJ})\\ & \qquad \wedge \ast (D \beta - a_k \epsilon_{kl} D\gamma_l - \epsilon_{kl} (\tfrac12 c_{kK} - a_k b_K) \eta^{KL} D c_{lL})  \\ & + \tfrac12 \int \beta DB_i \wedge DG^i  \ .
\end{aligned} \end{equation}
For later convenience, we collect all relevant covariant derivatives from \eqref{eq:cov_metric}, \eqref{eq:cov_B}, \eqref{eq:Cfields_covder}, \eqref{eq:covder_beta} and \eqref{eq:covder_gamma}: the covariant derivatives of the vectors are given by
\begin{equation}\label{eq:covder_vectors}\begin{aligned}
 D G^i = & \diff G^i + \tfrac12 t^i G \cdot \epsilon \cdot G \ ,\\
 DB_i = & \diff B_i  - \epsilon_{ij}t^k G^j \wedge B_k   \ , \\
 D C_{12} = & \diff C_{12} + C_i t^i + C_{12} \wedge G \cdot \epsilon \cdot t   \ , \\
 D C_{I} = & \diff C_{I} + C_i t^i_I + T^J_{iI} C_J \wedge G^i  \ ,
 \end{aligned} \end{equation}
 and the covariant derivatives of the scalars read
\begin{equation}\label{eq:covder_scalars}\begin{aligned}
 D\rho_2 = & \diff \rho_2 - G \cdot \epsilon \cdot t \ , \\
 D\rho_4 = & \diff \rho_4 +G \cdot \epsilon \cdot t \ , \\
 D \tau = & \diff \tau - ((1 ,\, \tau) \cdot G)  ((1 ,\, \tau) \cdot t) \ , \\
 D \zeta^a_I = & \diff \zeta^a_I - G^i \tilde T^J_{iI} \zeta^a_J \ , \\
 D b_{12} = & \diff b_{12} + b_{12} t \cdot \epsilon \cdot G - B_i t^i \ , \\
 D b_{I} = & \diff b_{I} - B_i t^i_I  - b_J T^J_{iI} G^i \ , \\
 D c_{iI} = & \diff c_{iI} + \epsilon_{ij} t^j_I C_{12} - C_J T^J_{iI} + \epsilon_{ij} G^j c_{kI} t^k  - c_{iJ} T^J_{jI} G^j \ , \\
 D a_{i} = & \diff a_{i} + \epsilon_{ij} G^j t^k a_k   \ , \\
 D\gamma_i = & \diff \gamma_i - \epsilon_{ij} t^j \gamma \cdot G + \epsilon_{ij} t^j_I \eta^{IJ} \tilde C_J + \epsilon_{ij} t^j \tilde C_{12} + B_i t^j_I \eta^{IJ} c_{jJ} \ , \\
 D \beta = & \diff \beta - \tfrac12  c_{iI} \eta^{IJ} ( \epsilon_{ij} T^K_{jJ} C_K + t^i_j C_{12}) \ .
\end{aligned} \end{equation}

%%%%%%%%%%%%%%%%%%%%%%%%%%
\subsection{$N=4$ gauged supergravity}
%%%%%%%%%%%%%%%%%%%%%%%%%%
\label{sec:supergravity}

The reduced action that we found above should fit into the constrained framework of $N=4$ gauged supergravity. For the case $t^i_I=0$, this was already shown in \cite{Danckaert:2011ju}. In this section, we will read off the embedding tensor components from the action derived above and thus determine the gauge group of the theory. One can subsequently check that the action derived in Section~\ref{sec:reduction} falls into the class of $N=4$ gauged supergravities.

The vectors and scalars of our description map to the $N=4$ vectors and scalars of \cite{Schon:2006kz}. The correct identification in the standard $N=4$ frame has been worked out in \cite{Danckaert:2011ju}. The electric and magnetic vectors $V^{M+}$ and $V^{M-}$, given by
\begin{equation}\begin{aligned}
 V^{M+} = & (G^i, \tilde B^{\bar \imath}, A, \tilde C_{12}, \eta^{IJ}C_J) \ ,  \\ V^{M-} =& (B^i , \tilde G^{\bar \imath}, C_{12}, \tilde A, \eta^{IJ}\tilde C_J) \ ,
\end{aligned}\end{equation}
are in the fundamental representation of $SO(6,n)$ defined by the the metric $\eta_{MN}$
\begin{equation}
 \eta_{MN} = \left( \begin{aligned}
 & 0  && \delta_{i \bar \jmath} && 0 && 0 && 0 \\
 & \delta_{\bar \imath j} && 0  && 0 && 0 && 0 \\
 & 0  && 0 && 0 && 1 && 0 \\
 & 0  && 0 && 1 && 0 && 0 \\
 & 0  && 0 && 0 && 0 && \eta_{IJ} \,
 \end{aligned} \right) \ .
\end{equation}
We can now read off the embedding tensor components as they appear in the formulation of \cite{Schon:2006kz}. The $N=4$ axiodilaton $\sigma$ is given by $\sigma = \tfrac{1}{2} ( - b_{12} + \iu \e^{-\rho_2})$. From its kinetic term we find the gauging \cite{Danckaert:2011ju}
\begin{equation}\label{eq:embed_tensor_xi}
\xi_{+M} = - \epsilon_{ij} t^j \ , \qquad \xi_{-M} = 0 \ ,
\end{equation}
so that
\begin{equation}
D\sigma  = \diff \sigma + V^{M(\alpha} \epsilon^{\beta)\gamma} \xi_{\gamma M} t_{\alpha \beta}= \diff \sigma + (t \cdot \epsilon \cdot G) \sigma + \tfrac12 t \cdot B \ ,
\end{equation}
where $t_{\alpha \beta}$ are the generators of $SL(2)$ so that $t_{22}$ and $t_{(12)}$ generate real shifts and rescalings of $\sigma$.

The $SO(6,n)$ matrix $M_{MN}$ collecting all other scalar fields has been worked out in \cite{Danckaert:2011ju}. We can read off the covariant derivative of its scalar fields from \eqref{eq:cov_metric}, \eqref{eq:cov_B}, \eqref{eq:Cfields_covder}, \eqref{eq:covder_beta} and \eqref{eq:covder_gamma}. We thereby find the following for the gauged Killing vectors of $N=4$:
\begin{equation} \label{eq:Killingvectors}\begin{aligned}
 k_{i+} = & - \epsilon_{ij} t^j (\frac{\partial}{\partial \rho_{2}} + b_{12} \frac{\partial}{\partial b_{12}}  - \frac{\partial}{\partial \rho_4} + \gamma_k \frac{\partial}{\partial \gamma_k} +\sigma \frac{\partial}{\partial \sigma}) - \tilde T^J_{iI} \zeta^a_J \frac{\partial}{\partial \zeta^a_I} \\ & - T^J_{iI} b_J \frac{\partial}{\partial b_I}
  + t^j ( \epsilon_{ik} g^\tau_{jl} +\epsilon_{il} g^\tau_{jk}- \epsilon_{ij} g^\tau_{kl} )\frac{\partial}{\partial g^\tau_{kl}} - \epsilon_{ij} (t \cdot a) \frac{\partial}{\partial a_j}
  \\ & - (\epsilon_{ij} t^k \delta^J_I + \delta_j^k T^J_{iI}) c_{kJ} \frac{\partial}{\partial c_{jI}}   \ , \\
k_{6+} = & - t^i \epsilon_{ij} \frac{\partial}{\partial \gamma_j} \ , \\
 k_{I+} = &  - \eta_{IK} T^K_{iJ}  (\frac{\partial}{\partial c_{iJ}} - \tfrac12 \epsilon_{ij} \eta^{JL} c_{jL} \frac{\partial}{\partial \beta}) \ , \\
 k_{i-} = & - t^i \frac{\partial}{\partial b_{12}} - t^i_I \frac{\partial}{\partial b_I} + t^j_I \eta^{IJ} c_{jJ} \frac{\partial}{\partial \gamma_i} +\tfrac12 t^i \frac{\partial}{\partial \sigma} \ , \\
 k_{5-} = & - t^i_I (\epsilon_{ij} \frac{\partial}{\partial c_{jI}} + \tfrac12 c_{iI} \frac{\partial}{\partial \beta}) \ , \\
k_{I-} = & - t^i_I \epsilon_{ij} \frac{\partial}{\partial \gamma_j} \ .
\end{aligned} \end{equation}
The general covariant derivative hence reads
\begin{equation}
 D = \diff + V^{M\alpha} k_{M\alpha}= \diff + G^i k_{i+} + \tilde  C_{12} k_{6+}  + C_I \eta^{IJ} k_{J+} + B^i k_{i-}+ C_{12} k_{5-} + \tilde C_I \eta^{IJ} k_{J-} \ .
\end{equation}
The only non-vanishing entries of the totally anti-symmetric embedding tensor component $f_{\alpha [MNP]}$ are thus
\begin{equation}\label{eq:embed_tensor_f} \begin{aligned}
f_{+ij\bar \imath} = & - \tfrac12 \epsilon_{ij} \delta_{\bar \imath k} t^k \ , \\
f_{+i56} = & \tfrac12 \epsilon_{ij} t^j \ , \\
f_{+iIJ} = & - T^K_{iI} \eta_{KJ} \ , \\
f_{- i I 5} = & \epsilon_{ij} t^j_I \ .
\end{aligned}\end{equation}
The embedding tensor components in the last line are those that did not appear in \cite{Danckaert:2011ju}. They correspond to magnetic gaugings in the electric-magnetic duality frame of choice.
We find that $\xi_{\alpha M}$ and $f_{\alpha [MNP]}$ given in \eqref{eq:embed_tensor_xi} and \eqref{eq:embed_tensor_f} obey
\begin{equation}\begin{aligned}
 \xi_{\alpha M} \xi_{\beta N} \eta^{MN} \, = &\, 0 \ , \\ \xi_{\alpha M} \xi_{\beta N} \epsilon^{\alpha \beta} \,= &\, 0 \ , \\
 \xi_{\alpha M} f_{\beta NPQ} \eta^{MN} \,= &\, 0 \ ,  \\
 3 f_{\alpha R[MN} f_{\beta PQ]S} \eta^{RS} + 2 \xi_{(\alpha|M} f_{\beta) NPQ} \,=& \, 0 \ , \\
 \epsilon^{\alpha \beta}( f_{\alpha MNR} f_{\beta PQS} \eta^{RS} - \eta^{RS} \xi_{\alpha R} f_{\beta S [ M[P} \eta_{Q]N]} - \eta_{\alpha [M} f_{N][PQ] \beta} + \xi_{\alpha[P} f_{Q][MN] \beta}) \,=& \, 0 \ ,
\end{aligned}\end{equation}
using the conditions \eqref{eq:quadratic_constraints}. This implies the usual quadratic constraints on the embedding tensor \cite{Schon:2006kz}.

The algebra of Killing vectors $k_M$ can be easily computed from \eqref{eq:Killingvectors}, using \eqref{eq:quadratic_constraints}. The Killing vectors $(k_{i-}, k_{5-}, k_{6+}, k_{I+}, k_{I-})$ commute with each other. Only the Lie brackets with $k_i$ are non-vanishing. They are given by
\begin{equation}\begin{aligned}
\ [ k_{i+},  k_{j+}] =& - \epsilon_{ij} t^k  k_{k+} \ , \qquad \qquad  \qquad   [ k_{i+}, k_{j-}] = \quad \epsilon_{ik} t^k k_{j-} \ , \\
 [ k_{i+}, k_{6+}] =& \quad \epsilon_{ij} t^j k_{6+}  \ , \qquad \qquad  \qquad \  [ k_{i+}, k_{5-}] = - 2 \epsilon_{ij} t^j k_{5-} \ , \\
 [ k_{i+}, k_{I+}] =& - \epsilon_{ij} t^j k_{I+} - T^J_{iI} k_{J+} \ , \qquad   [ k_{i+}, k_{I-}] = \quad \epsilon_{ij} t^j k_{I-} \ .
\end{aligned}\end{equation}
Therefore, the algebra is solvable. It consists of a semi-direct product of a two-dimen\-sional solvable algebra spanned by $k_{i+}$ with an Abelian algebra spanned by the other vectors. Note that in fact, $t^i k_{i+}$ commutes with all Killing vectors except the $k_{i+}$ and $k_{I+}$. These facts are also manifest in the covariant derivative of the vector fields in  \eqref{eq:cov_metric}, \eqref{eq:cov_B} and \eqref{eq:Cfields_covder}, as all non-Abelian parts of the field strength are proportional to $G^i$.

We could have performed a similar reduction of M-theory to five dimensions. The resulting five-dimensional $N=4$ gauged supergravity would exhibit the same gaugings as the ones discussed above. However, the global symmetry group would be $\mathbb{R}_+ \times SO(5,n-1)$. The above gaugings must hence necessarily be contained in this subgroup of $Sl(2) \times SO(6,n)$.

In Appendix~\ref{sec:SUSYvar}, we find an expression for the four-dimensional gravitino mass matrix in terms of the internal geometry of $SU(2)$ structures. It is block-diagonal if we set the internal field strengths to zero. In this case, each block is proportional to $(\sigma_a)^{ij} P^a$ with
\begin{equation}
P^a = \int_6 \e^{\phi-\rho_2-\rho_4} \bar K\wedge ( K \wedge \diff \bar K \wedge J^a + \epsilon_{abc} J^b \wedge \diff J^c) \ .
\end{equation}
We furthermore argue there that the theory admits a supersymmetric vacuum if the forms
\begin{equation} \label{eq:CYSU2structure}
J = J^3 + \tfrac12 \iu K \wedge \bar K  \ , \qquad  \Omega = K \wedge (J^1 + \iu J^2 ) \ ,
\end{equation}
satisfy the Calabi-Yau condition
\begin{equation}\label{CY_condition}
\diff J = 0 \ , \qquad \diff \Omega = 0 \ .
\end{equation}

%%%%%%%%%%%%%%%%%%%%%%%%%%
\section{Calabi-Yau manifolds with vanishing Euler number}
%%%%%%%%%%%%%%%%%%%%%%%%%%
\label{sec:CYEulerzero}
%%%%%%%%%%%%%%%%%%%%%%%%%%
\subsection{$SU(2)$ structures on Calabi-Yau manifolds}
%%%%%%%%%%%%%%%%%%%%%%%%%%
\label{sec:CY}

In the following, we want to study Calabi-Yau threefolds $Y$ with vanishing Euler number. By the Hopf theorem, Euler number zero implies the existence of a nowhere vanishing vector field $\hat K^1$ on the space.\footnote{The explicit construction of such a nowhere vanishing vector field on $Y$ might be difficult.} For a Calabi-Yau threefold, which has $SU(3)$ holonomy, this implies the reduction of the structure group to $SU(2)$, as follows. The complex structure $I$ on the Calabi-Yau relates $\hat K^1$ to a second nowhere vanishing vector $\hat K^2= I \hat K^1$, which is everywhere linearly independent of $\hat K^1$. They can hence be combined into a complex vector $K$ satisfying the first two equations in \eqref{eq:K_compatible}. This already provides the structure needed to define an almost product structure $P$ via \eqref{eq:almost_product_structure}. Since $K$ is a holomorphic vector with respect to $I$, we can define $J^a$, $a=1,\ldots,3$, in terms of the K\"ahler form $J$ and the holomorphic three-form $\Omega$ of the Calabi-Yau manifold via the relations \eqref{eq:CYSU2structure}. These satisfy the third condition in \eqref{eq:K_compatible}. Moreover, we can rescale $\Omega$ such that \eqref{eq:JaJb} is fulfilled. This thus completes the definition of an $SU(2)$-structure on $Y$.

In general, we cannot expect all moduli of a Calabi-Yau manifold to decompose into $SU(2)$ singlets and triplets; some of them might have an $SU(2)$ doublet component. These moduli will lie outside of the reduction ansatz outlined in Section~\ref{sec:cons_trunc} and will thus not appear in the effective action that we have derived for compactification on manifolds with $SU(2)$ structure. Instead, the $SU(2)$ doublet components will be part of $N=4$ massive gravitino multiplets that must be coupled to the $N=4$ gauged supergravity that we have derived in Section \ref{sec:reduction}. We will comment on these modes further below.

For a given Calabi-Yau manifold, it is non-trivial to find a reduction ansatz which closes under exterior differentiation, i.e.\ a set of forms that satisfy \eqref{eq:truncation_ansatz}. This set must be constructed on a case by case basis. It is not clear that this is always possible. In Section~\ref{sec:Enriques}, we provide such a set of expansion forms in the case of the Enriques Calabi-Yau. Further examples of Calabi-Yau manifolds of vanishing Euler characteristic can be found in \cite{Donagi:2008ht,Schulz:2012uj}. It would be interesting to construct analogous forms in these cases as well.

Let us now assume that a reduction ansatz satisfying the requirements of Section~\ref{sec:cons_trunc} exists and discuss its special form for the case of a Calabi-Yau manifold. A first requirement on the charges that appear in \eqref{eq:truncation_ansatz} comes from the fact that for proper Calabi-Yau spaces, the holonomy group is exactly $SU(3)$ and the first cohomology of $Y$ must therefore be zero. This translates into the fact that the exterior derivative on any linear combination of the nowhere vanishing one-forms $v^i$ is non-zero.\footnote{Note that on a compact manifold $Y$ a nowhere vanishing one-form can never be exact, as any function on $Y$ acquires its extremal points.} This implies that either $t^i \epsilon_{ij} t^j_I \ne 0$ or $t^1_I$ and $t^2_I$ are linearly independent. In other words, the rank of the matrix $(t^i, t^i_I)$ is maximal (i.e.\ two). The almost product structure of Calabi-Yau manifolds of Euler number zero hence cannot be integrable. As announced in the introduction, they thus lie outside of the ansatz of \cite{ReidEdwards:2008rd,Louis:2009dq,Danckaert:2011ju}. This implies furthermore that the charges of the corresponding $N=4$ gauged supergravity will always include some magnetic gaugings, cf.\ \eqref{eq:Killingvectors} and \eqref{eq:embed_tensor_f}.

Now let us determine the number of harmonic forms included in the reduction ansatz of Section~\ref{sec:cons_trunc}. Since $(t^i, t^i_I)$ has rank two, two linear combinations of the forms $(v^1 \wedge v^2,\omega^I)$ are exact. Furthermore, the rank of the matrix
\begin{equation} \label{eq:hatT_def}
 \hat T^{\hat I}_{iI} = (-\epsilon_{ij} t^j_I, T^J_{iI} )
\end{equation}
determines the number of closed forms among the two-forms $(v^1 \wedge v^2, \omega^I)$. It is $n+1-{\rm rk} (\hat T)$, giving rise to $n-1-{\rm rk} (\hat T)$ cohomology classes. This is the number of massless fields we expect to find parametrizing the $N=2$ K\"ahler moduli space. Similarly, there are ${\rm rk} (\hat T)$ exact three-forms within the $2n$ three-forms of the truncation ansatz, and $2n-{\rm rk} (\hat T)$ closed three-forms. This gives $2(n-{\rm rk} (\hat T))$ third-cohomology classes within the truncation ansatz. From this counting, we see that the number of massless modes parametrizing the complexified K\"ahler and complex structure moduli spaces agree. In Section~\ref{sec:4to2}, we will re-encounter this condition in the general analysis of supersymmetry breaking from $N=4$ to $N=2$.

In the following, we want to discuss the Calabi-Yau condition for an $SU(2)$-structure geometry of the form \eqref{param_forms}. The Calabi-Yau condition should define the $N=2$ vacuum condition inside the $N=4$ scalar field space. It says that the holomorphic three-form and the K\"ahler form of the Calabi-Yau threefold, given in \eqref{eq:CYSU2structure}, should be closed. This implies
\begin{equation}\label{N=2_conditions_forms}
 \diff (J^1 + \iu J^2 )\wedge K + (J^1 + \iu J^2 )\wedge \diff K = 0 \ , \qquad
\diff J^3 + \tfrac12 \iu \diff (K \wedge \bar K) = 0 \ .
\end{equation}
In terms of four-dimensional fields, the above relations can be rewritten as
\begin{equation} \label{eq:N=2_conditions}\begin{aligned}
k_i t^i_J \eta^{JI} \zeta^+_I & = 0 \ , \\
k_i \epsilon^{ij} T^I_{jK} \eta^{KJ}  \zeta^+_J  & = 0 \ , \\
\e^{\rho_4/2} T^J_{iI} \zeta^3_J  & = (k \cdot \epsilon \cdot k)_{ij} t^j_I  \ ,
\end{aligned}\end{equation}
where we have used \eqref{eq:quadratic_constraints} to simplify the second equation.
Using the parametrization \eqref{matrix_k}, we find
\begin{equation} \label{eq:N=2_conditions_exp}\begin{aligned}
(t^1_J + \tau t^2_J) \eta^{JI} \zeta^+_I & = 0\ , \\
( T^J_{1K}+\tau T^J_{2K})\eta^{KI}  \zeta^+_I & = 0 \ , \\
 \e^{\rho_4/2} T^J_{iI} \zeta^3_J  & = \epsilon_{ij} t^j_I \e^{\rho_2} \ .
\end{aligned}\end{equation}

Equations \eqref{eq:N=2_conditions_exp} describe the $N=2$ conditions for the scalars coming from the metric. For the $N=2$ vacuum, the ten-dimensional form fields should be closed as well,
\begin{equation} \label{N=2_conditions_pforms}
\diff B = 0\ , \qquad \diff C_1 = 0 \ , \qquad \diff C_3 = 0 \ .
\end{equation}
Inserting the parametrization \eqref{form_fields_expansion_IIA}, this yields the following conditions on the scalar fields\footnote{We also list the conditions on the two-forms $C_i$ here as they can be dualized to scalars.}
\begin{equation}\label{eq:N=2_conditions_BC}\begin{aligned}
 t^i_I b_{12} + \epsilon^{ij} T^J_{jI} b_J & = 0 \ , \\
 a_i & = 0 \ , \\
 C_i & = 0 \ , \\
 t^i_J \eta^{JI} c_{iI} & = 0 \ , \\
 (t^i \delta^I_J + \epsilon^{ij} T^I_{jJ} ) c_{iI} & = 0 \ .
\end{aligned}\end{equation}
Similarly, we find for the vector fields
\begin{equation}\label{eq:N=2_conditions_BC_vectors}\begin{aligned}
 G^i & = 0 \ , \\
 B_i & = 0 \ , \\
 t^i_I C_{12} + \epsilon^{ij} T^J_{jI} C_J & = 0 \ .
 \end{aligned}\end{equation}
We will discuss this in further detail in Section \ref{sec:4to2general}. In particular, we will find that these equations fix the values of all vector fields that acquire a mass due to the Higgs mechanism, while the modes `eaten' correspond to the exact forms on $Y$.

%%%%%%%%%%%%%%%%%%%%%%%%%%
\subsection{The Enriques Calabi-Yau}
%%%%%%%%%%%%%%%%%%%%%%%%%%
\label{sec:Enriques}

An example of a Calabi-Yau threefold of Euler number zero has been given in \cite{Ferrara:1995yx}. It is constructed by quotienting $K3 \times T^2$ by a freely acting involution $\tau$. $\tau$ acts as the Enriques involution on $K3$ and as the standard involution on the torus.\footnote{For a review of the the Enriques involution on K3, see \cite{Barth:1984}.}
The resulting manifold, also referred to as the Enriques Calabi-Yau, has very special properties. It is self-mirror and thereby does not have experience any worldsheet instanton corrections at the two-derivative level \cite{Ferrara:1995yx}. There is however a non-trivial pattern of higher-derivative corrections for this background (graviphoton-curvature couplings), some of which have been computed in \cite{Klemm:2005pd}.

We begin by considering the harmonic forms on the Enriques Calabi-Yau $Y = (K3\times T^2)/\mathbb{Z}_2^\tau$. For their construction, let us consider the orbifold limit $T^4/\mathbb{Z}_2^\sigma$ of $K3$. $\sigma$ here is the standard involution on the four-torus. If we introduce coordinates $x^a$, $a=1,\dots, 6$, on $T^6$ with periodicity $x^a = x^a +1$, the involutions $\sigma$ and $\tau$ are given by \cite{Gopakumar:1996mu}
\begin{equation}  \label{ex_involutions}
\begin{aligned}
\sigma: \qquad \qquad & (x^1,x^2,x^3,x^4,x^5,x^6) \to (-x^1,-x^2,-x^3,-x^4,x^5,x^6) \ , \\
\tau: \qquad \qquad & (x^1,x^2,x^3,x^4,x^5,x^6) \to (x^1,x^2+1/2,-x^3,-x^4+1/2,-x^5,-x^6) \ .
\end{aligned}
\end{equation}
The involution $\sigma$ has sixteen fixed points. Blowing up the resulting orbifold singularities yields sixteen exceptional divisors $E_{(i,j,k,l)}$, $i,j,k,l= 0, \frac{1}{2}$, with the subscript denoting the $x^i$-coordinates of the fixed point. By considering the linear combinations
\be  \label{ex_lin_comb}
E_{(i,j,k,l)} \pm E_{(i,j+\frac{1}{2},k,l +\frac{1}{2})} \ ,
\ee
we obtain two sets of eight cycles $E_\pm^{\alpha}$, with the subscript denoting  the transformation behavior under the involution $\tau$. The intersection matrix of both $E_+^{\alpha}$ and $E_-^{\alpha}$ is minus twice the Cartan matrix of the exceptional Lie group $E_8$.

The harmonic forms on $Y$ can be read off from \eqref{ex_involutions}. They consist of all anti-symmetric combinations of the $\diff x^a$ that are even under both $\sigma$ and $\tau$. There are no harmonic one-forms on $Y$, a necessary condition for a proper Calabi-Yau manifold. The harmonic two-forms are given by combinations of $\diff x^1 \wedge \diff x^2$, $\diff x^3 \wedge \diff x^4$, $\diff x^5 \wedge \diff x^6$, which define volume forms of sub-two-tori  of the original  unquotiented $T^6$, and the (1,1) forms $E^+_{\alpha}$ dual to the exceptional cycles. The harmonic three-forms are given by those combinations of the one-forms that have one leg on each such two-torus, i.e.\ $\diff x^1 \wedge \diff x^3 \wedge \diff x^5$ etc., as well as $E^-_{\alpha}\wedge \diff x^5 $ and $E^-_{\alpha}\wedge \diff x^6 $.
Thus, $Y$ has $h^{1,1} = 11$ and $h^{2,1}= 11$, and its Euler number vanishes, as announced.

The massless spectrum of a conventional Calabi-Yau compactification on $Y$ gives an ungauged $N=2$ theory with moduli space
\begin{equation}
 M_{N=2} = \frac{Sl(2,\mathbb{R})}{SO(2)} \times \frac{SO(2,10)}{SO(2)\times SO(10)} \times \frac{SO(4,12)}{SO(4)\times SO(12)} \ .
\end{equation}
The first two factors together carry a special K\"ahler structure, the last factor a quaternionic K\"ahler structure.
As $Y$ is obtained as a quotient of $K3\times T^2$, $M$ is naturally a submanifold of the $N=4$ moduli space associated to this compactification manifold,
\begin{equation}
M_{K3 \times T^2} = \frac{Sl(2,\mathbb{R})}{SO(2)} \times \frac{SO(6,22)}{SO(6)\times SO(22)} \ .
\end{equation}

Now let us describe the construction of expansion forms on $Y$. The first step is to define two nowhere vanishing one-forms $v^i$, $i=1,2$, on $Y$ that are linearly independent at any point, yielding a nowhere vanishing two-form
\begin{equation}
\vol_2 \equiv  v^1 \wedge v^2 \ .
\end{equation}
We start by constructing such one-forms on $T^6$. If they are well-defined on the quotient and survive the blow-up, they sit in the image of the pullback map $T^*Y \rightarrow T^* T^6$, hence give nowhere vanishing one-forms on $Y$. We define
\begin{equation}
v^1 = \sin(2\pi x^2) \diff x^3 + \cos(2\pi x^2) \diff x^5 \ , \qquad
v^2 = \sin(2\pi x^2) \diff x^4 + \cos(2\pi x^2) \diff x^6 \ ,
\end{equation}
so that
\begin{equation}\begin{aligned}
\vol_2 = & \tfrac12 (\diff x^3 \wedge \diff x^4 + \diff x^5 \wedge \diff x^6) - \tfrac12 \cos(4\pi x^2) (\diff x^3 \wedge \diff x^4 - \diff x^5 \wedge \diff x^6) \\  & + \tfrac12 \sin(4\pi x^2) (\diff x^3 \wedge \diff x^6 - \diff x^4 \wedge \diff x^5) \ ,
\end{aligned} \end{equation}
which is clearly nowhere vanishing. The one-forms $v^i$ span the two-dimensional component $T^*_2 Y$ of $T^* Y$ at each point over the base $Y$. We need to work a little harder to describe the orthogonal complement $T^*_4 Y$ explicitly, as it is non-trivially fibered over $Y$. We begin therefore by considering the quotient $\tilde Y = T^6/\mathbb{Z}_2^\tau$, where we can specify a global basis for $T^*_4 \tilde Y$, spanned by
\begin{equation}
u^1 = \cos(2\pi x^2) \diff x^3 - \sin(2\pi x^2) \diff x^5 \ , \qquad
u^2 = \cos(2\pi x^2) \diff x^4 - \sin(2\pi x^2) \diff x^6 \ ,
\end{equation}
and $\diff x^1$ and $\diff x^2$. This shows that $\tilde Y$ is a twisted six-torus, with $T^* \tilde Y$ spanned by $\diff x^i$, $u^i$ and $v^i$, $i=1,2$. We can define a product structure on $\tilde Y$ corresponding to the split $T^* \tilde Y = T^*_2 \tilde Y \oplus T^*_4 \tilde Y$ via
\begin{eqnarray}
P_{\tilde Y}  &= &  dv^1 \partial_{v^1} + dv^2 \partial_{v^2} - dx^1 \partial_1 - dx^2 \partial_2 - du^1 \partial_{u^1} + du^2 \partial_{u^2}  \\
&=&  - \diff x^1 \partial_1 - \diff x^2 \partial_2 + \cos(4\pi x^2) (\diff x^5 \partial_5+\diff x^6 \partial_6-\diff x^3 \partial_3-\diff x^4 \partial_4) \nonumber \\ & &+\sin(4\pi x^2 )(\diff x^5 \partial_3+\diff x^3 \partial_5+\diff x^6 \partial_4+\diff x^4 \partial_6) \ .
\end{eqnarray}
Now note that $P_{\tilde Y}$ descends to a map $T^* Y \rightarrow T^* Y$ after quotienting by $\sigma$, thus defining the sought after almost product structure on $Y$. Since the $\diff x^i$ and $u^i$ are odd under $\sigma$, the identity structure of $\tilde Y$ is enlarged to $SU(2)$ after this quotienting (and the subsequent blow-up).

The torsion of $\tilde Y$ is specified by the algebra
\begin{equation}\label{ex_twisted_torus_algebra}
\begin{aligned}
\diff v^1 =& \quad 2 \pi \diff x^2\wedge u^1 \ , \\
\diff v^2 =& \quad 2 \pi \diff x^2\wedge u^2 \ , \\
\diff u^1 =& - 2 \pi \diff x^2\wedge v^1 \ , \\
\diff u^2 =& -2 \pi \diff x^2\wedge v^2 \ ,
\end{aligned}
\end{equation}
with $\diff x^i$ being closed. The algebra spanned by these one-forms and their wedge products is manifestly closed under the operations $\diff$, $\wedge$, and $\ast$. The preimage of this algebra on $\oplus_i \Lambda^i T^*Y$ inherits this property. It is spanned by the (preimage of the) one-forms $v^1$ and $v^2$, as well as the six (preimages of the) two-forms obtained from all possible wedge products among the 1-forms $\diff x^i$ and $u^i$, $i=1,2$. We have thus succeeded in constructing a set of forms on $Y$ that satisfy the conditions of our reduction ansatz outlined in section \ref{sec:cons_trunc}.

The ansatz so far does not contain the harmonic two-forms $E^\alpha_+$ dual to the exceptional divisors on $Y$. These are all located at $x^2=0,\tfrac12$, where $v^1=\diff x^5$ and $v^2=\diff x^6$. Therefore, they are elements of $T^*_4 Y$. Being in addition closed and anti-self-dual, they can be safely included as generators in our reduction ansatz. The ansatz misses one additional harmonic form, $\diff x^3 \wedge \diff x^4 - \diff x^5 \wedge \diff x^6$. This form lies in the preimage of a linear combination of two-forms on $\tilde Y$ which involves $v^1 \wedge u^2 - v^2 \wedge u^1$, hence does not respect the $T_2^* Y \oplus T_4^* Y$ split of the cotangent space. Under penalty of having to include massive gravitini, as explained in section \ref{sec:cons_trunc}, we must hence exclude this harmonic form from our reduction ansatz. In total, the two-forms in our ansatz are hence spanned by $v^1 \wedge v^2 \in \Lambda^2 T^*_2 Y$ and
\begin{equation}\begin{aligned}
(\omega)^{I=1,\ldots,14} = ( & \diff x^1 \wedge u^1 -\diff x^2 \wedge u^2, \diff x^1 \wedge u^2 + \diff x^2 \wedge u^1, u^1\wedge u^2 +\diff x^1 \wedge \diff x^2, \\ & \diff x^1 \wedge u^1 +\diff x^2 \wedge u^2, \diff x^1 \wedge u^2 - \diff x^2 \wedge u^1,u^1\wedge u^2- \diff x^1 \wedge \diff x^2, E^\alpha_+ ) \ ,
\end{aligned}\end{equation}
$\omega^I \in \Lambda^2 T^*_4 Y$. We have here organized the $\omega^I$ in terms of three self-dual and eleven anti-self-dual two-forms. They satisfy $\omega^I \wedge \omega^J = \eta^{IJ} \vol_4^{(0)}$ with
\begin{equation} \label{inter_m}
 (\eta)^{IJ} = \left(\begin{aligned}
   (\delta)^{ij} && 0 && 0\\
  0 && - (\delta)^{ij} && 0\\
  0 && 0 && 2(I_8)^{\alpha \beta}
  \end{aligned} \right) \ .
\end{equation}
Here, $i,j = 1, \dots, 3$, $\alpha, \beta= 1, \ldots, 8$, and $I_8$ is the negative definite Cartan matrix of the Lie group $E_8$. In this basis, the torsion algebra can be computed from \eqref{ex_twisted_torus_algebra} to be
\begin{equation}\label{ex_torsion_algebra}
\begin{aligned}
\diff v^1 =&  \quad \pi (\omega^2 - \omega^5) \ , \\
\diff v^2 =&  - \pi (\omega^1 - \omega^4) \ , \\
\diff \omega^1 =\diff \omega^4 =& \quad \pi v^1 \wedge (\omega^3 -\omega^6) \ , \\
\diff \omega^2 =\diff \omega^5 =& \quad \pi v^2 \wedge (\omega^3 -\omega^6) \ , \\
\diff \omega^3 =\diff \omega^6=& - \pi v^1\wedge (\omega^1 - \omega^4) - \pi v^2\wedge (\omega^2 - \omega^5) \ , \\
\end{aligned}
\end{equation}
with all other derivatives being zero.

The scalar field space of a type II compactification on this background is
\begin{equation}\label{moduli_space_N=4}
M_{N=4} = \frac{Sl(2,\mathbb{R})}{SO(2)} \times \frac{SO(6,14)}{SO(6)\times SO(14)} \ .
\end{equation}
From \eqref{ex_torsion_algebra} we can read off the gaugings to be
\begin{equation}
 (t)^{i=1,2}_{I=1,\ldots,14} = \pi \left(\begin{aligned}
   0 && 1 && 0&& 0&& -1 && 0&& 0_\alpha \\
   -1 && 0 && 0 && 1 && 0 && 0 && 0_\alpha
  \end{aligned} \right) \ ,
\end{equation}
as well as
\begin{equation} \label{T1_ECY}
 (T)^I_{1J} = \pi \left(\begin{aligned}
   0 && 0 && 1 && 0&& 0 && - 1 && 0_\alpha\\
  0 && 0 && 0&& 0&& 0 && 0 && 0_\alpha\\
   -1 && 0 && 0 && 1 && 0 && 0 && 0_\alpha\\
   0 && 0 && 1 && 0&& 0 && - 1 && 0_\alpha\\
  0 && 0 && 0&& 0&& 0 && 0 && 0_\alpha\\
   -1 && 0 && 0 && 1 && 0 && 0&& 0_\alpha\\
   0_\beta && 0_\beta && 0_\beta && 0_\beta && 0_\beta && 0_\beta&& 0_{\alpha\beta}
  \end{aligned} \right) \ ,
\end{equation}
and
\begin{equation}  \label{T2_ECY}
 (T)^I_{2J} = \pi \left(\begin{aligned}
  0 && 0 && 0&& 0&& 0 && 0 && 0_\alpha\\
   0 && 0 && 1 && 0&& 0 && - 1 && 0_\alpha\\
   0 && -1 && 0 &&  0 && 1 && 0 && 0_\alpha\\
  0 && 0 && 0&& 0&& 0 && 0 && 0_\alpha\\
   0 && 0 && 1 && 0&& 0 && - 1 && 0_\alpha\\
   0 && -1 && 0 &&  0 && 1 && 0&& 0_\alpha\\
   0_\beta && 0_\beta && 0_\beta && 0_\beta && 0_\beta && 0_\beta&& 0_{\alpha\beta}
  \end{aligned} \right) \ .
\end{equation}
In Section~\ref{sec:Enriques4to2} we will discuss the spontaneous supersymmetry breaking to $N=2$ for this example.

%%%%%%%%%%%%%%%%%%%%%%%%%%
\section{Spontaneous breaking of $N=4$ to $N=2$}
%%%%%%%%%%%%%%%%%%%%%%%%%%
\label{sec:4to2}

The $SU(2)$-structure reduction we performed above specialized to any Calabi-Yau background should yield an $N=4$ gauged supergravity which exhibits $N=2$ Minkowski vacua. The supersymmetry conditions translate into \eqref{eq:N=2_conditions_exp} and \eqref{eq:N=2_conditions_BC}. Spontaneous partial supersymmetry breaking in $N=4$ supergravity has already been discussed in \cite{Ferrara:1983gn,Wagemans,Tsokur:1994gr,Andrianopoli:2002rm,Dall'Agata:2009gv,Dibitetto:2011gm,HLS}. Our results give new examples of such a breaking. Before we discuss them in more detail, let us review general facts on spontaneous supersymmetry breaking to $N=2$. See \cite{HLS} for a similar (but independently performed) analysis.

When $N=4$ supergravity is spontaneously broken to $N=2$, the fields in the $N=4$ gravity and vector multiplets rearrange non-trivially into $N=2$ multiplets. In particular, two of the four gravitini must acquire a mass and form massive gravitino multiplets. There are two kinds of massive gravitino multiplets in $N=2$ supergravity, a long multiplet and a short one that transforms as a doublet under a central charge \cite{Ferrara:1983gn}. Let us first focus on the case that the massive gravitini assemble in long massive multiplets. These multiplets each have four (massive) vectors, six spinors and four scalars. There can be additional massive fields in $m_v$ massive vector multiplets, consisting of a massive vector, four spinors and five scalars. The massless fields of $N=2$ supergravity assemble as usual into a gravity multiplet as well as $n_v$ vector and $n_h$ hypermultiplets. In descending from $N=4$, one can expect correlations between the numbers $n_v$ and $n_h$ of $N=2$ multiplets, as (massless) $N=4$ gauged supergravity knows only vector multiplets. Indeed, as detailed in Table \ref{tab:4to2longmultiplets}, for this pattern of supersymmetry breaking, the number of hypermultiplets of the effective $N=2$ theory exceeds the number of vector multiplets by one, i.e.\ $n_v=n_h-1$. We found the same relation from our truncation ansatz in Section~\ref{sec:CY}. This is incidentally also the relation that holds for regular compactifications on Calabi-Yau manifolds of Euler number zero.
\begin{table}[ht!]
\centering
\begin{tabular}{rl|cr}
{\bf Massless} &  {\bf $N=4$ multiplets:}\\
&gravity multiplet &\quad& ({\bf 2}), 4 ({\bf 3/2}), 6 ({\bf 1}), 4 ({\bf 1/2}), 2 ({\bf 0}) \\
$n_v + m_v+3$ & vector multiplets && ({\bf 1}), 4 ({\bf 1/2}), 6 ({\bf 0}) \\ \hline
{\bf Massless} &{\bf $N=2$ multiplets:}\\
&gravity multiplet && ({\bf 2}), 2 ({\bf 3/2}), \ \ ({\bf 1}) \hspace{2.64cm} \\
$n_v$ &vector multiplets && ({\bf 1}), 2 ({\bf 1/2}), 2 ({\bf 0}) \\
$n_v+1$ &hypermultiplets && 2 ({\bf 1/2}), 4 ({\bf 0}) \\ \hline
{\bf Massive} & {\bf $N=2$ multiplets:}\\
2 &gravitino multiplets && ({\bf 3/2} +{\bf 1/2}), 4 ({\bf 1}+{\bf 0}), 6 ({\bf 1/2}), 4 ({\bf 0}) \\
$m_v$&vector multiplets && ({\bf 1}+{\bf 0}), 4 ({\bf 1/2}), 5 ({\bf 0})
\end{tabular}
\caption{\small{The $N=4$ gravity and vector multiplet degrees of freedom rearrange in massless and massive $N=2$ multiplets. The spins of the various (real) fields are displayed in bold-face. Here the gravitino multiplets are long multiplets. There is one more massless hypermultiplet than massless $N=2$ vector multiplets. \label{tab:4to2longmultiplets}}}
\end{table}
If the massive gravitini assemble in short multiplets that form a doublet under a central charge, the multiplets are smaller. Each of the two gravitino multiplets contains two massive vectors and one spinor. Moreover, their masses must be equal. The counting of fields, displayed in Table~\ref{tab:4to2shortmultiplets}, now gives a different result: the number of massless $N=2$ vector multiplets must be larger by two than the number of massless hypermultiplets. We thus do not expect this breaking pattern to appear for the $N=4$ gauged supergravities related to $\chi=0$ Calabi-Yau backgrounds.
\begin{table}[ht!]
\centering
\begin{tabular}{rl|cr}
{\bf Massless} &  {\bf $N=4$ multiplets:}\\
&gravity multiplet &\quad& ({\bf 2}), 4 ({\bf 3/2}), 6 ({\bf 1}), 4 ({\bf 1/2}), 2 ({\bf 0}) \\
$n_h + m_v +1$ & vector multiplets && ({\bf 1}), 4 ({\bf 1/2}), 6 ({\bf 0}) \\ \hline
{\bf Massless} &{\bf $N=2$ multiplets:}\\
&gravity multiplet && ({\bf 2}), 2 ({\bf 3/2}), \ \ ({\bf 1}) \hspace{2.64cm} \\
$n_h +2 $ &vector multiplets && ({\bf 1}), 2 ({\bf 1/2}), 2 ({\bf 0}) \\
$n_h$ &hypermultiplets && 2 ({\bf 1/2}), 4 ({\bf 0}) \\ \hline
{\bf Massive} & {\bf $N=2$ multiplets:}\\
2 &gravitino multiplets && ({\bf 3/2} +{\bf 1/2}), 2 ({\bf 1}+{\bf 0}), \ \ ({\bf 1/2}) \hspace{1cm}  \\
$m_v$&vector multiplets && ({\bf 1}+{\bf 0}), 4 ({\bf 1/2}), 5 ({\bf 0})
\end{tabular}
\caption{\small{The $N=4$ gravity and vector multiplet degrees of freedom rearrange in massless and massive $N=2$ multiplets. The spins of the various (real) fields are displayed in bold-face. Here the gravitino multiplets are short multiplets. There are two more massless $N=2$ vector multiplets than massless hypermultiplets. \label{tab:4to2shortmultiplets}}}
\end{table}

%%%%%%%%%%%%%%%%%%%%%%%%%%
\subsection{General Calabi-Yau manifolds of vanishing Euler number}
%%%%%%%%%%%%%%%%%%%%%%%%%%
\label{sec:4to2general}

We now study $N=4 \to N=2$ supersymmetry breaking for general $N=4$ gauged supergravities coming from $SU(2)$ structure reductions of Calabi-Yau spaces. We begin by considering the metric sector. There are two isometries acting on the scalars coming from the ten-dimensional metric. These read
\begin{equation} \begin{aligned}
\delta_i \zeta^a_I & = - \tilde T^J_{iI} \zeta^a_J \ , \\
\delta_i \rho_2 & = - \epsilon_{ij} t^j \ , \\
\delta_i \rho_4 & = \epsilon_{ij} t^j \ , \\
\delta_i \tau & = - (1, \tau)_i (t^1 + \tau t^2) \ .
\end{aligned} \end{equation}
Thus,
\be  \label{trans_J}
\delta_i J = \delta_i ( e^{\frac{\rho_4}{2}} \zeta_I^3 \omega^I + e^{\rho_2} v^1 \wedge v^2 ) = - \epsilon_{ij} e^{\rho_2} dv^j \ ,
\ee
and
\be
\delta_i \Omega = \delta_i  \left( \frac{e^{\frac{\rho_2 + \rho_4}{2}}}{\sqrt{\Im \tau}} (v^1 + \tau v^2) \wedge \zeta_J^+ \omega^J   \right) = \frac{e^{\frac{\rho_2 + \rho_4}{2}}}{\sqrt{\Im \tau}} (1, \tau)_i \zeta_J^+ \diff \omega^J - i \Im (1 , \tau)_i t^i \Omega  \ .
\ee
The isometries hence act by adding exact pieces to $J$ and $\Omega$, and by rescaling $\Omega$. As the metric is determined by the cohomology class of $J$
and projectively by the cohomology of $\Omega$, these directions in field space are unphysical. Correspondingly, when the $G_i$, the fields which gauge these isometries, acquire a mass, these directions are modded out of the scalar field space.

To count degrees of freedom, it will be useful to introduce the vectors
\begin{equation}
 \hat \zeta^3_{\hat I} = ( \e^{\rho_2}, \e^{\rho_4/2} \zeta^3_I) \ ,
\end{equation}
and
\begin{equation}
 \hat t^i_{\hat I} = ( t^i, t^i_I) \ .
\end{equation}
In terms of these, the isometries act as shifts
\begin{equation}\label{eq:zeta3shifts}
 \delta_i \hat \zeta^3_{\hat I} = - \e^{\rho_2} \epsilon_{ij} \hat t^j_{\hat I} \ .
\end{equation}
Similarly, we can define
\begin{equation} \label{eq:def_hatzeta+}
\hat \zeta^+_{iI} =  (k^1_i + \iu k^2_i) \zeta^+_{I} \ ,
\end{equation}
so that the gauged isometries act as
\begin{equation}\label{eq:zeta+shifts}
\delta_j \hat \zeta^+_{iI} = (\epsilon_{ij} t^k \delta^J_I - \delta^k_i \tilde T^J_{jI}) \hat \zeta^+_{kJ} \ .
\end{equation}
The closure conditions \eqref{eq:N=2_conditions} on $J$ and $\Omega$ in terms of these variables read
\begin{equation}\label{eq:N=2_conditions_general} \begin{aligned}
 \hat T^{\hat I}_{iI} \hat \zeta^3_{\hat I} & = 0 \ , \\
 \hat T^{\hat I}_{iI} \epsilon_{ij} \eta^{IJ} \hat \zeta^+_{jJ} & = 0 \ .
\end{aligned} \end{equation}
The matrix $\hat T^{\hat I}_{iI}$ was defined in \eqref{eq:hatT_def}. Note that the compatibility of these equation with the gauged isometries \eqref{eq:zeta3shifts} and \eqref{eq:zeta+shifts} is guaranteed by \eqref{eq:quadratic_constraints}.

Next, we want to understand how the constraints \eqref{zeta_ab}, which can be written as
\begin{equation}\label{eq:zeta3zeta3}
\zeta^3_I \eta^{IJ} \zeta^3_J = 1 \ ,
\end{equation}
\begin{equation}\label{eq:zeta+zeta+}
 \zeta^+_I \eta^{IJ}  \zeta^+_J =0 \ , \qquad  \zeta^+_I \eta^{IJ}  \zeta^-_J =2  \ ,
\end{equation}
\begin{equation}\label{eq:zeta+zeta3}
 \zeta^3_I \eta^{IJ} \zeta^+_J = 0 \ ,
\end{equation}
are compatible with the factorization of the $N=2$ moduli space. Equation \eqref{eq:zeta+zeta3} apparently couples $J$ and $\Omega$. Notice however that this equation imposes two real constraints, which is the number of isometries acting on $\zeta_I^a$. The coupling of $\zeta^3$ with $\zeta^+_I$ can thus be absorbed by these isometries; $J$ and $\Omega$ are only coupled via exact contributions, which are not physical. The K\"ahler structure is thus parametrized by $n-{\rm rk} (\hat T)-1$ real variables: the $n+2$ real parameters $\hat \zeta^3$ subject to the linear constraint that is the first condition in \eqref{eq:N=2_conditions_general}, the quadratic constraint \eqref{eq:zeta3zeta3}, and the two isometries (\ref{eq:zeta+shifts}) . Let us preform the analogous counting for the complex structure moduli space: one complex direction is fixed by fixing the gauge freedom \eqref{eq:zeta+shifts}. Furthermore, \eqref{eq:zeta+zeta+} together with $U(1)$ rotations of $\hat \zeta^+$ fix two more complex degrees of freedom. Naively \eqref{eq:N=2_conditions_general} would count for ${\rm rk} (\hat T)$ complex conditions. However, since $\hat \zeta^+$ is defined as the direct product \eqref{eq:def_hatzeta+}, there is an additional local U(1) symmetry that reduces the number of independent equations in \eqref{eq:N=2_conditions_general} by one. Therefore, we find that $\hat \zeta^+$ depends on $n-{\rm rk} (\hat T)-1$ complex degrees of freedom.

$\hat \zeta^+_{iI}$ forms a holomorphic symplectic vector with respect to the symplectic pairing $\epsilon_{ij} \eta^{IJ}$ that is subject to the second condition in \eqref{eq:N=2_conditions_general}. It coordinatizes a special-K\"ahler space with the K\"ahler potential
\begin{equation}\label{eq:Kahlerpot_cc}
 K = - \log \iu \Omega \wedge \bar{\Omega} =  - \log ( \iu \hat \zeta^+_{iI} \epsilon_{ij} \eta^{IJ} \hat \zeta^-_{jJ}) \ ,
\end{equation}
where $\hat \zeta^-_{iI}$ is the complex conjugate of $\hat \zeta^+_{iI}$. The expression for the K\"ahler potential in terms of the holomorphic three-form $\Omega$ is valid in terms of projective coordinates $\hat \zeta$. We consequently disregard the second constraint of (\ref{eq:zeta+zeta+}). Note that interpreting $\hat \zeta_{iI}$ as projective coordinates allows us to scale $k_i$ and $\zeta_I$ independently.

Let us now understand the form of the holomorphic prepotential. We will rescale $k^1_1 + \iu k^2_1$ to be a constant, leaving $k^1_2 + \iu k^2_2$ as a free complex parameter. Next, we perform a linear coordinate transformation such that, retaining the name of the variables, the first constraint of (\ref{eq:zeta+zeta+}) takes the form $ \zeta_1^+ \zeta_2^+ + \zeta_M \tilde \eta^{MN} \zeta_N = 0$, with $M,N=1, \ldots,n-2,$ and $\tilde \eta$ having signature $(1,n-3)$. Rescaling $\zeta$ to fix $\zeta_1^+=1$ yields $\zeta_2^+$ as a quadratic expression in the remaining $n-2$ variables. The tensor product \eqref{eq:def_hatzeta+} thus gives rise to a symplectic vector with one constant component, $n-1$ linear components, $n-1$ quadratic components, and one cubic component. This is the hallmark of a cubic prepotential. It reads
\begin{equation}\label{eq:prepSTUzeta}
{\cal F}_{\Omega} = \tfrac12 \tau (\zeta_M \tilde \eta^{MN}\zeta_N) \ .
\end{equation}
Before imposing the Calabi-Yau condition (\ref{eq:N=2_conditions_general}), the $\hat \zeta$ thus parametrize the special-K\"ahler space
\begin{equation}\label{eq:moduliSl2Sl2n}
M_{{\rm SK},0} = \frac{Sl(2)}{SO(2)} \times \frac{SO(2,n-2)}{SO(2) \times SO(n-2)} \ .
\end{equation}
Imposing the linear constraint (\ref{eq:N=2_conditions_general}) on $\hat \zeta_{iI}^+$ leads to a K\"ahler subspace that we will now argue to be special-K\"ahler as well. If we consider the linear constraint in the form \eqref{eq:N=2_conditions}, they form a set of linear equations on $\zeta_{I}^+$ with coefficients that might depend linearly on $\tau$. If the coefficients depend non-trivially on $\tau$, the number of coordinates in $\zeta_{I}^+$ simply reduces by ${\rm rk} (\hat T)$ and the number of components in $\hat \zeta_{iI}^+$ by $2\,{\rm rk} (\hat T)$, while the general form of the special-K\"ahler structure is unaltered. In this case, the $N=2$ moduli space takes the form of $M_{{\rm SK},0}$ with $n$ replaced by $n-{\rm rk} (\hat T)$, cf.\ \eqref{eq:N=2_conditions_general}. In the more generic case that $\tau$ does appear non-trivially in \eqref{eq:N=2_conditions}, some coordinates will be eliminated while others will depend linearly on $\tau$. Therefore, the holomorphic prepotential will no longer retain the factorized form \eqref{eq:prepSTUzeta}. It will however remain cubic. The scalar field space will in general not correspond to a symmetric space though it will remain a homogeneous space, all of which have been classified in \cite{classification}.
This determines the $N=2$ moduli space of complex structures. In particular, we have shown that $M_{{\rm SK}}$ is special-K\"ahler with cubic prepotential.

We now turn to the form fields. Their expansion in \eqref{form_fields_expansion_IIA} involves three types of internal forms: these internal differential forms split into those that are exact, closed but not exact and those that are not closed. If an expansion form is closed, the corresponding four-dimensional field turns out to be massless, as no potential terms are generated. If the internal form is not closed, then a potential term is generated. In the case of four-dimensional scalars, this is just a scalar potential that gives a mass to exactly these fields. In the case of four-dimensional vectors, they instead show up in the covariant derivative of some massless scalars (which were expanded in the corresponding exact form of one degree higher). The vector acquires a mass during the super-Higgs mechanism and eats up the scalar during this process. By this, all remaining fields of the effective $N=2$ theory come from expansion in non-trivial cohomology classes, as it is expected from the standard Calabi-Yau reduction.
Let us now discuss this in detail.

We begin with the $B$ field. From (\ref{eq:Killingvectors}), we read off the action of the two isometries gauged by $G^i$ on the internal components of $B$,
\be
\delta_i b_{12} = - \epsilon_{ij} t^j b_{12} \ , \quad \delta_i b_I = - T_{iI}^J b_J \ .
\ee
Invoking the Calabi-Yau constraints (\ref{eq:N=2_conditions_BC}), this yields
\be
\delta_i B^{\rm int} = \delta_i (b_{12} v^1 \wedge v^2 + b_i \omega^I ) = - \epsilon_{ij} b_{12} \diff v^j  \ ,
\ee
which combines nicely with the transformation (\ref{trans_J}) of $J$ to
\be
\delta_i (B^{\rm int} + i J) = - \epsilon_{ij} ( b_{12} + i e^{\rho_2}) \diff v^j  \ .
\ee
Introducing the variable $b_{\hat I} = (b_{12} , b_I)$, we see that the Calabi-Yau constraint (\ref{eq:N=2_conditions_BC}) can be written as
\begin{equation}
 \hat T^{\hat I}_{iI} \hat b_{\hat I}  = 0 \ ,
\end{equation}
which has the same form as \eqref{eq:zeta3shifts}. Hence, both the isometries and the constraints are holomorphic in the variable $\hat b + \iu \hat \zeta^3$.

Analogously to the complex structure moduli space, $\hat b + \iu \hat \zeta^3$ parametrizes the special K\"ahler space of complexified K\"ahler moduli, that is the subspace defined by complexification of \eqref{eq:N=2_conditions_general} within $M_{{\rm SK},0}$, given in \eqref{eq:moduliSl2Sl2n}.

It remains to discuss the fate of the scalars coming from the Ramond-Ramond fields in the spontaneous $N=4 \to N=2$ supersymmetry breaking. The conditions \eqref{eq:N=2_conditions_BC} and \eqref{eq:N=2_conditions_BC_vectors} eliminate all modes that correspond to non-closed forms on $Y$.
In particular, we find
\begin{equation} \label{eq:N2cond_RR}\begin{aligned}
 a_i =& 0 \ , \\
 \hat T^{\hat I}_{iI} \epsilon_{ij} \eta^{IJ} c_{jJ} =& 0 \ , \\
 \hat T^{\hat I}_{iI} C_{\hat I} =& 0 \ , \\
 \hat t^i_{\hat I} \tilde C^{\hat I} =& 0 \ ,
\end{aligned}\end{equation}
where we have defined
\begin{equation}
C_{\hat I} = ( C_{12}, C_I) \ , \qquad  \tilde C_{\hat I} = ( \tilde C_{12}, \tilde C_I) \ ,  \qquad \hat t^i_{\hat I} = (t^i, t^i_I) \ .
\end{equation}
The scalars corresponding to the shift isometries
\begin{equation}
\delta_{\hat I} c_{iI} = \hat T^{\hat I}_{iI}
\end{equation}
are eaten by the vectors $C_{\hat I}$ to give them a mass. Furthermore, the scalars $\gamma_i$ are eaten due to the Higgs mechanism that gives a mass to the magnetic vectors $\hat t^i_{\hat I} \tilde C^{\hat I}$. Equivalently, one could say that the vectors corresponding to the $t^i_{\hat I}$ directions are eaten by the two-forms $C_i$ to acquire a mass.

The remaining vector fields are those $C_{\hat I}$ that satisfy \eqref{eq:N2cond_RR} and that are invariant under the shifts
\begin{equation}
\delta_i C_{\hat I}= t^i_{\hat I} \ .
\end{equation}
Therefore, there is exactly one vector for each complex scalar $\hat b_{\hat I} + \iu \hat \zeta^3_{\hat I}$ that remains in the $N=2$ theory. Together, they form $N=2$ vector multiplets. Note that the $N=2$ vector multiplet moduli space is not necessarily a direct product, in contrast to the case of purely electric gaugings \cite{HLS}.

Similarly, the remaining scalars $c_{jJ}$ that obey \eqref{eq:N2cond_RR} form a real symplectic vector under the symplectic product $\epsilon_{ij} \eta^{IJ}$. Together with $\phi$ and $\beta$, they build the classical c-map \cite{cmap} over the special-K\"ahler space spanned by $\hat \zeta^+_{iI}$, with the K\"ahler potential \eqref{eq:Kahlerpot_cc}.

In total, we find the standard structure of classical $N=2$ supergravity, with a cubic prepotential. However, we in general cannot incorporate the full $N=2$ massless spectrum of an ordinary Calabi-Yau compactification in our $N=4$ description without including couplings to $N=4$ massive gravitino multiplets.

In ordinary Calabi-Yau reductions, one obtains an $N=2$ supergravity theory with cubic prepotential as well. This prepotential is modified away from cubic form by worldsheet instanton corrections. The $N=4$ context of our setup is protected against such quantum effects. We hence need to confront the question of how these corrections arise upon breaking to $N=2$.\footnote{An initial guess that Calabi-Yau manifolds with vanishing Euler number experience no worldsheet instanton corrections is correct in the case of the Enriques Calabi-Yau, but not true in general.} We expect to be able to incorporate them in the couplings of the very modes we excluded from our ansatz by not considering $N=4$ massive gravitino multiplets. The couplings of these fields to $N=4$ supergravity have not been studied. They are not fully constrained by $N=4$ supersymmetry, and can hence experience quantum corrections which should translate into the expected worldsheet instanton contributions to the prepotential of the $N=2$ theory. We conjecture that these corrections will only involve the $SU(2)$ doublet directions in scalar field space.

\subsection{The Enriques Calabi-Yau}
\label{sec:Enriques4to2}
For the Enriques Calabi-Yau, the conditions \eqref{eq:N=2_conditions_exp} read
\begin{equation}\label{eq:N=2_cond_ECY}\begin{aligned}
\zeta^+_{2+} + \tau \zeta^+_{1+}  & = 0\ , \\
\zeta^+_{3+}  & = 0 \ , \\
 \tilde \zeta^3_{i+}  & = 0 \quad {\rm for} \ i=1,2 \ , \\
 \tilde \zeta^3_{3+}  & = \e^{\rho_2} \ ,
\end{aligned}\end{equation}
where we have defined $\zeta^a_{i\pm} = \zeta^a_i \pm \zeta^a_{i+3}$, $i=1,2,3$, and $\tilde \zeta^3_\cdot = \e^{\rho_4/2} \zeta^3_\cdot$. We will use the same conventions for the two-forms $\omega^I$ below. As $t^i=0$, the two gauged isometries in the metric scalar directions are generated by
\begin{equation}
 \delta_i \zeta^a_I = T_{iI}^J \zeta^a_J \ .
\end{equation}
Plugging in (\ref{T1_ECY}) and (\ref{T2_ECY}) for $T_{iI}^J$ yields
\ba \label{eq:gaugedIsom_zeta}
\delta_i \zeta_{j+}^a = 0 \ , && \delta_i \zeta_{j-}^a = -2 \pi \zeta_{3+}^a \delta_{ij}\ , \\
\delta_i \zeta_{3+}^a =0 \ , && \delta_i \zeta_{3-}^a= 2 \pi \zeta_{i+}^a \ .
\ea
As expected, these isometries hence preserve the Calabi-Yau conditions (\ref{eq:N=2_cond_ECY}). Their action amounts to adding exact forms $\omega^{i-}$ and $\omega^{3-} \wedge v^i$, $i=1,2$, to $J$ and $\Omega$ respectively.

The quadratic constraints \eqref{zeta_ab} on $\zeta^a$ take the form
\begin{equation}\label{eq:cond_zeta}\begin{aligned}
\zeta^+_{1+} (\zeta^+_{1-} -\tau \zeta^+_{2-})  +\zeta^+_\alpha I_8^{\alpha \beta} \zeta^+_\beta &= 0 \ , \\
\Re( \zeta^+_{1+}(\zeta^-_{1-} - \tau \zeta^-_{2-}) +\zeta^+_\alpha I_8^{\alpha \beta} \zeta^-_\beta ) & = 1 \ , \\
 \zeta^+_{i+} \tilde \zeta^3_{i-} + \e^{\rho_2} \zeta^+_{3-} +\zeta^+_\alpha I_8^{\alpha \beta} \tilde \zeta^3_\beta & = 0  \ , \\
\e^{\rho_2} \tilde \zeta^3_{3-} + \tilde \zeta^3_\alpha I_8^{\alpha \beta} \tilde \zeta^3_\beta & > 0 \ ,
\end{aligned}\end{equation}
where the intersection matrix $\eta_{IJ}$ of the two forms is given by (\ref{inter_m}). From the first equation, we see that $\zeta^+_{1+}$ must be non-zero. In the following, we rescale $\zeta^+ \to \tilde \zeta^+$ (corresponding to a rescaling of $\Omega$, which does not change the complex structure) such that $\tilde \zeta^+_{1+} = 1$. The first constraint in (\ref{eq:cond_zeta}) then simplifies to
\begin{equation}\label{eq:cond_zeta+}\begin{aligned}
\tilde \zeta^+_{1-} - \tau \tilde \zeta^+_{2-}  + \tilde \zeta^+_\alpha I^{\alpha \beta} \tilde \zeta^+_\beta & = 0 \ , \\
\Im \tau \Im \tilde \zeta^-_{2-} +\Im \tilde \zeta^+_\alpha I^{\alpha \beta} \Im \tilde \zeta^-_\beta  & >0 \ .
\end{aligned}\end{equation}
Furthermore, we gauge-fix the isometries \eqref{eq:gaugedIsom_zeta} such that $\zeta^+_{3-}=0$. We thus obtain a nice product representation of $\Omega$ on the covering space $K3 \times T^2$ (the resolution of $T^6 / \IZ_2^\sigma$),
\begin{equation}\label{eq:Omega_ECY}
\Omega = (\diff x^5 + \tau \diff x^6) \wedge \left((\diff x^3 + \tau \diff x^4) \wedge ( \diff x^1 + \zeta^+_{2-} \diff x^2)+ \zeta^+_\alpha E^\alpha_- \right) \ .
\end{equation}
We have here used the equality $v^i \wedge E^\alpha_+ =\diff x^{5} \wedge E_-^\alpha$, as at the support of the two components of $E^{\alpha}_+$, see (\ref{ex_lin_comb}), $\cos (2 \pi x^2 )= \pm1$.

Similarly, the conditions on $\tilde \zeta^3$ read
\begin{equation}\label{eq:cond_zeta3}\begin{aligned}
\tilde \zeta^3_{1-} - \tau \tilde \zeta^3_{2-} & = -  \tilde \zeta^+_\alpha I^{\alpha \beta} \tilde \zeta^3_\beta   \ , \\
\e^{\rho_2} \tilde \zeta^3_{3-}  & > - \tilde \zeta^3_\alpha I^{\alpha \beta} \tilde \zeta^3_\beta \ .
\end{aligned}\end{equation}
We can write $J$ as
\begin{equation}\label{eq:J_ECY}
 J = \e^{\rho_2} (\diff x^3 \wedge \diff x^4 + \diff x^5 \wedge \diff x^6) + \tilde \zeta^3_{3-} \diff x^1 \wedge \diff x^2 + \zeta^3_\alpha  \tilde E^\alpha  \ ,
\end{equation}
where we have defined
\begin{equation}
 \tilde E^\alpha_+ = E^\alpha_+ + \tfrac{1}{2\pi \Im \tau}  I^{\alpha \beta} \Im (\tilde \zeta^+_\beta (\diff v^1 - \bar \tau \diff v^2)) \ ,
\end{equation}
which differ from $E^\alpha_+$ by an exact piece.\footnote{Note that this additional exact piece is needed for the closure of the differential algebra of the expansion forms, a property the conventional expansion in harmonic forms lacks.}

From the form \eqref{eq:Omega_ECY} and \eqref{eq:J_ECY} we see that the moduli space of $\Omega$ and $J$ completely decouple, as expected for the complex structure and K\"ahler moduli space of a Calabi-Yau manifold. However, in the case of both moduli spaces, we do not reproduce the full result of the standard Calabi-Yau reduction. For instance, $\Omega$ should span the special-K\"ahler space
\begin{equation}
M_\Omega = \frac{Sl(2,\mathbb{R})}{SO(2)} \times \frac{SO(2,10)}{SO(2)\times SO(10)} \ .
\end{equation}
Though we can guess from \eqref{eq:Omega_ECY} the required cubic form of the full holomorphic prepotential, one of the massless modes is missing in our expression \eqref{eq:Omega_ECY} for the holomorphic three-form $\Omega$: both $\diff x^4$ and $\diff x^6$ exhibit the same complex parameter $\tau$ as coefficient. We therefore only reproduce a special K\"ahler subspace of codimension one. Similarly, one massless mode is missing in our expression \eqref{eq:J_ECY} for the K\"ahler form $J$, corresponding to the deformation in the direction of the two-form $\diff x^3 \wedge \diff x^4 - \diff x^5 \wedge \diff x^6$. We already noted in section (\ref{sec:Enriques}) that this harmonic form is excluded from our reduction ansatz. It does not respect the $T_2^* Y \oplus T_4^* Y$ split of the cotangent space, thus giving rise, as mentioned repeatedly above and explained in \cite{Triendl:2009ap}, to $N=4$ gravitino multiplets which lie beyond the $N=4$ supergravity action considered in this paper.

What remains is the discussion of the degrees of freedom coming from the form fields. Nothing is gained in this sector by restricting to the Enriques Calabi-Yau example. Suffice it to repeat that the form field $B_2$ pairs with $J$ in the standard way, enhancing the K\"ahler moduil space to a special-K\"ahler space with cubic prepotential. Moreover, the Ramond-Ramond fields form a fibration over the complex structure moduli so that we find a quaternion-K\"ahler space in the image of the c-map \cite{cmap}.

%%%%%%%%%%%%%%%%%%%%%%%%%%
\section{Discussion}
%%%%%%%%%%%%%%%%%%%%%%%%%%
\label{sec:conclusions}

In this work, we dimensionally reduced the ten-dimensional action of type IIA on $SU(2)$ structure manifolds. We showed that this action is consistent with $N=4$ gauged supergravity and derived the embedding tensor components that specify the gaugings.
The gauge group of this gauged supergravity is solvable and in general contains a large number of commuting isometries that transform non-trivially under a pair of isometries that are gauged under the Kaluza-Klein vectors.
In principle, one should be able to assemble all degrees of freedom of these $SU(2)$ structure backgrounds not captured by our ansatz in massive multiplets of $N=4$ supergravity.

Calabi-Yau spaces of Euler number zero constitute a distinguished class of $SU(2)$ structure manifolds. These backgrounds famously admit $N=2$ supersymmetry. The $N=4$ gauged supergravity obtained upon compactification on these manifolds hence exhibits at least one vacuum that breaks supersymmetry only partially, yielding new examples of spontaneous $N=4 \to N=2$ supersymmetry breaking, extending the discussion of \cite{Ferrara:1983gn,Wagemans,Tsokur:1994gr,Andrianopoli:2002rm,Dall'Agata:2009gv,Dibitetto:2011gm,HLS}. Integrating out the fields that acquire a mass upon supersymmetry breaking to $N=2$ results in the standard $N=2$ supergravity of Calabi-Yau reductions, up to those fields that have a doublet component under the $SU(2)$ structure group. These $SU(2)$ doublets correspond to $N=4$ massive gravitino multiplets in four dimensions, which must be included in order to recover the full $N=2$ moduli space of Calabi-Yau compactifications. The  multiplet structure of the massive gravitino multiplet was worked out long ago \cite{Ferrara:1983gn}. However, its dynamics has not yet been studied,  and there is still much to be understood about its role in $N=4$ supergravity.

Since $N=4$ gauged supergravity does not allow for any quantum corrections to the massless multiplets (at the two-derivative level) and since we find cubic holomorphic prepotentials from the super-Higgs mechanism, the $N=2$ effective actions we obtain upon supersymmetry breaking cannot incorporate quantum corrections. This is in accord with the absence of perturbative quantum corrections in Euler number zero Calabi-Yau compactifications. As worldsheet instanton corrections are not absent in such compactifications, we conjecture that these quantum corrections to the $N=2$ holomorphic prepotential of the K\"ahler moduli space can be encoded fully in the couplings of massive gravitino multiplets to $N=4$ gauged supergravity. This provides strong motivation to study massive gravitino multiplets in gauged supergravity in the future.

The discussion of quantum corrections to the holomorphic prepotential of the complex structure moduli space is completely analogous. Our analysis shows that before $SU(2)$ doublets are included, this prepotential is also cubic, consistent with mirror symmetry.\footnote{The Euler numbers of mirror Calabi-Yau manifolds differ by sign only.} The $SU(2)$ doublet deformations will fiber over these modes, giving the full prepotential. Therefore, the Picard-Fuchs equations should drastically simplify in a parametrization of the holomorphic three-form compatible with the $SU(2)$ structure. As in the K\"ahler sector, all non-perturbative corrections, here $g_s$ corrections due to D-brane and NS5 instantons, should come from massive gravitino multiplets that incorporate the $SU(2)$ doublet degrees of freedom.

The enhancement of supersymmetry from $N=2$ to $N=4$ should also be visible from a different point of view. Even if these additional supercharges are broken by the background, one should still be able to find the related conserved currents in the $N=2$ supergravity. Their existence might be related to the absence of the quantum corrections that we discussed above.

In the long run, discussing quantum corrections of Calabi-Yau spaces of vanishing Euler number in terms of their $SU(2)$ structure might also enable one to understand quantum corrections for general $G$-structure backgrounds better, as some of the techniques could carry over to the case where no supercharges are preserved. One related idea is to find the correct twisted cohomology  on the internal space that keeps track of all modes of the $N=4$ supergravity. Though first steps into this direction have been undertaken in \cite{Schulz:2012uj} for orbifold examples, the general  definition of such a cohomology  still seems to be missing.

We shall conclude with two comments concerning related constructions for theories with higher dimension or less supersymmetry.

If the Calabi-Yau manifold $Y$ is elliptically fibered, the four-dimensional $N=2$ theory has an interpretation in six dimensions in terms of a $(1,0)$ theory. Equivalently, we can think of such a theory as coming from an F-theory compactification on $Y$ \cite{vafa}. When the Euler number of $Y$ vanishes,  not only the irreducible part of the gravitational anomaly vanishes, as required for anomaly cancellation, but the  {\it reducible} part, i.e.\ the coefficient of $(\mbox{tr} R^2)^2$ \cite{fms} does as well (see \cite{WT} for a recent review). It would be interesting to understand this vanishing of gravitational anomalies from the point of view of hidden extra supersymmetry in the framework presented in this work.

Finally, one could hope to derive an $N=2$ gauged supergravity with a vacuum preserving $N=1$ supersymmetry by  considering the heterotic string on a Calabi-Yau manifold of Euler number zero. Though less supersymmetry and the inclusion of the gauge bundle might complicate the analysis, the discussion of quantum corrections to the holomorphic prepotentials should carry over to this case. Moreover, since these Calabi-Yau spaces give non-trivial examples of $SU(2)$ structure manifolds, one could in principle hope that such a study could be useful for understanding heterotic strings on general $G$-structure backgrounds.

%%%%%%%%%%%%%%%%%%%%%%%%%%
\section*{Acknowledgments}
%%%%%%%%%%%%%%%%%%%%%%%%%%
We would like to thank Hans Jockers, Jan Louis, Ilarion Melnikov, Daniel Park, Danny Martinez-Pedrera, Paul Smyth, Gautier Solard and Edward Witten for useful discussions. This work was supported in part by the ANR grant 08-JCJC-0001-0 and the ERC Starting Grants 259133 -- ObservableString and 240210 - String-QCD-BH.

\appendix

%%%%%%%%%%%%%%%%%%%%%%%%%%
\section{Appendix}
%%%%%%%%%%%%%%%%%%%%%%%%%%
\label{app:Ricci}
In this appendix we compute the Ricci scalar $r_{10}$. For this we must first determine the connection from \eqref{10d_connection} for the ansatz \eqref{eq:truncation_ansatz}.

%%%%%%%%%%%%%%%%%%%%%%%%%%
\subsection{Useful identities}
%%%%%%%%%%%%%%%%%%%%%%%%%%
Let us first collect some identities that will be useful in the following computation.

From \eqref{matrix_k}, we find that the matrix $k$ satisfies the identities
\begin{equation}\begin{aligned}
  k^T \cdot \epsilon \cdot k  \,=\, & k\cdot \epsilon \cdot k^T   \,=\, \e^{\rho_2} \epsilon \ ,\\
 k^{-1} \,=\, & - \e^{-\rho_2} \epsilon \cdot k^T \cdot \epsilon \ ,\\
 k^T \cdot k  \,=\, & \e^{\rho_2} g^{\tau} \ ,
 \end{aligned} \end{equation}
 where
\begin{equation}
 g^{\tau} = (\Im \tau)^{-1} \left( \begin{aligned} 1 && \Re \tau \\ \Re \tau && |\tau|^2 \end{aligned} \right) \ .
\end{equation}
We will need the following expressions for the four-dimensional derivatives of $k^i_j$:
\begin{equation}\label{diffK}\begin{aligned}
( (\diff k)\cdot k^{-1})_{\rm sym}= & \tfrac12 \diff \rho_2 {\bf 1} - \tfrac12 \sigma^3 (\Im \tau)^{-1} \diff \Im \tau  + \tfrac12 \sigma^1 (\Im \tau)^{-1}\diff \Re \tau  \ , \\
{\rm tr}( \epsilon \cdot (\diff k)\cdot k^{-1}) = & -(\Im \tau)^{-1}\diff \Re \tau  \ ,
 \end{aligned} \end{equation}
where $\sigma^1$ and $\sigma^3$ are the standard Pauli matrices.

From \eqref{zeta_ab} we find
\begin{equation}\label{delzeta}
(\partial_\mu \zeta^a_I) \eta^{IJ} \zeta^b_J = 0 \ ,
\end{equation}
which means that over four-dimensional spacetime, the $J^a$ do not rotate into each other and therefore really move in $\frac{SO(3,n)}{SO(3) \times SO(n)}$.
The decomposition of $\omega^I$ into representations of $SU(2)$ reads
\begin{equation}
\omega^I = \zeta^{a\, I} \zeta^a_J \omega^J + (\delta^I_J - \zeta^{a\, I} \zeta^a_J) \omega^J \ .
\end{equation}
The latter term is invariant under the $I^a$. We hence find
\begin{equation}\label{SU2rot_omega}
 \tfrac12 (I^a)^\alpha_\beta (\omega^I)_{\beta \gamma} e^\alpha \wedge e^\gamma = \e^{-\rho_4/2} \epsilon^{abc} \zeta^{b\, I} J^c \ .
\end{equation}
Furthermore, we have
\begin{equation}\label{Iomega}
 (\omega^I)_{\alpha \beta }  (I^a)^\alpha_\beta  = 4 \e^{\rho_4/2} \zeta^{a\, I} \ .
\end{equation}

%%%%%%%%%%%%%%%%%%%%%%%%%%
\subsection{Connection}
%%%%%%%%%%%%%%%%%%%%%%%%%%
\label{app:connection}
From the first equation in \eqref{10d_connection}, we see that $\gamma_\mu^i (\hat K_j)$ and $\lambda_\mu^\alpha(\hat e_\beta)$  must be symmetric in $i$ and $j$ and $\alpha$ and $\beta$, respectively.
It is straight-forward to solve for the first two equations. For the third, we need to use $\diff J^a = (I^a)^\alpha_\beta \diff e^\alpha \wedge e^\beta$ as well as \eqref{param_forms} and \eqref{delzeta}. We find
\begin{equation} \label{10d_dJ}\begin{aligned}
 0 = & \tfrac12 \diff \rho_4 \wedge J^a + \e^{\rho_4/2} \diff \zeta^a_I \wedge \omega^I + \e^{\rho_4/2} \zeta^a_I T^I_{jJ}(k^{-1})^j_i K^i \wedge \omega^J - \e^{\rho_4/2} \zeta^a_I T^I_{iJ} G^i \wedge \omega^J  \\ &  - (I^a)^\alpha_\beta \tau^i_{\alpha \gamma} K^i \wedge e^\beta \wedge e^\gamma + 2 \epsilon^{abc} \phi^b \wedge J^c + (I^a)^\alpha_\beta \lambda^\alpha_\mu \wedge e^\mu \wedge e^\beta  \ .
\end{aligned}\end{equation}
Now let us first solve the second equation in \eqref{10d_connection}. For this, we recall that $\epsilon_{ij} \gamma_\mu^i (\hat K_j) = 0$ and \eqref{diffK}. This gives us
\begin{equation}\begin{aligned}
\tau^i_{\alpha\beta} e^\alpha \wedge e^\beta = & k^i_j t^j_I \omega^I \ , \\
\phi^0 = & - \e^{-\rho_2} t \cdot k^T \cdot K  - \frac{\diff \Re \tau}{2\Im \tau} + \tfrac12 t \cdot g^\tau \cdot G \ , \\
\gamma^i_\mu \wedge e^{\mu} = &  - k^i_j D G^j + \tfrac12 K^i \wedge D \rho_2 - \tfrac12 (\sigma^3)^i_j (\Im \tau)^{-1} K^j \wedge D \Im \tau \\ & + \tfrac12 (\sigma^1)^i_j (\Im \tau)^{-1} K^j \wedge D \Re \tau \ ,
\end{aligned}\end{equation}
where we have defined the covariant derivatives
\begin{equation}\begin{aligned}
 D G^i = & \diff G^i + \tfrac12 t^i G \cdot \epsilon \cdot G \ , \\
 D\rho_2 = & \diff \rho_2 - G \cdot \epsilon \cdot t \ , \\
 D\Re \tau = & \diff \Re \tau + (G \cdot \epsilon \cdot k^{-1} \cdot \sigma^1 \cdot k \cdot t)\Im \tau  \ , \\
 D\Im \tau = & \diff \Im \tau + (G \cdot \epsilon \cdot k^{-1} \cdot \sigma^3 \cdot k \cdot t)\Im \tau   \ .
\end{aligned}\end{equation}
The latter two can be merged into the covariant derivative of the complex scalar $\tau$ as
\begin{equation}
D \tau = \diff \tau - ((1 ,\, \tau) \cdot G)  ((1 ,\, \tau) \cdot t) \ .
\end{equation}
Next, we make the ansatz
\begin{equation}\begin{aligned}
\tau^i_{\alpha\beta}= & \tfrac12 k^i_j t^j_I \omega^I_{\alpha \beta} + \tau^i_0 \delta_{\alpha \beta} + (\tau^i_a)_{\alpha \gamma} (I^a)^\gamma_\beta \ , \\
\lambda^\alpha_\mu = & \lambda^\alpha_{(\mu \nu)} e^\nu + \lambda^0_\mu e^\alpha + (\lambda^a_\mu)_{\alpha \gamma} (I^a)^\gamma_\beta e^\beta \ ,
\end{aligned}\end{equation}
where $\tau^i_a = \tau^i_{a\,I} (\delta^I_J - \zeta^{a\, I} \zeta^a_{\ J})\omega^J$. We have used that $\lambda_\mu^\alpha(\hat e_\beta)$ should be symmetric in $\alpha$ and $\beta$ as well as the fact that \eqref{10d_connection} implies that $\lambda^\alpha_\mu (\hat e^\nu)$ is symmetric in $\mu$ and $\nu$. We then solve \eqref{10d_dJ} with help of \eqref{SU2rot_omega} and find
\begin{equation}\begin{aligned}
\tau^i_{\alpha\beta}  = &  \tfrac12 k^i_j t^j_I \omega^I_{\alpha \beta} + \tfrac14 \e^{-\rho_2} \epsilon_{ij} k^j_k t^k \delta_{\alpha \beta} \\ &  +  \tfrac12 \e^{\rho_4/2} (k^{-1})_i^j \zeta^a_K \tilde T^K_{jI} (\delta^I_J - \zeta^{a\, I} \zeta^a_{\ J})\omega^J_{\alpha \gamma}(I^a)^\gamma_\beta  \ , \\
\phi^a = & - \tfrac12 \e^{-\rho_4 /2} \zeta^{aI} t^j_I k^i_j K^i + \tfrac14 \epsilon^{abc} \zeta^b_I \zeta^{c\,J} \tilde T^I_{iJ}(k^{-1})^i_j K^j -\tfrac14 \epsilon^{abc} \zeta^b_I \zeta^{c\,J} \tilde T^I_{iJ}G^i      \ , \\
\lambda^\alpha_\mu \wedge e^{\mu} = &  \tfrac14 e^\alpha \wedge (\diff \rho_4 - t \cdot \epsilon \cdot G )\\ &  + \tfrac12 \e^{\rho_4/2} (I^a)^\beta_\alpha \omega^I_{\beta \gamma}  e^\gamma \wedge (\diff \zeta^a_I - G^i \tilde T^J_{iK} \zeta^a_J(\delta^K_I - \zeta^{b K} \zeta^b_I) )  \ .
\end{aligned}\end{equation}
Finally, we solve the first equation in \eqref{10d_connection}. This gives us
\begin{equation}\begin{aligned}
\lambda^\alpha_{(\mu\nu)}  =\, \, & 0  \ , \qquad  \qquad
\gamma^i_{(\mu}(\hat e_{\nu)}) = 0 \ , \\
\omega^\mu_\nu =\, \, & \tilde \omega^\mu_\nu + D_{[\mu} G_{\nu]} \cdot k^T \cdot K  \ ,
\end{aligned}\end{equation}
where $\tilde \omega$ is the connection of the four-dimensional metric. This determines the ten-dimensional connection up to the $SU(2)$ component $\theta$. We will see that $\theta$ drops out of all four-dimensional expressions. All other components are given by
\begin{equation} \label{10d_conn_expl}\begin{aligned}
\omega^\mu_\nu =\, \, & \tilde \omega^\mu_\nu + D_{[\mu} G_{\nu]} \cdot k^T \cdot K \ , \\
\gamma^i_\mu = &  k^i_j D_{[\mu} G^j_{\nu]} e^\nu + \tfrac12 D_\mu \rho_2 K^i  - \tfrac12 (\Im \tau)^{-1} D_\mu \Im \tau (\sigma^3)^i_j  K^j  \\ & + \tfrac12 (\Im \tau)^{-1} D_\mu \Re \tau (\sigma^1)^i_j K^j  \ , \\
\lambda^\alpha_\mu  = &  \tfrac14 e^\alpha D_\mu \rho_4  + \tfrac12 \e^{\rho_4/2} (I^a)^\beta_\alpha \omega^I_{\beta \gamma}  e^\gamma (\delta^J_I - \zeta^b_I \zeta^{b J})D_\mu \zeta^a_J \ , \\
\phi^0 = & - \e^{-\rho_2} t \cdot k^T \cdot K  - \frac{\diff \Re \tau}{2\Im \tau} + \tfrac12 t \cdot g^\tau \cdot G \ , \\
\phi^a = & (\tfrac14 \epsilon^{abc} \zeta^b_I \zeta^{c\,J} \tilde T^I_{jJ}(k^{-1})^j_i - \tfrac12 \e^{-\rho_4 /2} \zeta^{aI} t^j_I k^i_j) K^i -\tfrac14 \epsilon^{abc} \zeta^b_I \zeta^{c\,J} \tilde T^I_{iJ}G^i     \ , \\
\tau^i_{\alpha}  = &  \tfrac12 k^i_j t^j_I \omega^I_{\alpha \beta} e^\beta + \tfrac14 \e^{-\rho_2} \epsilon_{ij} k^j_k t^k e^\alpha \\ &  +  \tfrac12 \e^{\rho_4/2} (k^{-1})_i^j \zeta^a_K \tilde T^K_{jI} (\delta^I_J - \zeta^{b\, I} \zeta^b_{\ J})\omega^J_{\alpha \gamma}(I^a)^\gamma_\beta e^\beta  \ ,
\end{aligned}\end{equation}
where the covariant derivatives are defined as
\begin{equation}\begin{aligned}
 D\rho_2 = & \diff \rho_2 - G \cdot \epsilon \cdot t \ , \\
 D\rho_4 = & \diff \rho_4 +G \cdot \epsilon \cdot t \ , \\
 D \tau = & \diff \tau - ((1 ,\, \tau) \cdot G)  ((1 ,\, \tau) \cdot t) \ , \\
 D \zeta^a_I = & \diff \zeta^a_I - G^i \tilde T^J_{iI} \zeta^a_J \ , \\
 D G^i = & \diff G^i + \tfrac12 t^i G \cdot \epsilon \cdot G \ .
\end{aligned}\end{equation}

%%%%%%%%%%%%%%%%%%%%%%%%%%
\subsection{Torsion and curvature of the $SU(2)$ connection}
%%%%%%%%%%%%%%%%%%%%%%%%%%
In order to compute the Ricci scalar, we derive a relation between the curvature and the torsion tensor of the $SU(2)$ connection.
Since $\eta_{+i}$ are invariant under the $su(2)$ in which $\theta$ lives, we have
\begin{equation}
 0 = R^\theta_{\alpha \beta \gamma \delta} \gamma^{\gamma \delta} \eta_{+i} \ .
\end{equation}
Contracting this equation with $\gamma^\beta$ gives
\begin{equation}
 0 =  R^\theta_{\alpha \beta \gamma \delta} \gamma^{\beta \gamma \delta} \eta_{+i} - \operatorname{Ric}^\theta_{\alpha \beta} \gamma^{\beta} \eta_{+i} \ .
\end{equation}
We see that if $\theta$ were torsion-free, the Riemann tensor would obey the Bianchi identity and the first term would be zero, thereby imposing that the connection be Ricci-flat. In general, the above equation gives a tool to compute the Ricci curvature in terms of torsion classes. We can contract the above equation with $\bar \eta^i_{+} \gamma^\rho$ and find
\begin{equation}
\operatorname{Ric}^\theta_{\alpha \beta} =  R^\theta_{\alpha \gamma \delta \rho} \epsilon^{\gamma \delta \rho \beta} \ .
\end{equation}
The failure to satisfy the Bianchi identity is measured by $R^\theta \wedge e = \nabla T = \diff T + \theta  T$ (as can be seen from the definition of the torsion tensor $T$), therefore
\begin{equation} \label{RicSU2_torsion}
\operatorname{Ric}^\theta_{\alpha \beta} =  (\diff T^\alpha+\theta^\alpha_{\lambda} \wedge T^\lambda)(\hat e_\gamma, \hat e_\delta, \hat e_\rho ) \epsilon^{\gamma \delta \rho \beta} \ .
\end{equation}
Note that only the component of the $SU(2)$-covariant derivative of the torsion two-form inside $\Lambda^3 T^*_4$ appears.

Now let us evaluate the $SU(2)$ Ricci scalar by computing the $SU(2)$ torsion tensor and its covariant derivative by use of \eqref{10d_conn_expl}. We find
\begin{equation}\label{SU2curvature}\begin{aligned}
r^\theta=&  \epsilon^{\alpha \beta \gamma \delta} \nabla^\theta T^\alpha (\hat e_\beta, \hat e_\gamma, \hat e_\delta)  =  -  \epsilon^{\alpha \beta \gamma \delta}  ((I^a)^\alpha_{\beta} \phi^a_i + \tau^i_{\alpha \beta}) k^i_j t^j_I \omega^I_{\gamma \delta} \\ = & - \epsilon^{\alpha \beta \gamma \delta}( (I^a)^\alpha_{\beta}(\tfrac14 \epsilon^{abc} \zeta^b_I \zeta^{c\,J} \tilde T^I_{jJ}(k^{-1})^j_i - \tfrac12 \e^{-\rho_4 /2} \zeta^{aI} t^j_I k^i_j) + \tfrac12 k^i_j t^j_I \omega^I_{\alpha \beta} )k^i_j t^j_K \omega^K_{\gamma \delta}
\\ = & - \e^{-\rho_4/2} \epsilon^{abc}\zeta^{a\,K} \zeta^b_I \zeta^{c\,J} \tilde T^I_{jJ} t^j_K  - 2 \e^{\rho_2-\rho_4} t_I \cdot g^\tau \cdot t_J (\eta^{IJ} -\zeta^{aI}\zeta^{aJ}) \ .
\end{aligned}\end{equation}

%%%%%%%%%%%%%%%%%%%%%%%%%%
\subsection{Ricci scalar}
%%%%%%%%%%%%%%%%%%%%%%%%%%
Our next aim is to compute the ten-dimensional Ricci scalar $r_{10}$ from
\begin{equation}
r_{10} = {\rm Ric}_{\mu \mu} + {\rm Ric}_{ii} + {\rm Ric}_{\alpha \alpha} \ .
\end{equation}
For the Ricci tensor, we have
\begin{equation}\begin{aligned}
{\rm Ric}_{\mu \nu} = & R_{\mu \lambda \nu \lambda}+  R_{\mu i\nu i} + R_{\mu \alpha \nu\alpha} \ , \\
{\rm Ric}_{\mu i} = & R_{\mu \nu i \nu}+  R_{\mu j i j} + R_{\mu \alpha i\alpha} \ , \\
{\rm Ric}_{\mu \alpha} = & R_{\mu \nu \alpha \nu}+  R_{\mu i \alpha i} + R_{\mu \beta \alpha \beta} \ , \\
{\rm Ric}_{ij} = & R_{i\mu j\mu}+  R_{ikjk} + R_{i\alpha j\alpha}\ , \\
{\rm Ric}_{i\alpha} = & R_{i \mu \alpha \mu}+  R_{i j \alpha j} + R_{i \beta \alpha \beta}   \ , \\
{\rm Ric}_{\alpha \beta} = & R_{\alpha \mu \beta \mu}+  R_{\alpha k \beta k} + R_{\alpha \gamma \beta \gamma} \ ,
\end{aligned}\end{equation}
so that
\begin{equation}
r_{10} = R_{\mu \nu \mu \nu} + 2 R_{\mu i \mu i} + 2 R_{\mu \alpha \mu \alpha} + R_{ijij} +  2 R_{i\alpha i\alpha} + R_{\alpha \beta \alpha \beta} \ .
\end{equation}
We will now compute the components via
\begin{equation}
R = \diff \Omega + \Omega \wedge \Omega \ .
\end{equation}
For instance
\begin{equation}\begin{aligned}
R_{\mu \nu \mu \nu} = &(\diff \omega^\mu_\nu)(\hat e_\mu, \hat e_\nu) + (\omega^\mu_\lambda \wedge \omega^\lambda_\nu)(\hat e_\mu, \hat e_\nu) + (\lambda^\mu_\alpha \wedge \lambda^\alpha_\nu)(\hat e_\mu, \hat e_\nu) + (\gamma^\mu_i \wedge \gamma^i_\nu)(\hat e_\mu, \hat e_\nu) \\
= & \hat r_4 + \e^{\rho_2} D_{[\mu} G_{\nu]} \cdot g^\tau \cdot D^{[\mu} G^{\nu]} \ ,
\end{aligned}\end{equation}
where $\hat r_4$ is the Ricci scalar of the four-dimensional metric up to a conformal factor, which we will rescale below.
Similarly, we have
\begin{equation}\begin{aligned}
R_{i \mu i \mu} = & \big( \diff \gamma^i_\mu + \gamma^i_\nu \wedge \omega^\nu_\mu + \epsilon_{ij} \phi^0 \wedge \gamma^j_\mu + \tau^i_\alpha \wedge \lambda^\alpha_\mu \big) (\hat K_i, \hat e_\mu)\\
=& -\nabla^\mu D_\mu \rho_2 - \tfrac12 (D^\mu \rho_2) (D_\mu \rho_2)- \frac{(D^\mu \tau) (D_\mu \bar \tau)}{2 (\Im \tau)^2} - \e^{\rho_2}  D_{[\mu} G_{\nu]} \cdot g^\tau \cdot D^{[\mu} G^{\nu]} \ , \\
R_{\alpha \mu \alpha \mu} = & \big( \diff \lambda^\alpha_\mu + \lambda^\alpha_\nu \wedge \omega^\nu_\mu + \tau^\alpha_i \wedge \gamma^i_\mu + (I^a)^\alpha_\beta \phi^a \wedge \lambda^\beta_\mu + \theta^\alpha_\beta \wedge \lambda^\beta_\mu \big) (\hat e_\alpha, \hat e_\mu) \\
= & - \nabla^\mu D_\mu \rho_4 - \tfrac14 D_\mu \rho_4 D^\mu \rho_4  \ ,\\
R_{ijij} = & (\epsilon_{ij} \diff \phi^0 + \gamma^i_\mu \wedge \gamma^\mu_j + \tau^i_\alpha \wedge \tau^\alpha_j)(\hat K_i,\hat K_j )\\
=& -2 \e^{-\rho_2} t \cdot g^\tau \cdot t - \tfrac12  (D_\mu \rho_2)(D^\mu \rho_2) + \tfrac12 (\Im \tau)^{-2} (D_\mu \tau)(D^\mu \bar \tau) \ .
\end{aligned}\end{equation}
For the component $R_{i\alpha i \alpha}$, we compute from \eqref{10d_connection} the following expressions
\begin{equation}\begin{aligned}
 \diff e^\alpha (\hat e_i, \hat e_\alpha) = & - \tau^i_\alpha (\hat e_\alpha) = -2 \e^{-\rho_2} \epsilon_{ij} k^j_k t^k \ ,\\
 \diff (\omega^I_{\alpha\beta}e^\beta) (\hat e_i, \hat e_\alpha)= & \omega^I_{\alpha\beta} (\diff e^\beta) (\hat e_i, \hat e_\alpha) = - \omega^I_{\alpha\beta} ((I^a)^\beta_\alpha \phi^a(\hat e_i) + \tau^i_\beta (\hat e_\alpha) )\\
  = &  \tfrac14 \e^{\rho_4/2} \epsilon^{abc} \zeta^{a\,I} \zeta^b_J \zeta^{c\,K} \tilde T^J_{jK}(k^{-1})^j_i  - \tfrac12 k^i_j t^j_J (\eta^{JI}- \zeta^{b\,J}\zeta^{b\,I} ) \ ,\\
 \diff (\omega^I_{\alpha\gamma} (I^a)^\gamma_{\beta} e^\beta) (\hat e_i, \hat e_\alpha)= &  \omega^I_{\alpha\gamma} (I^a)^\gamma_{\beta} (\diff  e^\beta) (\hat e_i, \hat e_\alpha) = \omega^I_{\alpha\gamma} (I^a)^\gamma_{\beta} \tau^i_\beta (\hat e_\alpha) \\
 =& - \tfrac12 \e^{\rho_4/2} (k^{-1})^j_i \zeta^a_K \tilde T^K_{j\, J} (\delta^J_I - \zeta^{b\,J} \zeta^b_I)\ .
\end{aligned}\end{equation}
We find
\begin{equation}
\diff \tau^i_\alpha (\hat e_i, \hat e_\alpha) = (I^a)^\alpha_\beta \tau^i_{\alpha \beta}  \phi^a_i - \tau^i_{\alpha \beta} \tau^i_{\beta \alpha} \ .
\end{equation}
The curvature component $R_{i\alpha i \alpha}$ then reads
\begin{equation}\begin{aligned}
R_{i\alpha i \alpha} = &  (\diff \tau^i_\alpha + \tau^i_\beta \wedge (I^a)^\beta_\alpha \phi^a + \epsilon_{ij} \phi^0 \wedge \tau^j_\alpha + \gamma^i_\mu \wedge \lambda^\mu_\alpha)(\hat e_i, \hat e_\alpha) \\
= & - \tau^i_{\alpha \beta} \tau^i_{\beta \alpha }  + \epsilon_{ij} \phi^0_i \tau^j_{\alpha\alpha} + \gamma^i_{\mu i} \lambda^\mu_{\alpha\alpha} \\
 = & \tfrac34 \e^{-\rho_2} t \cdot g^\tau \cdot t + 2 \e^{\rho_2-\rho_4} t_I \cdot g^\tau \cdot t_J H^{JI}
 - D_\mu \rho_2 D^\mu \rho_4\\ &
+\e^{- \rho_2} (\eta^{IJ} - \zeta^{b\, I}\zeta^{b\, J})\zeta^{a}_K \zeta^{a}_L \tilde T^K_I \cdot (g^\tau)^{-1} \cdot \tilde T^L_J  \ .
\end{aligned}\end{equation}
Finally, we can use \eqref{SU2curvature} to determine
\begin{equation}\begin{aligned}
 R_{\alpha \beta \alpha\beta} = & r^\theta + \big( \diff \phi^a (I^a)^\alpha_\beta +\epsilon^{abc} \phi^b \wedge \phi^c (I^a)^\alpha_\beta - \tau^i_\alpha \wedge \tau^i_\beta - \lambda^\alpha_\mu \wedge \lambda^\beta_\mu \big)(\hat e_\alpha, \hat e_\beta) \\
 = & - 4 \e^{\rho_2-\rho_4}   \zeta^{a\, I} t_I \cdot g^\tau \cdot t_J \zeta^{a\, J} - \tfrac34 \e^{-\rho_2} t \cdot  g^\tau \cdot t \\ &
  - \e^{- \rho_2} (\eta^{IJ} - \zeta^{b\, I}\zeta^{b\, J})\zeta^{a}_K \zeta^{a}_L \tilde T^K_I \cdot (g^\tau)^{-1} \cdot \tilde T^L_J
 -\tfrac34 D_\mu \rho_4 D^\mu \rho_4 \\&
  +   D_\mu \zeta^a_I (\eta^{IJ} - \zeta^{b\, I}\zeta^{b\, J}) D^\mu \zeta^a_J   \ .
\end{aligned}\end{equation}
Summing up alI components, we arrive at the following expression for the ten-dimensional Ricci scalar
\begin{equation}\begin{aligned}
r_{10} = & \hat r_4 - \e^{\rho_2} D_{[\mu} G_{\nu]} \cdot g^\tau \cdot D^{[\mu} G^{\nu]} - 2\nabla^\mu D_\mu \rho_2 - \tfrac32(D^\mu \rho_2) (D_\mu \rho_2)  \\ & - \tfrac12 (\Im \tau)^{-2} (D_\mu \tau)(D^\mu \bar \tau) - 2 \nabla^\mu D_\mu \rho_4 -\tfrac{5}{4} D_\mu \rho_4 D^\mu \rho_4 -\tfrac54  \e^{-\rho_2} t \cdot g^\tau \cdot t \\ & -4 \e^{\rho_2-\rho_4} t_I \cdot g^\tau \cdot t_J \eta^{IJ}
+\e^{- \rho_2} (\eta^{IJ} - \zeta^{b\, I}\zeta^{b\, J})\zeta^{a}_K \zeta^{a}_L \tilde T^K_I \cdot (g^\tau)^{-1} \cdot \tilde T^L_J  \\ & - 2 D_\mu \rho_2 D^\mu \rho_4     +  D_\mu \zeta^a_I (\eta^{IJ} - \zeta^{b\, I}\zeta^{b\, J}) D^\mu \zeta^a_J \ .
\end{aligned}\end{equation}

%%%%%%%%%%%%%%%%%%%%%%%%%%
\section{Fermions and supersymmetry}
%%%%%%%%%%%%%%%%%%%%%%%%%%
\label{sec:SUSYvar}

In this appendix, we want to derive the gravitino mass matrix of the four-dimensional theory, as it appears in the supersymmetry variation of the gravitini. For this, we will first identify the four-dimensional gravitini and fermions in terms of ten-dimensional fermionic fields. Subsequently, we will derive the vacuum contribution to the supersymmetry variations in terms of the internal geometry. Similar discussions can be found in \cite{Grana:2005ny,MartinezPedrera:2009zz}.

In type IIA supergravity, the ten-dimensional gravitini $\Psi_M^{(10)\,i}$ and the dilatini $\chi^{(10)\,i}$ form the vector $\otimes$ spinor representation $\hat \Psi_M^{(10)\,i}$ given by
\begin{equation}
 \hat \Psi_M^{(10)\,i} = \Psi_M^{(10)\,i} + \tfrac18 \Gamma_M \chi^{(10)\,i} \ .
\end{equation}
We can expand the internal and external components of $\hat \Psi_M^{(10)\,i}$ as
\begin{equation}\begin{aligned}
 \hat \Psi_\mu^{(10)\,1} \ =\ & \epsilon_{jk} \Psi_{\mu\,+}^{j} \otimes \eta^k_+ + \epsilon_{jk} \Psi_{\mu\,-}^{j} \otimes \eta^k_- + \tfrac12 \epsilon_{jk} \gamma_\mu \chi^{j}_+ \otimes \eta^k_+ + \tfrac12  \epsilon_{jk} \gamma_\mu \chi^{j}_- \otimes \eta^k_- \ , \\
 \hat \Psi_\mu^{(10)\,2} \ =\ & \epsilon_{jk} \Psi_{\mu\,-}^{j+2} \otimes \eta^k_+ + \epsilon_{jk} \Psi_{\mu\,+}^{j+2} \otimes \eta^k_- + \tfrac12  \epsilon_{jk} \gamma_\mu \chi^{j+2}_- \otimes \eta^k_+ + \tfrac12  \epsilon_{jk} \gamma_\mu \chi^{j+2}_+ \otimes \eta^k_- \ , \\
 \hat \Psi_j^{(10)\,1} \ =\ & \epsilon_{kl} \xi_{j\,+}^{k} \otimes \eta^l_+ + \epsilon_{kl} \xi_{j\,-}^{k} \otimes \eta^l_- \ , \\
 \hat \Psi_j^{(10)\,2} \ =\ & \epsilon_{kl} \xi_{j\,-}^{k+2} \otimes \eta^l_+ + \epsilon_{kl} \xi_{j\,+}^{k+2} \otimes \eta^l_- \ , \\
 \hat \Psi_a^{(10)\,1} \ =\ & \epsilon_{kl}  \zeta^\alpha_{I} \lambda_{\alpha \,+}^{k} \otimes \omega^I_{bc} \gamma^{bc} \gamma_a \eta^l_+ + \epsilon_{kl}  \zeta^{\hat a}_{I} \lambda_{{\hat a} \,-}^{k} \otimes \omega^I_{bc} \gamma^{bc} \gamma_a \eta^l_- \\ &
 + \tfrac14 \epsilon_{jk} \rho^{j}_+ \otimes \gamma_a \eta^k_+ + \tfrac14 \epsilon_{jk} \rho^{j}_- \otimes \gamma_a \eta^k_- \ , \\
 \hat \Psi_a^{(10)\,2} \ =\ & \epsilon_{kl}  \zeta^\alpha_{I} \lambda_{\alpha \,-}^{k+2} \otimes \omega^I_{bc} \gamma^{bc} \gamma_a \eta^l_+ + \epsilon_{kl}  \zeta^{\hat a}_{I} \lambda_{{\hat a} \,+}^{k+2} \otimes \omega^I_{bc} \gamma^{bc} \gamma_a \eta^l_- \\ &
 + \tfrac14 \epsilon_{jk} \rho^{j+2}_- \otimes \gamma_a \eta^k_+ + \tfrac14 \epsilon_{jk} \rho^{j+2}_+ \otimes \gamma_a \eta^k_-\ .
\end{aligned}\end{equation}
Here, $\mu$ is a space-time index, and $j$ and $a$ index the two- and four-dimensional component respectively of the tangent bundle of the internal manifold. The $\zeta^\alpha_{I}$ are defined such that
\begin{equation}
\zeta_{I}^{\alpha} \eta^{IJ} \zeta^a_{J} = 0 \ , \qquad \eta_{IJ} = \zeta^a_I \zeta^a_J - \zeta^\alpha_{I} \zeta^\alpha_{J} \ .
\end{equation}

From the above decomposition of the ten-dimensional fermions, we can read off the four-dimensional fields. In the following, we only write the component of positive chirality and drop the $+$ index. The four-dimensional gravitini $\Psi_\mu^a$ read \cite{Grana:2005ny}
\begin{equation} \label{gravitini_4d_10d}\begin{aligned}
 \Psi_\mu^{i} \ =\ & ({\bf 1} \otimes \bar \eta^i_+) \Psi^{(10)\, 1}_\mu + \tfrac12 (\gamma_\mu \otimes \bar \eta^i_+ \gamma^k) \Psi^{(10)\, 1}_k + \tfrac12 (\gamma_\mu \otimes \bar \eta^i_+ \gamma^a) \Psi^{(10)\, 1}_a \ ,\\
 \Psi_\mu^{i+2} \ =\ & ({\bf 1} \otimes \bar \eta_{-\, i}) \Psi^{(10)\, 2}_\mu + \tfrac12 (\gamma_\mu \otimes \bar \eta_{-\, i} \gamma^k) \Psi^{(10)\, 2}_k  + \tfrac12 (\gamma_\mu \otimes \bar \eta_{-\, i} \gamma^a) \Psi^{(10)\, 2}_a \ .\end{aligned} \end{equation}
Here, the second term ensures that the four-dimensional gravitini are traceless, i.e.\ $\gamma^\mu \Psi_\mu^a = 0$.
Furthermore, the spin-$1/2$ fermions read
\begin{equation} \label{fermions_4d_10d}\begin{aligned}
\chi^{i}\ =\ &\tfrac14 ({\bf 1}\otimes \bar \eta^i_+) \chi^{(10)\,1} - (1\otimes \bar \eta^i_+ \gamma^a) \Psi^{(10)\,1}_a - (1\otimes \bar \eta^i_+ \gamma^k) \Psi^{(10)\,1}_k \ ,\\
\chi^{i+2}\ =\ &\tfrac14 ({\bf 1}\otimes \bar \eta^i_-) \chi^{(10)\,2} - (1\otimes \bar \eta^i_- \gamma^a) \Psi^{(10)\,2}_a - (1\otimes \bar \eta^i_- \gamma^k) \Psi^{(10)\,2}_k \ ,\\
\xi_{k}^{i} \ =\ & ({\bf 1}\otimes \bar \eta^i_+) \Psi^{(10)\,1}_k +\tfrac18 (1\otimes \bar \eta^i_+ \gamma_k) \chi^{(10)\,1} \ , \\
\xi_{k+2}^{i} \ =\ & ({\bf 1}\otimes \bar \eta^i_-) \Psi^{(10)\,2}_k +\tfrac18 (1\otimes \bar \eta^i_- \gamma_k) \chi^{(10)\,2} \ , \\
\rho^{i} \ =\ &({\bf 1}\otimes \bar \eta^i_+ \gamma^a) \Psi^{(10)\,1}_a + \tfrac12 (1\otimes \bar \eta^i_+) \chi^{(10)\,1} \ , \\
\rho^{i+2} \ =\ &({\bf 1}\otimes \bar \eta^i_- \gamma^a) \Psi^{(10)\,2}_a + \tfrac12 (1\otimes \bar \eta^i_-) \chi^{(10)\,2} \ , \\
\lambda_{\alpha}^{i} \ =\ & \zeta^{\alpha}_{I} ({\bf 1}\otimes \bar \eta^i_+\gamma^a \omega^I_{bc} \gamma^{bc}) \Psi^{(10)\,1}_a \ , \\
\lambda_{\alpha}^{i+2} \ =\ & \zeta^{\alpha}_{I} ({\bf 1}\otimes \bar \eta^i_-\gamma^a \omega^I_{bc} \gamma^{bc}) \Psi^{(10)\,2}_a \ .
\end{aligned}\end{equation}
These fermions form the $4$ dilatini and $4n$ gaugini of $N=4$ supergravity. The dilatini could in principle be identified by identifying the linear combinations of the above fermions whose supersymmetry transformation involves a spacetime derivative of the complex scalar $\sigma = - \tfrac12 (b_{12} + \iu \e^{-\rho_2})$.

In the following, we will restrict our attention to the supersymmetry transformation of the gravitini, which in $N=4$ gauged supergravity reads
\begin{equation}
 \delta \Psi_\mu^a =   D_\mu \epsilon^a_+ + \tfrac{1}{3} A_1^{ab} \gamma_\mu \epsilon_{-\, b} + \dots \ .
\end{equation}
The dots indicate terms involving four-dimensional vector fields. Furthermore, the gravitino mass matrix $A_1$ is symmetric and has a a precise definition in terms of the four-dimensional embedding tensor and the sigma model vielbeins \cite{Schon:2006kz}. In the remainder of this Section, we will determine $A_1$ in terms of the internal geometry of $Y$.

The ten-dimensional supersymmetry parameters $\epsilon^{(10)\, i}$ are related to their four-dimensional counterparts $\epsilon^a$, $a=1,\dots, 4$, via
\begin{equation}
\epsilon^{(10)\, 1} = \epsilon^i \otimes \eta_{i+} + h.c. \ , \quad
\epsilon^{(10)\, 2} = \epsilon^{i+2} \otimes \eta_{i-} + h.c. \ .
\end{equation}
We can thus relate the four-dimensional to ten-dimensional supersymmetry variations by invoking the relation \eqref{gravitini_4d_10d}.
The supersymmetry variations of the ten-dimensional gravitini and dilatini in the Einstein frame read \cite{Bergshoeff:2001pv,Grana:2005ny}
\begin{align}
\label{10d_susy_gravitini}
 \delta \Psi^{(10)\, i}_M  = & D_M \epsilon^{(10)\, i} + \tfrac{1}{96} \e^{-\phi/2}(H_{NPQ} \Gamma_M^{NPQ}-9H_{MNP} \Gamma^{NP}) \Gamma^{11} \epsilon^{(10)\, i} \nonumber \\ &
 -\tfrac1{64} \e^{(5-2n)\phi/4} \sum_{n} \frac{1}{(2n)!}  ((2n-1)\Gamma_M^{M_1\dots M_{2n}} - 2n(9-2n)\delta_M^{M_1}\Gamma^{M_2\dots M_{2n}} ) \nonumber \\
 & \qquad \qquad \qquad \qquad F_{M_1 \dots M_{2n}} \Gamma_{11}^n (\sigma^1)^i_j \epsilon^{(10)\, j} \ , \\
\nonumber
 \delta \chi^{(10)\, i} = & (\partial_M \phi^{(10)}) \Gamma^M \epsilon^{(10)\, i}+ \tfrac{1}{96} H_{MNP} \Gamma^{MNP} \Gamma_{11} \epsilon^{(10)\, i} \\ & - \tfrac{1}{32}  \e^{(5-2n)\phi/4} \sum_{n} \frac{5-2n}{(2n)!} F_{M_1 \dots M_{2n}} \Gamma^{M_1\dots M_{2n}}  \Gamma_{11}^n (\sigma^1)^i_j \epsilon^{(10)\, j} \ , \label{10d_susy_dilatini}
\end{align}
where the $F_{2n}$ are the type IIA field strengths in the democratic formulation and the matrix $\Gamma_{11}$ is the chirality operator.
In the following, we only consider the scalar contribution that determine the matrix $A_1$.
From this and \eqref{gravitini_4d_10d}, we compute the four-dimensional gravitino variation to be
\begin{equation}\begin{aligned}
 \delta \Psi_\mu^i = & D_\mu \epsilon^i_+ + \tfrac12 (\bar \eta^i_+ (\gamma^m D_m + \tfrac{1}{24}\gamma^{mnp} H_{mnp} ) \eta_-^j) \gamma_\mu \epsilon_{-\, j}\\& + \tfrac18 \sum_{n} \tfrac{1}{(2n)!} (\bar \eta^i_+ \gamma^{m_1 \dots m_{2n}} F_{m_1 \dots m_{2n}} \eta_+^j) \gamma_\mu \epsilon_{-\, (2+j)}\ ,\\
 \delta \Psi_\mu^{(2+i)} = & D_\mu \epsilon^{(2+i)}_+ + \tfrac12 (\bar \eta_{-\, i} (\gamma^m D_m + \tfrac{1}{24}\gamma^{mnp} H_{mnp}) \eta_{+\, j}) \gamma_\mu \epsilon_{-\, (2+j)}\\&+ \tfrac18 \sum_{n} \tfrac{1}{(2n)!} (\bar \eta_{-\, i} \gamma^{m_1 \dots m_{2n}} F_{m_1 \dots m_{2n}} \eta_{-\, j}) \gamma_\mu \epsilon_{-\, j}\ ,
\end{aligned}\end{equation}
where the indices $m,m_i,n$ and $p$ run over the internal coordinates of $Y$.
 We see that the components of $A_1$ are given by
\begin{equation}\begin{aligned}
A_1^{ij}= & \tfrac34 \int_6 \e^{\phi} \bar \eta^i_+ (\gamma^m D_m + \tfrac{1}{24}\gamma^{mnp} H_{mnp} ) \eta_-^j \ ,\\
A_1^{i(2+j)}= & \tfrac{3}{32} \int_6 \e^{2\phi} \sum_{n} \tfrac{1}{(2n)!} \bar \eta^i_+ \gamma^{m_1 \dots m_{2n}} F_{m_1 \dots m_{2n}} \eta_+^j \ ,\\
A_1^{(2+i)j}= & \tfrac{3}{32} \int_6 \e^{2\phi} \sum_{n} \tfrac{1}{(2n)!} \bar \eta^i_- \gamma^{m_1 \dots m_{2n}} F_{m_1 \dots m_{2n}} \eta_-^j \ ,\\
A_1^{(2+i)(2+j)}= & \tfrac34 \int_6 \e^{\phi}  \bar \eta_{-\, i} (\gamma^m D_m + \tfrac{1}{24}\gamma^{mnp} H_{mnp} ) \eta_{+\, j} \ ,
\end{aligned}\end{equation}
where the additional factor of $\e^{\phi}$ comes from the Weyl rescaling \eqref{eq:Weyl}.
Note that the components are related by $A_1^{(2+i)(2+j)} = \bar A_1^{ij}$ and $A_1^{i(2+j)} = A_1^{(2+i)j}$.

We can now use the spinor bilinears \eqref{spinor_bilinears} to express $A_1^{ij}$ in terms of internal forms. More precisely, we have
\begin{equation}\begin{aligned}
\diff (\Psi_-)^j_i  =   [\gamma^m , D_m \eta_-^j \bar \eta_{-\, i}]_+  = &  (\gamma^m D_m \eta_-^j) \bar \eta_{-\, i} + (\gamma^m \eta_-^j) (D_m \bar \eta_{-\, i})\\ & + ( D_m \eta_-^j) (\bar \eta_{-\, i}\gamma^m) + \eta_-^j (D_m  \bar \eta_{-\, i}\gamma^m) \ .
\end{aligned}\end{equation}
Taking the product with the spinor bilinear $(\Psi_0)^{kl}$, this reads
\begin{equation}
 \e^{-\rho_2-\rho_4}\langle (\Psi_0)^{kl}, \diff (\Psi_-)^j_i \rangle = \delta^k_i \bar \eta^l_+ \gamma^m D_m \eta^j_- + \epsilon^{lj} \bar K^m (D_m \bar \eta_{-\, i}) \eta^k_- \ .
\end{equation}
Similarly, we find
\begin{equation}
\tfrac{1}{48} \bar \eta^i_+ \gamma^{mnp} H_{mnp}  \eta_-^j =  \e^{-\rho_2-\rho_4}\langle (\Psi_0)^{k(i} , H_3 \wedge (\Psi_-)^{j)}_k \rangle \ .
\end{equation}
This gives us an expression for $A_1^{ij}$ in terms of forms,
\begin{equation}
A_1^{ij} = \tfrac32 \int_6\e^{\phi-\rho_2-\rho_4} \langle (\Psi_0)^{k(i} , (\diff-H_3 \wedge) (\Psi_-)^{j)}_k \rangle \ .
\end{equation}
Inserting \eqref{spinor_bilinears} for the spinors bilinears and writing $A_1^{ij} = \tfrac{3}{4} \iu (\sigma_a)^{ij} P^a$, we find
\begin{equation} \label{eq:prep}
P^a =  \int_6 \e^{\phi-\rho_2-\rho_4} \bar K\wedge ( K \wedge \diff \bar K \wedge J^a + \epsilon_{abc} J^b \wedge \diff J^c -H_3 \wedge J^a) \ .
\end{equation}
These terms are in fact (up to complex conjugation) the only $SU(2)$-covariant one-derivative expressions that can be built out of $H_3$, $K$ and the $J^a$.
Similarly, we find for the off-diagonal entries 
\begin{equation}
A_1^{i(2+j)} = \tfrac34 \int_6 \e^{2\phi-(\rho_2+\rho_4)/2} \langle (\Psi_+)^i_j , F \rangle \ .
\end{equation}

Using \eqref{eq:prep}, we can now discuss the amount of supersymmetry the vacuum preserves depending on the internal geometry. We will here focus on the case of vanishing internal field strengths $F$ and $H_3$. Then, $A_1$ is block-diagonal and one can discuss each $N=2$ subsector individually. Since the two components of $A_1$ are identical up to complex conjugation, we can only have $N=4$, $N=2$ or non-supersymmetric vacua.\footnote{In the presence of Ramond-Ramond fluxes, vacua preserving $N=1$ or $N=3$ are no longer excluded.} For an $N=4$ Minkowski vacuum, we need both $K$ and the $J^a$ to be closed, and we find that the manifold is $K3 \times T^2$. If we want to have at least an $N=2$ supersymmetric vacuum, the discussion is rather similar to $N=2 \to N=1$ supersymmetry breaking as discussed in \cite{Louis:2009xd}. A requirement for an $N=2$ Minkowski vacuum therefore is that $A_1^{ij}$, given in \eqref{eq:prep}, should be of rank one. Supersymmetry should then impose that the variations of $A_1^{ij}$ have similar properties.

If we restrict ourselves to the supersymmetry related to $\eta_1$, then $P^-= \bar A_1^{11}$ becomes (up to a K\"ahler prefactor) the holomorphic superpotential of an $SU(3)$ structure defined by $\eta_1$, as first proposed in \cite{Gurrieri:2002wz}, with the definitions
\begin{equation} \label{eq:CYSU2structure_app}
J = J^3 + \tfrac12 \iu K \wedge \bar K  \ , \qquad  \Omega = K \wedge (J^1 + \iu J^2 ) \ .
\end{equation}
For this supersymmetry to be unbroken, the variations of $\bar A_1^{11}$ with respect to the spacetime scalars must hence be set to zero. This is equivalent to the Calabi-Yau conditions
\begin{equation}\label{CY_condition_app}
\diff J = 0 \ , \qquad \diff \Omega = 0 \ .
\end{equation}
Spontaneous partial supersymmetry breaking requires furthermore that $P^3$ and its analogues in the supersymmetry variations of the spin-$1/2$ particles vanish in the vacuum. From \eqref{eq:prep}, we find that
\begin{equation}
P^3 = \int_6 \e^{\phi-\rho_2-\rho_4} \bar \Omega \wedge {\cal L}_K \Omega + J \wedge J \wedge \diff \bar K \ .
\end{equation}
This vanishes on a Calabi-Yau manifold. From \eqref{fermions_4d_10d}, one can deduce that the terms appearing in the other fermion variations are not, as one might expect, variations of $P^3$ (which do not vanish). They are instead also proportional to \eqref{CY_condition_app}, thus ensuring that we find partial supersymmetry breaking on Calabi-Yau backgrounds.


\begin{thebibliography}{10}

\bibitem{Hitchin}
N.~Hitchin, ``The geometry of three-forms in six and seven
dimensions,'' J.\ Diff.\ Geom.\ {\bf 55} (2000), no.3 547 [arXiv: math.DG/0010054].

\bibitem{Hitchin:2001rw}
N. Hitchin, ``Stable forms and special metrics,''
in ``Global Differential Geometry: The Mathematical Legacy of Alfred
Gray'', M.Fernandez and J.A.Wolf (eds.),
Contemporary Mathematics {\bf 288}, American Mathematical Society,
Providence (2001) [arXiv:math.DG/0107101].

\bibitem{CS}
S.\ Chiossi and S.\ Salamon, ``The Intrinsic Torsion of $SU(3)$ and $G_2$
Structures,'' in \emph{Differential geometry, Valencia, (2001)}, pp. 115,
[arXiv: math.DG/0202282].

\bibitem{waldram}
J.~P.~Gauntlett, D.~Martelli, S.~Pakis and D.~Waldram,
  ``G-structures and wrapped NS5-branes,''
  Commun.\ Math.\ Phys.\  {\bf 247} (2004) 421
  [arXiv:hep-th/0205050].

\bibitem{Kaste:2003dh}
  P.~Kaste, R.~Minasian, M.~Petrini and A.~Tomasiello,
  ``Nontrivial RR two form field strength and SU(3) structure,''
  Fortsch.\ Phys.\  {\bf 51} (2003) 764
  [hep-th/0301063].

\bibitem{Gauntlett:2003cy}
  J.~P.~Gauntlett, D.~Martelli and D.~Waldram,
  ``Superstrings with intrinsic torsion,''
  Phys.\ Rev.\  D {\bf 69} (2004) 086002
  [arXiv:hep-th/0302158].

\bibitem{Hitchin:2004ut}
  N.~Hitchin,
  ``Generalized Calabi-Yau manifolds,''
  Quart.\ J.\ Math.\ Oxford Ser.\  {\bf 54} (2003) 281
  [arXiv:math/0209099].

\bibitem{Gualtieri:2003dx}
  M.~Gualtieri,
  ``Generalized complex geometry,''
  Ph.D.\ Thesis
  [arXiv:math/0401221].

\bibitem{Gurrieri:2002wz}
  S.~Gurrieri, J.~Louis, A.~Micu and D.~Waldram,
  ``Mirror symmetry in generalized Calabi-Yau compactifications,''
  Nucl.\ Phys.\ B {\bf 654} (2003) 61
  [hep-th/0211102].

\bibitem{Grana:2005ny}
  M.~Grana, J.~Louis and D.~Waldram,
  ``Hitchin functionals in N = 2 supergravity,''
  JHEP {\bf 0601} (2006) 008
  [arXiv:hep-th/0505264].\\
  M.~Grana, J.~Louis and D.~Waldram,
  ``SU(3) x SU(3) compactification and mirror duals of magnetic fluxes,''
  JHEP {\bf 0704} (2007) 101
  [arXiv:hep-th/0612237].

\bibitem{Grana:2004bg}
  M.~Grana, R.~Minasian, M.~Petrini and A.~Tomasiello,
  ``Supersymmetric backgrounds from generalized Calabi-Yau manifolds,''
  JHEP {\bf 0408} (2004) 046
  [hep-th/0406137].

\bibitem{Witt}
F.~Witt, ``Generalised $G_2$-manifolds'',
Commun.\ Math.\ Phys.\ {\bf 265} (2006) 275
[math.DG/0411642].\\
F.~Witt, ``Special metric structures and closed forms'',
Oxford University DPhil thesis (2004) [arXiv:math.DG/0502443].

\bibitem{Jeschek:2004wy}
  C.~Jeschek and F.~Witt,
  ``Generalised G(2)-structures and type IIB superstrings,''
  JHEP {\bf 0503} 053 (2005)
  [arXiv:hep-th/0412280].

\bibitem{Grana:2005sn}
  M.~Grana, R.~Minasian, M.~Petrini and A.~Tomasiello,
  ``Generalized structures of N=1 vacua,''
  JHEP {\bf 0511} 020 (2005)
  [arXiv:hep-th/0505212].


\bibitem{Candelas:1990rm}
  P.~Candelas, X.~C.~De La Ossa, P.~S.~Green and L.~Parkes,
  ``A Pair of Calabi-Yau manifolds as an exactly soluble superconformal theory,''
  Nucl.\ Phys.\ B {\bf 359} (1991) 21.
  %%CITATION = NUPHA,B359,21;%%


\bibitem{Antoniadis:1997eg}
  I.~Antoniadis, S.~Ferrara, R.~Minasian and K.~S.~Narain,
  ``R**4 couplings in M and type II theories on Calabi-Yau spaces,''
  Nucl.\ Phys.\ B {\bf 507} (1997) 571
  [hep-th/9707013].

\bibitem{Antoniadis:2003sw}
  I.~Antoniadis, R.~Minasian, S.~Theisen and P.~Vanhove,
  ``String loop corrections to the universal hypermultiplet,''
  Class.\ Quant.\ Grav.\  {\bf 20} (2003) 5079
  [hep-th/0307268].

\bibitem{ethomas}
E.~ Thomas, ``Vector fields on manifolds,'' Bull. Amer. Math. Soc. 75 (1969), 643-
683.



\bibitem{Bovy:2005qq}
  J.~Bovy, D.~L\"ust and D.~Tsimpis,
  ``N = 1,2 supersymmetric vacua of IIA supergravity and SU(2) structures,''
  JHEP {\bf 0508} (2005) 056
  [arXiv:hep-th/0506160].

\bibitem{ReidEdwards:2008rd}
  R.~A.~Reid-Edwards and B.~Spanjaard,
  ``N=4 Gauged Supergravity from Duality-Twist Compactifications of String Theory,''
  JHEP {\bf 0812} (2008) 052
  [arXiv:0810.4699 [hep-th]].

\bibitem{Lust:2009zb}
  D.~Lust and D.~Tsimpis,
  ``Classes of AdS(4) type IIA/IIB compactifications with SU(3) x SU(3) structure,''
  JHEP {\bf 0904} (2009) 111
  [arXiv:0901.4474 [hep-th]].

\bibitem{Triendl:2009ap}
  H.~Triendl and J.~Louis,
  ``Type II compactifications on manifolds with SU(2) x SU(2) structure,''
  JHEP {\bf 0907} (2009) 080
  [arXiv:0904.2993 [hep-th]].

\bibitem{Louis:2009dq}
  J.~Louis, D.~Martinez-Pedrera and A.~Micu,
  ``Heterotic compactifications on SU(2)-structure backgrounds,''
  JHEP {\bf 0909} (2009) 012
  [arXiv:0907.3799 [hep-th]].

\bibitem{Danckaert:2011ju}
  T.~Danckaert, J.~Louis, D.~Martinez-Pedrera, B.~Spanjaard and H.~Triendl,
  ``The N=4 effective action of type IIA supergravity compactified on SU(2)-structure manifolds,''
  JHEP {\bf 1108} (2011) 024
  [arXiv:1104.5174 [hep-th]].

\bibitem{Schulz:2012uj}
  M.~B.~Schulz,
  ``A class of Calabi-Yau threefolds as manifolds of SU(2) structure,''
  arXiv:1206.4027 [hep-th].

\bibitem{deWit:2005ub}
  B.~de Wit, H.~Samtleben and M.~Trigiante,
  ``Magnetic charges in local field theory,''
  JHEP {\bf 0509} (2005) 016
  [hep-th/0507289].

\bibitem{Schon:2006kz}
  J.~Schon and M.~Weidner,
  ``Gauged N=4 supergravities,''
  JHEP {\bf 0605} (2006) 034
  [arXiv:hep-th/0602024].

\bibitem{Cassani:2012pj}
  D.~Cassani, P.~Koerber and O.~Varela,
  ``All homogeneous N=2 M-theory truncations with supersymmetric AdS4 vacua,''
  JHEP {\bf 1211} (2012) 173
  [arXiv:1208.1262 [hep-th]].

\bibitem{MartinezPedrera:2009zz}
  D.~M.~Martinez Pedrera,
  ``Low-energy supergravities from heterotic compactification on reduced
  structure backgrounds,''
  CITATION = DESY-THESIS-2009-037;%%

\bibitem{Louis:2009xd}
  J.~Louis, P.~Smyth and H.~Triendl,
  ``Spontaneous N=2 to N=1 Supersymmetry Breaking in Supergravity and Type II String Theory,''
  JHEP {\bf 1002} (2010) 103
  [arXiv:0911.5077 [hep-th]].

\bibitem{Donagi:2008ht}
  R.~Donagi, P.~Gao and M.~B.~Schulz,
  ``Abelian Fibrations, String Junctions, and Flux/Geometry Duality,''
  JHEP {\bf 0904} (2009) 119
  [arXiv:0810.5195 [hep-th]].

\bibitem{Ferrara:1995yx}
  S.~Ferrara, J.~A.~Harvey, A.~Strominger and C.~Vafa,
  ``Second quantized mirror symmetry,''
  Phys.\ Lett.\  B {\bf 361} (1995) 59
  [arXiv:hep-th/9505162].

\bibitem{Barth:1984}
  W.~Barth, C.~Peters and A.~Van~de~Ven, ``Compact complex surfaces,'' Ergeb.\ Math.\
  Grenzgeb.\ {\bf (3)} 4, Springer-Verlag, Berlin, 1984.

\bibitem{Klemm:2005pd}
  A.~Klemm and M.~Marino,
  ``Counting BPS states on the enriques Calabi-Yau,''
  Commun.\ Math.\ Phys.\  {\bf 280} (2008) 27
  [hep-th/0512227].\\
  T.~W.~Grimm, A.~Klemm, M.~Marino and M.~Weiss,
  ``Direct Integration of the Topological String,''
  JHEP {\bf 0708} (2007) 058
  [hep-th/0702187 [hep-th]].

\bibitem{Gopakumar:1996mu}
  R.~Gopakumar and S.~Mukhi,
  ``Orbifold and orientifold compactifications of F - theory and M - theory to
  six-dimensions and four-dimensions,''
  Nucl.\ Phys.\  B {\bf 479} (1996) 260
  [arXiv:hep-th/9607057].

\bibitem{Bergshoeff:2001pv}
  E.~Bergshoeff, R.~Kallosh, T.~Ortin, D.~Roest and A.~Van Proeyen,
  ``New formulations of D = 10 supersymmetry and D8 - O8 domain walls,''
  Class.\ Quant.\ Grav.\  {\bf 18} (2001) 3359
  [arXiv:hep-th/0103233].

\bibitem{Ferrara:1983gn}
  S.~Ferrara and P.~van Nieuwenhuizen,
  ``Noether Coupling Of Massive Gravitinos To N=1 Supergravity,''
  Phys.\ Lett.\ B {\bf 127} (1983) 70.

\bibitem{Wagemans}
  M.~de Roo and P.~Wagemans,
  ``Partial Supersymmetry Breaking In N=4 Supergravity,''
  Phys.\ Lett.\ B {\bf 177} (1986) 352; \\
  P.~Wagemans,
  ``Breaking Of N=4 Supergravity To N=1, N=2 At Lambda = 0,''
  Phys.\ Lett.\ B {\bf 206} (1988) 241.

\bibitem{Tsokur:1994gr}
  V.~A.~Tsokur and Y.~.M.~Zinovev,
  ``Spontaneous supersymmetry breaking in N=4 supergravity with matter,''
  Phys.\ Atom.\ Nucl.\  {\bf 59} (1996) 2192
  [Yad.\ Fiz.\  {\bf 59N12} 2277 (1996)]
  [hep-th/9411104].

\bibitem{Andrianopoli:2002rm}
  L.~Andrianopoli, R.~D'Auria, S.~Ferrara and M.~A.~Lledo,
  ``Super Higgs effect in extended supergravity,''
  Nucl.\ Phys.\ B {\bf 640} (2002) 46
  [hep-th/0202116];\\
  L.~Andrianopoli, R.~D'Auria, S.~Ferrara and M.~A.~Lledo,
  ``Duality and spontaneously broken supergravity in flat backgrounds,''
  Nucl.\ Phys.\ B {\bf 640} (2002) 63
  [hep-th/0204145].

\bibitem{Dall'Agata:2009gv}
  G.~Dall'Agata, G.~Villadoro and F.~Zwirner,
  ``Type-IIA flux compactifications and N=4 gauged supergravities,''
  JHEP {\bf 0908} (2009) 018
  [arXiv:0906.0370 [hep-th]].

\bibitem{Dibitetto:2011gm}
  G.~Dibitetto, A.~Guarino and D.~Roest,
  ``Charting the landscape of N=4 flux compactifications,''
  JHEP {\bf 1103} (2011) 137
  [arXiv:1102.0239 [hep-th]].

\bibitem{HLS}
  C.~Horst, J.~Louis and P.~Smyth,
  ``Electrically gauged N=4 supergravities in D=4 with N=2 vacua,''
  arXiv:1212.4707 [hep-th].

\bibitem{cmap}
  S.~Cecotti, S.~Ferrara and L.~Girardello,
  ``Geometry of Type II Superstrings and the Moduli of Superconformal Field Theories,''
  Int.\ J.\ Mod.\ Phys.\ A {\bf 4} (1989) 2475.\\
  S.~Ferrara and S.~Sabharwal,
  ``Quaternionic Manifolds for Type II Superstring Vacua of Calabi-Yau Spaces,''
  Nucl.\ Phys.\ B {\bf 332} (1990) 317.


\bibitem{classification}
  D.~V.~Alekseevski\v\i, Math.\ USSR Izvestija 9 (1975) 297.\\
  B.~de Wit and A.~Van Proeyen,
  ``Special geometry, cubic polynomials and homogeneous quaternionic spaces,''
  Commun.\ Math.\ Phys.\  {\bf 149}, 307 (1992)
  [hep-th/9112027].

\bibitem{vafa}
  C.~Vafa,
  ``Evidence for F theory,''
  Nucl.\ Phys.\ B {\bf 469} (1996) 403
  [hep-th/9602022].\\
  D.~R.~Morrison and C.~Vafa,
  ``Compactifications of F theory on Calabi-Yau threefolds. 1,''
  Nucl.\ Phys.\ B {\bf 473} (1996) 74
  [hep-th/9602114].\\
    D.~R.~Morrison and C.~Vafa,
  ``Compactifications of F theory on Calabi-Yau threefolds. 2.,''
  Nucl.\ Phys.\ B {\bf 476} (1996) 437
  [hep-th/9603161].

\bibitem{fms}
  S.~Ferrara, R.~Minasian and A.~Sagnotti,
  ``Low-energy analysis of M and F theories on Calabi-Yau threefolds,''
  Nucl.\ Phys.\ B {\bf 474} (1996) 323
  [hep-th/9604097].

\bibitem{WT}
  W.~Taylor,
  ``TASI Lectures on Supergravity and String Vacua in Various Dimensions,''
  arXiv:1104.2051 [hep-th].

\end{thebibliography}
\end{document}